\documentclass[%
 reprint,
 amsmath,amssymb,
 aps,
 prb,
floatfix
]{revtex4-2}

\usepackage{dcolumn}
\usepackage{bm}
\usepackage{braket}
\usepackage{subcaption}
\usepackage{siunitx}
\usepackage{mathtools}
\usepackage{tikz} 
\usepackage[T1]{fontenc}

\begin{document}


\title{Decoherence of Coupled Flip-Flop Qubits Due to Charge Noise}

\author{John Truong}

\author{Xuedong Hu}%
 \email{xhu@buffalo.edu}
\affiliation{%
  Department of Physics\\
  SUNY at Buffalo\\
  jtruong@buffalo.edu
}
\date{\today}

\begin{abstract}

We study the decoherence effect of charge noise on a single flip-flop qubit and two dipole-coupled qubits~\cite{tosi_silicon_2017}. We find that a single flip-flop qubit is highly resistant to charge noise at its sweet spots. However, due to the proximity of the charge excited states to the flip-flop logical states, the presence of charge noise greatly reduces the fidelity of two-qubit operations. We identify leakage from the qubit Hilbert space as the main culprit for the reduced gate fidelity. We also explore different bias conditions to mitigate this decoherence channel.
\end{abstract}

\maketitle

\section{Introduction}

%

    Backed by the sophisticated fabrication technologies developed in the microelectronics industry, silicon is a promising host material for scalable quantum computers~\cite{divincenzo_physical_2000}.  Both charge and spin degrees of freedom can be employed as qubits in silicon.  While charge qubits have strong interactions and thus very fast gates, their fast decoherence~\cite{PhysRevB.69.113301,stavrou_charge_2005,PhysRevLett.105.246804,PhysRevLett.91.226804} and the general lack of pulse shaping technology in the terahertz regime make high-fidelity charge qubits in semiconductors extremely challenging.  On the other hand, decoherence times for individual spins are long in natural silicon, and are incredibly long in isotopically enriched $^{28}$Si~\cite{tyryshkin_electron_2012, zwanenburg_silicon_2013, pla_high-fidelity_2013,veldhorst_addressable_2014,pla_single-atom_2012,xiao_measurement_2010}, and exchange interaction in semiconductors is strong and can be electrically controlled~\cite{loss_quantum_1998, kane_silicon-based_1998, petta_coherent_2005}, making spins, whether electron \cite{loss_quantum_1998} or nuclear spins \cite{kane_silicon-based_1998} and whether quantum-dot-based or donor-based, an enticing qubit candidate.
%
%
    Recent years have seen tremendous experimental progress in the coherent control of spin qubits in silicon.  High-fidelity single qubit gates have been demonstrated in both quantum dot and donors~\cite{takeda_resonantly_2020,yoneda_quantum-dot_2018,asaad_coherent_2020,veldhorst_addressable_2014,sigillito_site-selective_2019,simmons_tunable_2011}. Two-qubit coupling and gating via exchange interaction have also been demonstrated~\cite{zajac_scalable_2016,watson_programmable_2018,veldhorst_two-qubit_2015}.  Furthermore, single-shot measurements of a single spin in silicon have been realized~\cite{morello_single-shot_2010, pla_high-fidelity_2013, zheng_rapid_2019, keith_single-shot_2019, zhao_single-spin_2019,west_gate-based_2019,crippa_gate-reflectometry_2019,gonzalez-zalba_probing_2015,hu_fast_2019}. 
    
    Some difficult challenges remain against a scalable spin-based quantum information processor, such as decoherence effects of charge noise~\cite{hu_charge-fluctuation-induced_2006,yoneda_quantum-dot_2018,struck_low-frequency_2019,huang_impact_2020}.  Another example is the lack of a viable means for long-range spin coupling and communication, which is highly desirable for a large scale multi-qubit device, while exchange interaction is short-ranged.
%
%
    Direct magnetic dipole coupling between spins is long-ranged, but is too weak for efficient information transfer.  A natural approach to allow long-range spin coupling is to hybridize it with charge and take advantage of the strong and long-ranged Coulomb interaction.  This approach has indeed been employed for capacitive coupling of singlet-triplet qubits~\cite{shulman_demonstration_2012}, and for enhancing spin-photon coupling in a cavity~\cite{mi_circuit_2017, mi_coherent_2018, samkharadze_strong_2018}, which would in turn allow long-range spin coupling mediated by cavity photons~\cite{borjans_split-gate_2020, borjans_long-range_2020}.
    %


    The flip-flop qubit proposed by Tosi et al. is an intriguing example of how to take advantage of spin-charge hybridization~\cite{tosi_silicon_2017}. The proposed hybrid qubit uses basis states where the donor nuclear and electron spins are pointing in opposing directions ($\ket{\uparrow\Downarrow}, \ket{\downarrow\Uparrow}$).  The donor electron is also allowed to tunnel to an interface quantum dot controlled by gate voltages.  The hybridization of the flip-flop spin states with the donor-dot charge qubit gives rise to coherent electrical control over the spin states~\cite{harvey-collard_coherent_2017,calderon_quantum_2006}.  Most importantly, spin-charge mixing allows long-distance qubit coupling via the long-ranged and tunable dipole-dipole coupling of the charge qubits, potentially satisfying the elusive scalability requirement for spin-based quantum computers.

    The benefits of a spin-charge hybrid qubit unfortunately comes with an inevitable degradation in its coherence properties. After all, charge qubits typically have short coherence times of a few nanoseconds~\cite{PhysRevB.69.113301, PhysRevLett.105.246804, Schoenfield2017,stavrou_charge_2005,li_conditional_2015} (some latest, well designed examples can reach above 100 ns~\cite{mi_strong_2017, thorgrimsson_extending_2017}), compared to spin coherence times upward of several seconds~\cite{zwanenburg_silicon_2013,veldhorst_addressable_2014, dehollain_single-shot_2014}. In the flip-flop qubit design, this issue seems to have been addressed with the second-order sweet spot, where charge-noise-induced single-qubit dephasing is greatly suppressed~\cite{tosi_silicon_2017}.  Further theoretical studies on the decoherence of the nuclear spin of the single-qubit system due to phonons and $1/f$ charge noise~\cite{boross_valley-enhanced_2016} and on a two-electron version of the system~\cite{hetenyi_hyperfine-assisted_2019,huang_spin_2018} have shown that a single flip-flop qubit is highly coherent.
    %
    %

    In this paper, we perform a comprehensive study of the effects of charge noise on the flip-flop qubit.  Our focus is especially on coupled qubits, which could open up new decoherence channels as evidenced in exchange-coupled single-spin qubits \cite{hu_charge-fluctuation-induced_2006, hung_hyperfine_2013}. In particular, we examine the effect of $1/f$ charge noise, which is known as a major source of decoherence in solid state qubit systems~\cite{RevModPhys.53.497, paladino_1/f_2014, PhysRevLett.97.076803, PhysRevB.100.165305}. We find that for a single-qubit, as expected, the second-order sweet spot provides great coherence times, in excess of $\SI{1}{ms}$ over a reasonable range of applied electric fields within control precision.  However, once multiple qubits are coupled together via the dipole-dipole interaction, in the parameter regime optimized for single-qubit coherence, multi-qubit coherence times are greatly reduced due to leakage in the charge qubit sector.  One way to minimize charge leakage is to increase the frequency detuning between the charge and flip-flop spin qubits, and we have indeed identified a multi-qubit parameter regime that promises excellent coherence properties at the expense of only a small increase in gate times.

    While our studies are focused on coupled flip-flop qubits, our results should be of general interest in the exploration of using less coherent objects (charge qubits in this case) to mediate coupling between highly coherent qubits (flip-flip spin states at the donor sites), with the goal of minimizing the decoherence effects of the mediator while enhancing the coupling between the highly coherent qubits.

    The rest of the paper is organized as follows. First we lay out the theoretical model and basis for the charge and flip-flop qubits. Then we introduce the dipole-dipole coupling and present an effective Hamiltonian for two flip-flop qubits. We follow that with a general solution for the time evolution of our system under the influence of a classical electrical noise. Lastly, we combine everything together and present results for the time evolution of the single and two-qubit density matrices.

\section{Model}

    The flip-flop qubit is a hybrid of a spin and a charge qubit~\cite{tosi_silicon_2017}. The physical system is a phosphorus donor implanted in an isotopically enriched silicon substrate close to a Si/SiO$_2$ interface. A surface metal gate induces an interface quantum dot (iQD), and is also used to tune the applied electric field in the growth($z$)-direction.  The bare flip-flop states for the electron and nuclear spins on the donor, $\ket{\uparrow\Downarrow}$ and $\ket{\downarrow\Uparrow}$, are isolated from the polarized states and are highly coherent.  However, as a pure spin qubit, albeit encoded in a two-spin state, it suffers from the same issues with respect to long-distance communication.  The solution provided by the flip-flop qubit design is to adjust the surface gate potentials so that the donor electron can tunnel to the iQD with a tunable magnitude.  The electron locating at the iQD or the donor would form the basis for a charge qubit.  With hyperfine interaction only present on the donor site, and the electron $g$-factor different between the donor and iQD sites, the spin and charge qubits are coupled by these magnetic inhomogeneities, which allow electrical control of the spin states.  The dressed flip-flop qubit is then defined as the lowest two eigenstates of this coupled spin-charge system.  Furthermore, the spin-charge mixing makes it possible to couple different flip-flop qubits together via the Coulomb interaction.  In this study we focus on coupling mediated by the electric dipole interaction. This, however, is not the only possibility. Another coupling mechanism that could be utilized is through a microwave cavity~\cite{tosi_silicon_2017}.
    %

    %

    In this section, we establish the effective Hamiltonian for single and coupled flip-flop qubits, clarify the spectrum, and examine how a qubit or two coupled qubits can be affected by electrical noises.

    \begin{figure}
    \centering
    \includegraphics[width=\columnwidth]{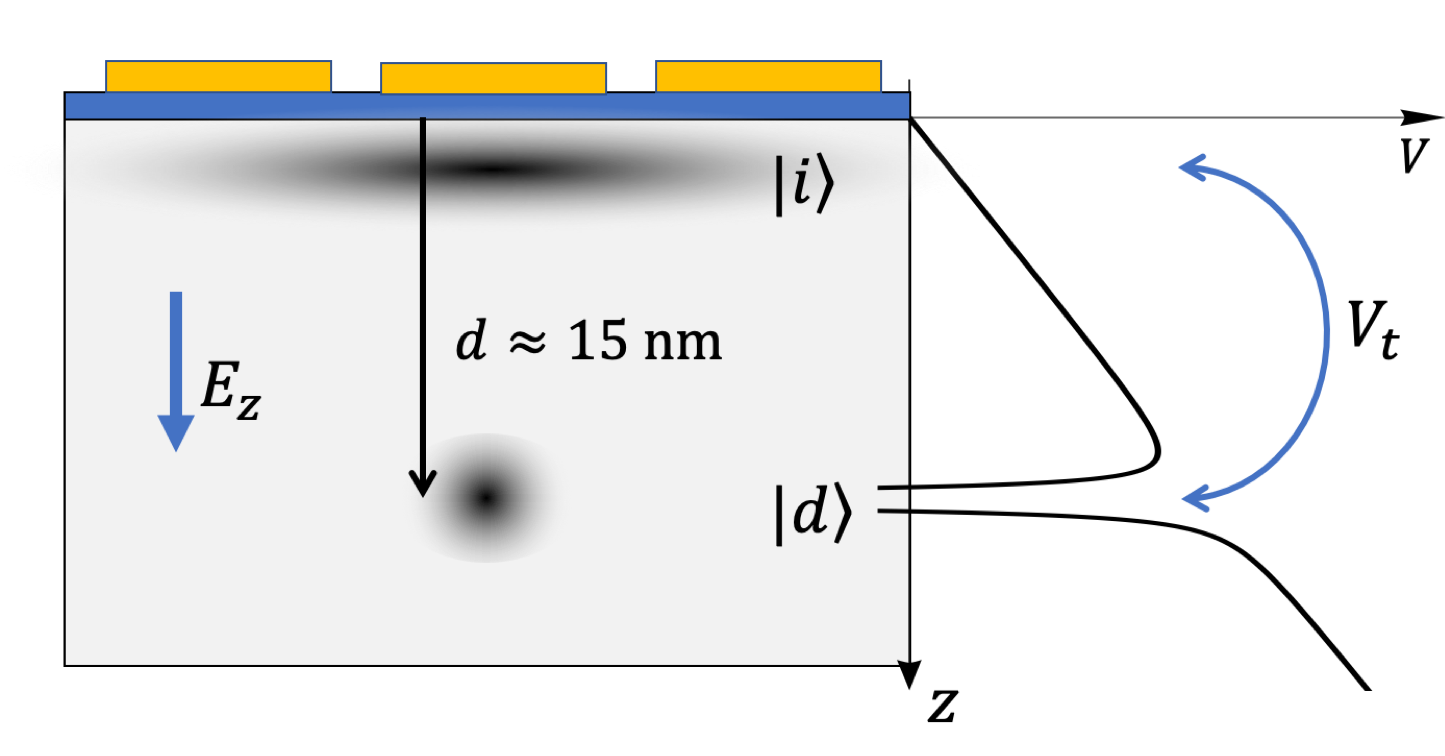}
    \caption{Schematic diagram of donor-dot hybrid system. A phosphorus donor is implanted into a silicon substrate. A metal top gate is used to move the electron position between the donor state and the interface state while nearby gates are used to shift the interface dot laterally, modulating the tunnel coupling between the two states.}
    \label{diagram}
\end{figure}
    \begin{figure*}[thb]
    \centering

    \begin{subfigure}{0.32\textwidth}
        \caption{}
        \includegraphics[width=\textwidth, keepaspectratio]{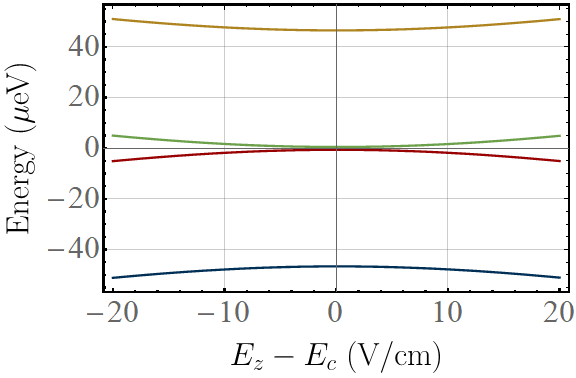}
    \end{subfigure}
    \begin{subfigure}{0.32\textwidth}
        \caption{}
        \includegraphics[width=\textwidth, keepaspectratio]{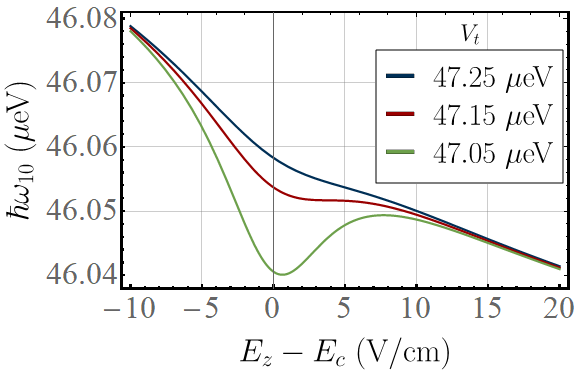}
    \end{subfigure}
    \begin{subfigure}{0.32\textwidth}
        \caption{}
        \includegraphics[width=\textwidth, keepaspectratio]{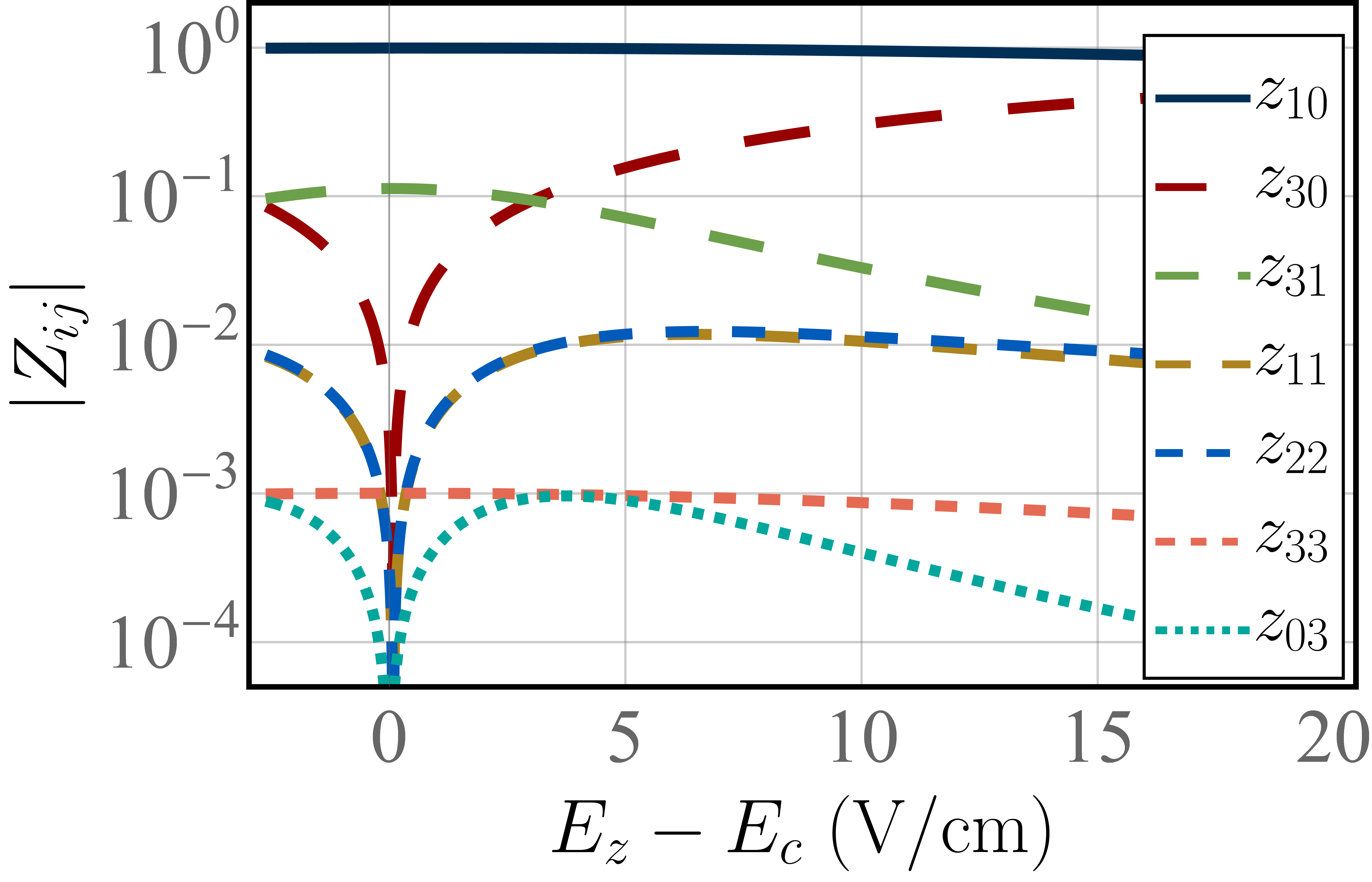}
    \end{subfigure}
    \caption{(a) The flip-flop qubit energy spectrum for $V_t = \SI{47.15}{\mu eV}$ and $B = \SI{0.796}{T}$. The two lower states constitute the logical basis $\ket{0}$ and $\ket{1}$ while the the upper two states (i.e. the charge excited states) form the leakage states $\ket{2}$ and $\ket{3}$. The energy splitting between states $\ket{1}$ and $\ket{2}$ is approximately equal to $\hbar(\omega_0 - \omega_B)$. (b) The energy difference between the two logical states. With certain operational parameters, the second derivative of the energy difference, $\partial^2\omega_{10}/\partial \epsilon^2$, becomes zero and we have a second-order sweet spot. This point should be highly resistant to fluctuations in the electric field detuning parameter, $\epsilon$. (c) Magnitudes for the largest coefficients of operator $Z$ in the dressed flip-flop basis. The flip-flop sweet spot occurs when $z_{03}+z_{33}=0$.}
    \label{1q-quantities}
\end{figure*}

    \subsection{Charge Qubit}

        With the donor and the iQD tunnel coupled, the donor electron forms a charge qubit based on these two sites.  In the electron position basis (iQD ground orbital state $\ket{i}$ and donor ground state $\ket{d}$), the charge qubit Hamiltonian is
        \begin{equation}
            H_\textrm{charge}^\textrm{id} = -\frac{1}{2}\epsilon Z + \frac{1}{2}V_tX
            \label{hamiltonian_pos_basis}
        \end{equation}
        where $Z=\ket{i}\bra{i} - \ket{d}\bra{d}$ and $X=\ket{i}\bra{d} + \ket{d}\bra{i}$ are the usual Pauli matrices in the position basis. The interface electric field causes a detuning $\epsilon=e(E_z-E_c)d$ between the two locations, where $E_c$ is the critical electric field when $\epsilon = 0$, so that the electron is shared equally between the donor and the iQD sites. We define the charge qubit basis to be the eigenbasis of the above Hamiltonian, yielding charge qubit states:
        \begin{subequations}
            \begin{align}
                \ket{g} &= \cos(\eta/2)\ket{i} - \sin(\eta/2)\ket{d}\\
                \ket{e} &= \sin(\eta/2)\ket{i} + \cos(\eta/2)\ket{d}
            \end{align}
        \end{subequations}
        with energy difference $\omega_0 = \sqrt{\epsilon^2 + V_t^2}$ and mixing angle $\tan\eta = V_t/\epsilon$. This qubit has a first order sweet spot at the anticrossing point $\epsilon = 0$ with respect to the detuning, where $\partial\omega_0/\partial \epsilon = 0$.

    \subsection{Flip-Flop Qubit}

        In the presence of the donor contact hyperfine interaction, the antiparallel electron and donor nuclear spin states $\ket{\uparrow\Downarrow}$ and $\ket{\downarrow\Uparrow}$ are decoupled from the two parallel-spin states, and form the bare flip-flop qubit.
        %

        %
        In the product basis between the charge qubit and the bare flip-flop qubit, and in the presence of an applied magnetic field, the total donor-iQD single-electron Hamiltonian is:
        \begin{equation}
            H=H_\textrm{charge}+H_B+\Delta H_B+H_A \,,
        \end{equation}
        where the individual terms are given by
        \begin{subequations}
            \begin{align}
                H_\textrm{charge} &= -\frac{1}{2}\omega_0 \sigma_z \,, \\
                              H_B &= -\frac{1}{2}\omega_B\tau_z \,, \\
                       \Delta H_B &= -\frac{1}{4}\Delta\omega_B\left(1+\cos\eta\sigma_z+\sin\eta\sigma_x\right)\tau_z \,, \\
                              H_A &= -\frac{1}{8}A\left(1-\cos\eta\sigma_z-\sin\eta\sigma_x\right)\left(1-2\tau_x\right) \,.
            \end{align}
        \end{subequations}
        Here $\sigma_i$ ($\tau_j$) are the Pauli operators in the charge qubit (bare flip-flop qubit) basis. $H_\textrm{charge}$ describes the bare charge qubit. $H_B$ is the total Zeeman energy if the electron is at the donor, and includes contributions from both the electron and nuclear spins. $\Delta H_B$ accounts for the positional dependence of the electron gyromagnetic ratio, which manifests itself as a difference in the electron Zeeman energy at the two locations, $\Delta\omega_B = \omega_{B,e}^\textrm{dot}-\omega_{B,e}^\textrm{donor}$. This difference is usually small. To compare with $H_B$, we can define a ratio $\Delta\omega_B/\omega_B = \Delta_\gamma\gamma_e/(\gamma_e+\gamma_n) \sim \Delta_\gamma$, where $\gamma_n/2\pi = \SI{17.23}{MHz/T}$ and $\gamma_e/2\pi = \SI{29.97}{GHz/T}$ are the nuclear and electron gyromagnetic ratios.  According to an atomistic calculation, $\Delta_\gamma < 0.7\%$~\cite{rahman_gate-induced_2009}, thus $\Delta\omega_B/\omega_B \lesssim 0.7\%$. Lastly, $H_A$ is the hyperfine interaction between the electron and nuclear spin at the donor site, with $A/h = \SI{117}{MHz}$.

        The flip-flop qubit is defined by the lowest two eigenstates of the total donor-iQD system, which are essentially the bare flip-flop basis dressed by the charge qubit. Under the condition that $|\omega_0-\omega_B| > A/4$, we can use nondegenerate perturbation theory to obtain the flip-flop eigenenergies and eigenstates, given in Appendix A.  In particular, the energies of the flip-flop qubit states at the lowest order (in terms of $A/(\omega_0 - \omega_B)$) are
        \begin{subequations}
            \begin{align}
                E_0 &= \frac{1}{2}(-\omega_0-\omega_B)-\frac{A}{8} \left(1-\cos \eta\right)-\frac{\Delta\omega_B}{4}\left(1+\cos \eta\right) \,, \\
                E_1 &= \frac{1}{2}(-\omega_0+\omega_B)-\frac{A}{8} \left(1-\cos \eta\right)+\frac{\Delta\omega_B}{4}\left(1+\cos \eta\right) \,.
            \end{align}
        \end{subequations}
        A typical single-flip-flop-qubit spectrum is plotted in Fig.2(a).  A notable feature of this spectrum is how close the first excited state (the excited state of the flip-flop qubit) is to the second excited state (a charge excited state) energetically.  As such, if a perturbation can couple these two states, it could lead to significant leakage for the qubit.  Another important feature is that the two qubit states are roughly parallel to each other when the interface electric field changes, because the basis states mostly consist of the bare flip-flop spin states.  However, the dressing from the charge qubit states does introduce an electric field dependence into the qubit energy splitting, as is shown in Fig.2(b).

        In Fig.2(b) we plot the flip-flop qubit energy splitting as a function of the interdot detuning. Clearly, the dressing by the charge qubit means that charge noise affecting detuning could cause dephasing to the flip-flop qubit.  However, as the figure shows, a sweet spot against charge noise can be achieved for the flip-flop qubit, when the derivative of the energy difference $\omega_{10}=E_1 - E_0$ is zero. This occurs when the following equation is satisfied:
        \begin{equation}
            \frac{\Delta\omega_B}{\omega_0} + \frac{A^2\omega_0^2\epsilon}{2(\omega_0^2-\omega_B^2)^2\omega_B} = 0 \,.
        \end{equation}
        Furthermore, a second-order sweet spot is possible as well when the second derivative is zero, as is the case in the figure when $V_t =\SI{47.15}{\mu eV}$ and $B = \SI{0.796}{T}$.  The sweet spot does disappear beyond a certain range of tunnel coupling and magnetic field.

        The flip-flop qubit energy spectrum shows that this is a qubit design that can resist charge-noise-induced dephasing with its sweet spot, so that as a single qubit it can be highly coherent, as long as whatever environmental coupling does not lead to leakage from the qubit subspace.

        In preparation for describing two-qubit coupling, we express the electron position operator $Z$ in the dressed flip-flop
        basis to obtain $Z = \sum_{jk}z_{jk}\sigma_{j}'\tau_{k}'$.  Here $\sigma_{j}'\tau_{k}'$ operates on the eigenbasis that spans the $4\times4$ dressed flip-flop Hilbert space. For example, $\sigma_x'=\ket{2}\bra{0}+\ket{0}\bra{2}+\ket{3}\bra{1}+\ket{1}\bra{3}$. In the far-detuned regime, $\sigma'_j$ would correspond to the Pauli matrices operating primarily on the charge states while $\tau'_k$ on the spin states.  Figure~\ref{1q-quantities}(c) gives the magnitudes of the coefficients of $Z_i$.

        Some particular coefficients to take note of are $z_{10}$ are $z_{31}$. The former allows pure charge excitations, while the latter provides spin operations while modifying the phase of the charge sector. We will show in the next section that these two together form the basis for two qubit operations when electric dipole coupling is introduced. The coefficient $z_{30}$ corresponds to single-qubit charge dephasing when electrical noise is introduced.  Despite being the second largest at times, it does not play a significant role in flip-flop qubit coupling since our primary interest is between the $\ket{0}$ and $\ket{1}$ states which are spin-dominated. Where the spin and charge qubits are mixed more, $z_{30}$ is small. It does however become important in determining the decoherence effect of noise on the system.


    \subsection{Electric Dipole Coupling of Flip-Flop Qubits}
    \label{subsection-dipole}

        The spin-charge mixing in a flip-flop qubit means that two such qubits can couple via the electric dipole interaction. By pulling the electron away from the donor via the applied electric field, we create a dipole moment for each qubit equal to $p_i = ed_i(1+Z_i)/2$, with $i=1,2$ referring to the two qubits, and $d$ the donor-dot separation. The dipole interaction Hamiltonian takes the form:
        \begin{equation}
            H_{dip} = \frac{p_1p_2}{4\pi\epsilon_r\epsilon_0r^3} = V_{dd}(1+Z_1+Z_2+Z_1Z_2) \,,
        \end{equation}
        with a dipole coupling strength $V_{dd} = e^2d_1d_2/16\pi\epsilon_r\epsilon_0r^3$, where $r$ is the separation of the two qubits measured between the corresponding donors. The first term in the Hamiltonian is an overall constant energy shift and can be ignored. The two single-qubit terms amount to a constant electric field shift applied at each donor site due to the other qubit irrespective of its state. These terms can be negated simply by increasing the applied electric field at each donor site by an amount $\Delta E_i = 2V_{dd}/ed_i$. We can therefore simplify the dipole interaction Hamiltonian to
        \begin{equation*}
            H_{dip} = V_{dd}Z_1 Z_2 \,.
        \end{equation*}
        Substituting in the definition for $Z_i$ and omitting negligible terms (keeping only the $z_{10}$ and $z_{31}$ terms), we obtain the final dipole interaction Hamiltonian in the single-qubit eigenbasis.  In the following calculations, we only consider symmetric operating parameters between the two donors ($\epsilon_1 = \epsilon_2$, etc.) to further simplify the two-qubit Hamiltonian in the $16\times16$ Hilbert space.  The corresponding effective Hamiltonian for the dipole coupling can thus be expressed as:
        \begin{multline}
            H_{dip} = V_{dd}\left( z_{31}^2 \sigma_{1z}'\tau_{1x}'\sigma_{2z}'\tau_{2x}' + z_{10}z_{31}\sigma_{1x}'\sigma_{2z}'\tau_{2x}' \right.\\\left.  +z_{10}z_{31}\sigma_{1z}'\tau_{1x}'\sigma_{2x}'+ z_{10}^2\sigma_{1x}'\sigma_{2x}' \right)
            \label{Hdip}
        \end{multline}
        To simplify the notation, we introduce the following coupling rates
        \begin{subequations}
            \begin{align}
                g_f &= V_{dd}z_{31}^2 \,, \\
                g_l &= V_{dd}z_{31}z_{10} \,, \\
                g_c &= V_{dd}z_{10}^2 \,.
            \end{align}
        \end{subequations}

        The dipole coupling Hamiltonian $H_{dip}$ forms the foundation for an iSWAP/XX gate between the two dressed flip-flop qubits, where the difference between the two gates is in the behavior of the $\ket{00}$ and $\ket{11}$ states. Due to the large energy difference between the two states compared to the coupling in parameter regimes we are interested in, their evolutions are constant when compared to the $\ket{01}$ and $\ket{10}$ states. This is then primarily an iSWAP gate and we will refer to it as such. The $g_f$ term gives rise to a direct swap between states $\ket{01}$ and $\ket{10}$.  The $g_l$ terms lead to charge leakage. They facilitate transitions $\ket{01}\leftrightarrow\ket{20}$ and $\ket{10}\leftrightarrow\ket{02}$ where the spin state of one qubit and the charge state of the other qubit flips. The $g_c$ term gives a swap between charge excitations $\ket{02}$ and $\ket{20}$. Under most operating parameters, $g_c \approx 10 g_l \approx 100 g_f$. By decreasing the detuning between the charge and flip-flop qubits, however, it is possible to increase $g_f$ enough so that $g_c \approx 10 g_f$.  Notice that while $g_f$ is the smallest among the couplings, the system is partially protected from leakage by the energy gap between states $\ket{1}$ (excited qubit state) and $\ket{2}$ (lowest energy excited state outside the qubit space) given roughly by $\omega_B - \omega_0$.
        %
        %
        %

        One interesting feature of the dipole coupling Hamiltonian $H_{dip}$ is the absence of Ising type ($ZZ$) interaction terms for the flip-flop qubits.  This term does exist in the full coupling Hamiltonian, though its magnitude is about $10^4$ times smaller than the $XX$ interaction included above, and too small for a useful quantum gate.  As long as the two flip-flop qubits are put in resonance, the Ising type coupling can be neglected.

        The form of $H_{dip}$ indicates that transitions between $\ket{01}$ and $\ket{10}$ consist of two main processes, the direct process due to the $g_f$ term, and an indirect one via the charge excited state.  Starting from $\ket{01}$, the indirect process to get to state $\ket{10}$ is then
        \begin{equation*}
            \ket{01} \longrightarrow \ket{20} \longrightarrow \ket{02} \longrightarrow \ket{10}
        \end{equation*}
        Since the direct process avoids excitations into the charge leakage states, it should be much more robust against charge noise when compared to the indirect process, as we will demonstrate below.

    \subsection{Charge Noise Coupling to a Flip-Flop Qubit}
        \begin{figure*}[bth]
    \centering
    \includegraphics[width=\textwidth]{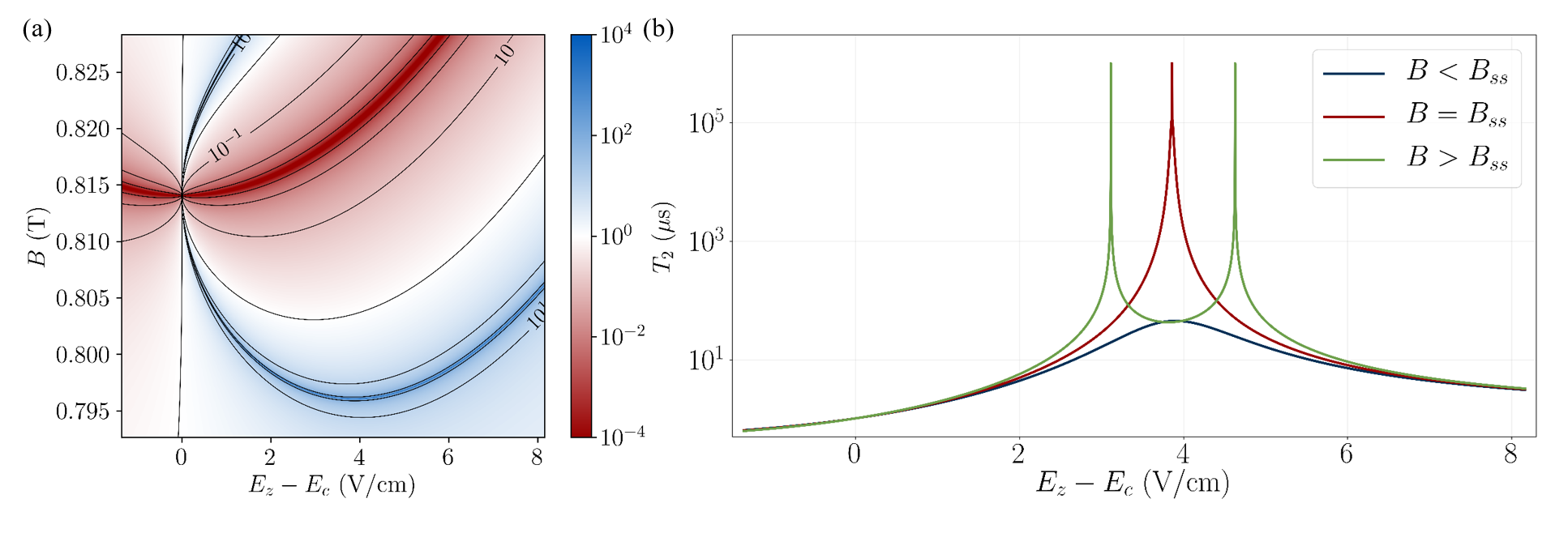}
    \caption{(a) Characteristic dephasing time for single flip-flop qubit calculating by solving $\left|\rho_{01}(T_2)/\rho_{01}(0)\right|=1/e$. The blue band marks the presence of the flip-flop sweet spot with the turning point at the bottom indicating the second-order sweet spot. The red line is where the charge splitting $\omega_0$ and the Zeeman splitting $\omega_B$ are degenerate. (b) Horizontal slices at, above, and below the second-order sweet spot. While the first-order sweet spots can yield equally excellent results compared to the second-order spot, they are much narrower in width, requiring higher experimental precision.}
    \label{single_qubit_dephase}
\end{figure*}

        The hybridization of spin and charge degrees of freedom opens up the flip-flop qubit to decoherence coming from electrical fluctuations. Two typical types of electrical fluctuations are those from lattice vibrations (phonons) and from background charge fluctuations.  The latter usually has a $1/f$ spectral form at low frequencies and a white spectrum at the intermediate to higher frequencies.  Since spin decoherence is typically dominated by pure dephasing~\cite{lutchyn_quantum_2008, huang_electron_2014, culcer_dephasing_2009},
        %
        %
        and dephasing is in turn determined by low-frequency noises, it is quite natural to find $1/f$ charge noise as the dominant source of decoherence for a single flip-flop qubit~\cite{hetenyi_hyperfine-assisted_2019}.  As such, we focus on the charge noise in this work, and consider a classical charge noise in the form:
        \begin{equation}
            H_n = \sum_i f_i(t)h_{i} \,,
        \end{equation}
        where the summation is over all independent ways the noise affects the system. The spectral properties of the noise are captured in the fluctuating dimensionless parameters $f_i(t)$, while $h_{i}$ contain information on both the strength of the noise and the form of the interaction between the noise and the system.  More specifically, $f_i(t)$ satisfies $\langle f_i(t)\rangle = 0$ and $\langle f_i(t_1)f_j(t_2)\rangle = \delta_{ij}\langle f_i(t_1)f_i(t_2)\rangle$.  In other words, different noise channels are uncorrelated with each other.  In this study, we consider $1/f$ noise with spectral density $S(\omega) = S_0/|\omega|$ for $\omega_l \le |\omega| \le \omega_h$, where $\omega_l$ ($\omega_h$) is the low(high)-frequency cutoff. The normalization constant $S_0$ is chosen so that the correlation function, $S(t)=\int_{-\infty}^{\infty}S(\omega)\exp(-i\omega t)d\omega = \langle f(0)f(t)\rangle$ is equal to $1$ when $t=0$.

        The dominant channel for electrical noise to affect the donor-dot charge qubit is via a fluctuating electric field in the $z$-direction (along the donor-dot axis for each qubit) that affects the charge qubit detuning:
        \begin{equation}
            h_{i} = \frac{1}{2}\omega_{ni}Z_i \,,
        \end{equation}
        with strength $\omega_{ni}\approx \SI{1}{\mu eV}$ corresponding to electric field fluctuations with rms strength of approximately $\SI{1}{V/cm}$~\cite{tosi_silicon_2017}. Omitting non-important coefficients of $Z_i$, we can write this noise operator as
        \begin{multline}
            h_{i} = \frac{1}{2}\omega_{ni}\left( z_{30}\sigma_{zi}'+z_{03}\tau_{zi}'+z_{33}\sigma_{zi}'\tau_{zi}' \right.\\
            \left.+z_{11}\sigma_{xi}'\tau_{xi}'+z_{22}\sigma_{yi}'\tau_{yi}' \right) \,.
        \label{eq_noise_hamiltonian}
        \end{multline}
        The first three terms cause dephasing of the qubit, while the last two result in leakage from the qubit space, specifically in charge excitation.
        %
        %

        %
        %

    \subsection{Time Evolution of Density Matrix}
    \label{time-evo}

        Incorporating the effect of charge noise on our our flip-flop qubit system, the total Hamiltonian can be formally written as
        \begin{equation*}
            H(t) = H_0 + \sum_i f_i(t) h_i \,.
        \end{equation*}
        The von-Neumann equation for the system density operator is given by
        \begin{equation}
            \frac{\partial \rho(t)}{\partial t} = -i\left[H(t),\rho(t)\right] \,.
        \end{equation}
        By means of a cumulant expansion up to the second order, the solution can be expressed as~\cite{yang_high-fidelity_2019,kubo_generalized_1962}
        \begin{multline}
            \langle\vec{\rho}(t)\rangle = \sum_{jk} e^{-i\omega_{jk}t}(R\otimes R)(\ket{j}\bra{j}\otimes\ket{k}\bra{k})\\\times e^{-\sum_iK_i(t)}(R^{-1}\otimes R^{-1})\vec{\rho}(0) \,.
            \label{general_rho}
        \end{multline}
        Additional details can be found in the appendix and in Ref.~\cite{yang_high-fidelity_2019}. Here $R$ is the rotation matrix that diagonalizes the noiseless system Hamiltonian $H_0$, and the $K_i(t)$ are matrices describing how the system is affected by the noise $f_i(t)h_i$. We further define a decay profile function $J(t,\omega_1,\omega_2)$ given by:
        \begin{equation}
            J(t,\omega_1,\omega_2) = \int_0^tdt_1\int_0^{t_1}dt_2 S(t_1-t_2)e^{i\omega_1t_1}e^{i\omega_2t_2} \,,
        \end{equation}
        where $S(t)$ is the time correlation function for the noise.

        For $1/f$ noise, zero-frequency noise ($\omega_1=\omega_2=0$) will be dominant in this function. Thus, we make an approximation by neglecting all non-zero $\omega_1$ and $\omega_2$ (see Appendix C for further details).
        %
        %
        %
        Under this approximation, $K_i(t)$ is diagonal and can be expressed as:
        \begin{equation}
            \bra{jk}K_i(t)\ket{jk} = J_i(t,0,0)\Gamma_{ijk}^2 \,.
        \end{equation}
        An exact expression for $K(t)$ is given in Appendix C. For $1/f$ noise in the long time limit ($1/\omega_h \ll t \ll 1/\omega_l$), $J(t,0,0)$ is approximately quadratic:
        \begin{equation}
            J_{1/f}(t,0,0) \approx \left(\frac{2-\gamma_E-\log(\omega_lt)}{2\log(\omega_h/\omega_l)}\right)t^2 \,,
        \end{equation}
        where $\gamma_e = 0.577$ is Euler's constant. We also define the dephasing rate between states $\ket{j}$ and $\ket{k}$ via noise channel $i$:
        \begin{equation}
            \Gamma_{ijk} = \bra{j}R^{-1}h_iR\ket{j}-\bra{k}R^{-1}h_iR\ket{k} \,.
        \end{equation}
        The solution for any element of the density matrix can then be written down as:
        \begin{multline}
            \rho_{ab}(t) = \sum_{jk}\sum_{mn}\exp\left(-i\omega_{jk}t-\sum_iJ_i(t,0,0)\Gamma_{ijk}^2\right)\\\times R_{aj}R^{-1}_{jm}R_{bk}R^{-1}_{kn}\rho_{mn}(0) \,.
            \label{rho}
        \end{multline}

        In the case where the system Hamiltonian $H_0$ is diagonal, the charge noise leads to pure dephasing among system eigenstates. When the system Hamiltonian is not diagonal, however, both dephasing and relaxation could be present depending on the form of the noise interactions $h_i$.
        %

        %
        %

        In order to extract a decoherence time from Equation~\ref{rho}, we extract the envelope of the fast oscillating off-diagonal element between two states, subtract its long time limit, and rescale it, yielding
        \begin{equation}
            F_{ab}(t) = \frac{\rho_{ab}(t)-\rho_{ab}(t\rightarrow \infty)}{\rho_{ab}(0)-\rho_{ab}(t\rightarrow \infty)}\,.
            \label{envelope}
        \end{equation}
        The coherence time $T$ can then be computed by setting $F_{ab}(T)$ equal to the desired threshold value, typically $1/e$.

\section{Results}
\subsection{Single Qubit Dephasing}

    Any qubit proposal has to meet the basic condition that a single qubit should be highly coherent. As we mentioned above, for a flip-flop qubit based on spin-charge hybridization, single-qubit coherence is most probably limited by charge noises, more specifically dephasing due to low-frequency charge noises reflected in the detuning between the donor and the iQD.  As such a single flip-flop qubit should have great coherence as long as it is kept at or near the second order sweet spot.
    
    Here we examine the robustness of the single-qubit coherence near the sweet spots.  Focusing on the electric field fluctuations along the donor-dot axis, the single-qubit Hamiltonian with noise is given by:
    \begin{equation}
        H_{1q}(t) = \sum_{j=0}^{3}E_j\ket{j}\bra{j} + \frac{1}{2}f(t)\omega_n Z \,.
    \end{equation}
    With the noiseless part of the Hamiltonian already diagonal, $R = I_4$, and Eq.~(~\ref{rho}) can be simplified as
    \begin{equation}
        \left|\frac{\rho_{ab}(t)}{\rho_{ab}(0)}\right| = \exp\left(-J(t,0,0)\Gamma_{ab}^2\right) \,.
        \label{1q_density}
    \end{equation}
    In other words, each matrix element experiences a Gaussian decay with a single rate. 
    
    Figure \ref{single_qubit_dephase} shows the dephasing time for a single flip-flop qubit between states ($\ket{0}$ and $\ket{1}$). The narrow blue band is the sweet spot for the qubit. The bottom of the band at the turning point where it bends upwards is the second-order sweet spot, where the qubit is particularly robust against fluctuations in the donor-dot detuning. At a sweet spot, $\Gamma_{01} = 0$, leading to perfect coherence within our second-order approximate solution to the von Neumann equation. While this enhanced coherence is present at both first and second-order sweet spots, the advantage of the second-order sweet spot is in its wider coherence peak, as shown in Fig.~\ref{single_qubit_dephase}(b), giving more experimental leeway when tuning system parameters.

    Figure \ref{single_qubit_dephase}(b) shows three different horizontal cuts of Fig.~\ref{single_qubit_dephase}(a) at different applied magnetic fields. When $B < B_{ss}$ [$B_{ss}$ is the magnetic field for the second order sweet spot, about 0.796 T with a tunnel coupling of 47.15 $\mu$eV in Fig.~\ref{single_qubit_dephase}(a)], there is no sweet spot (blue line), the coherence time is relatively short, in the order of tens of $\mu$s at its peak. When $B > B_{ss}$, there are two first-order sweet spots, corresponding to the two high coherence peaks.  The width of a peak can be used to characterize the robustness of the sweet spot.  For example, if targeting a $T_2\ge\SI{1}{ms}$, the two first-order peaks would have widths of $\Delta E_z\approx\SI{0.04}{V/cm}$. As we further tune the system such that $B = B_{ss}$, the two peaks merge to form a second-order sweet spot. At the same target $T_2$, this peak has a width of approximately $\SI{0.3}{V/cm}$, an order of magnitude improvement over the first-order peaks.
    
\subsection{Two Qubit Relaxation}

    When two flip-flop qubits are coupled via the electric dipole interaction, the total two-qubit Hamiltonian, including interaction with the reservoir, can be expressed as:
    \begin{equation}
        H_{2q}(t) = H_{01} + H_{02} + H_{dip} + \sum_{i=1}^2f_i(t) h_{i} \,,
    \end{equation}
    where $H_{01}$ and $H_{02}$ are for isolated qubits, while $H_{dip}$ is the dipole interaction.  Since $[H_{0i}, H_{dip}] \neq 0$, the environmental electrical fluctuations affect a two-qubit system differently from the individual qubits.  Here we first diagonalize the noiseless part of the Hamiltonian $H_{01} + H_{02} + H_{dip}$ (the corresponding rotation matrix is given in the appendix), then project the full Hamiltonian onto this eigenbasis.

    To simplify the notation, we define the energy detuning between the charge and flip-flop qubit as $\delta = E_2-E_1$, and two effective noise strengths as $\gamma_1 = \omega_n (z_{30}-z_{03})/2$ and $\gamma_2 = \omega_n (z_{11}+z_{22})/2$.  Here $\gamma_1$ is primarily responsible for dephasing in the charge leakage states. This can then lead to relaxation in the flip-flop states via the indirect process. $\gamma_2$, on the other hand, causes transitions directly between the flip-flop and charge excited states, and is thus a leakage channel from the qubit space.  Our results in most cases show that $\gamma_1$ is usually the more important source of decoherence, and is larger than $\gamma_2$ by about an order of magnitude. We assume identical noise strength on both donors and use the same strength of $\omega_n = \SI{1}{\mu eV}$ as in the single qubit case.

    \begin{figure}[htb]
    \centering
    \includegraphics[width=\columnwidth]{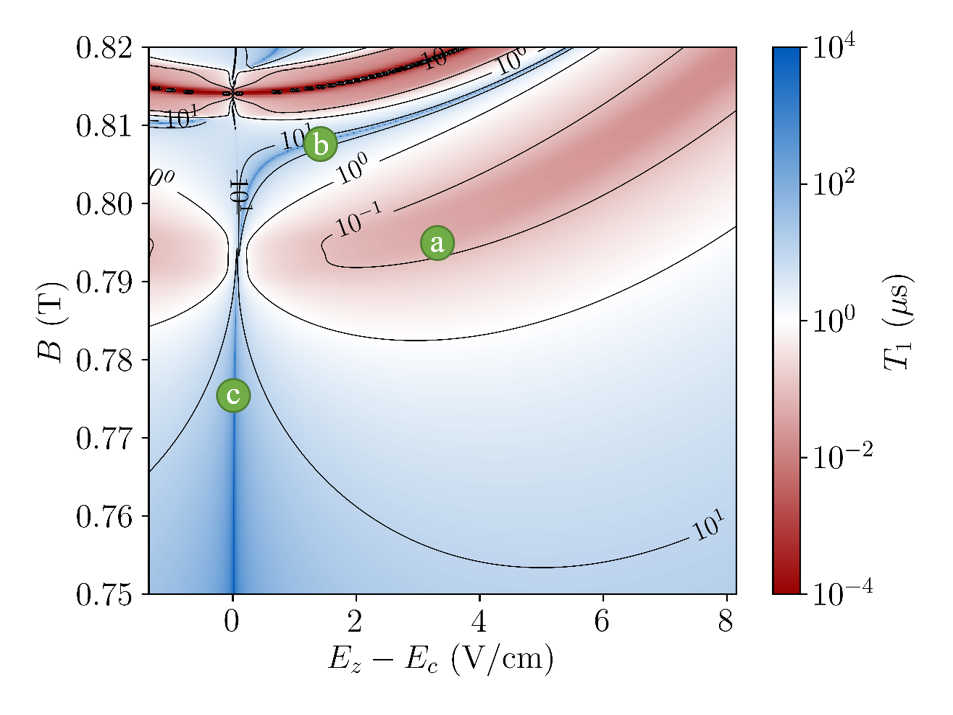}
    \caption{Relaxation time, $T_1$, for the two-qubit system computed by setting the envelope function (Eq.~\ref{envelope})  $F_{01,01}(T_1) = 1/e$. The tunnel coupling $V_t$ was set to $\SI{47.2}{\mu eV}$. We see a band that promises high coherence. In the area of marker b, this band is given by $2g_f\gamma_1 = g_l\gamma_2$. If we move along the band closer to marker a, it is better described by $2g_cg_l\gamma_1=\delta^2\gamma_2$.
    This band, particularly the horizontal part, can be a little misleading since it can hide some important features. 
    Since the overall time evolution is a summation of multiple terms (see Equation~\ref{rho} and Appendix E), less dominant terms that decay quickly are overshadowed by larger terms with slower decay rates. In this particular scenario, at marker b, the faster indirect transition process utilizing the leakage channels has an amplitude that is large enough to be important to the overall process, but small enough such that complete decay of this channel will not reduce the envelope function to below our cutoff threshold of $1/e$. Once we move over to marker c, this channel then becomes small enough to be negligible.
    The ideal operating point would then be to on the thin blue line while maintaining a large enough detuning between the charge and flip-flop qubits such that the faster decaying processes are negligible with respect to some target process fidelity.
    Time evolution plots are given in figure~\ref{2Q_Evolution} for the three marked operating points.}
    \label{RelaxTime}
\end{figure}
    \begin{figure}
    \centering
    \includegraphics[width=\columnwidth]{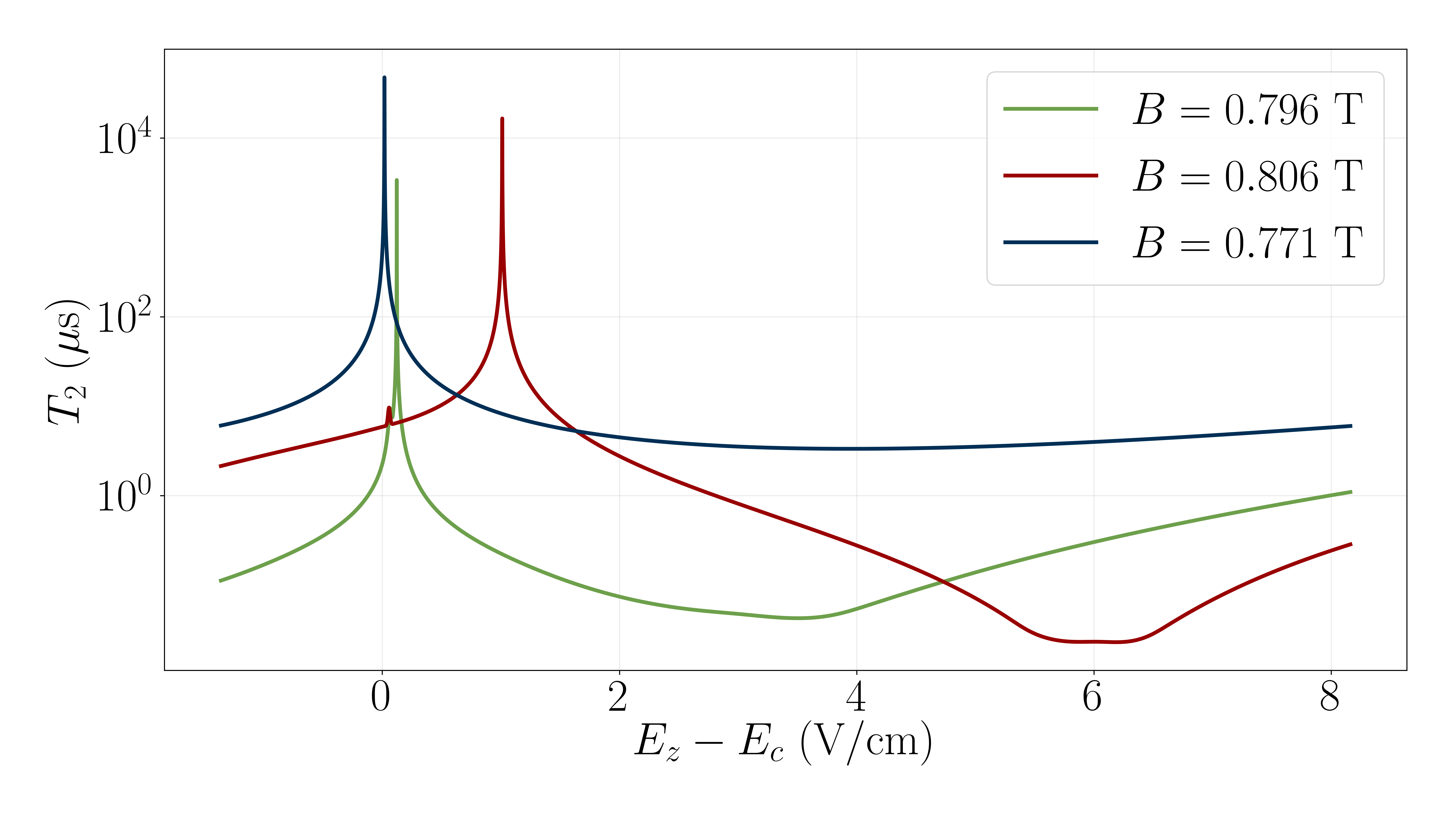}
    \caption{Horizontal cross cuts at the three indicated points in Figure~\ref{RelaxTime}.}
\end{figure}
    \begin{figure*}[thb]
    \centering
    \includegraphics[width=\textwidth]{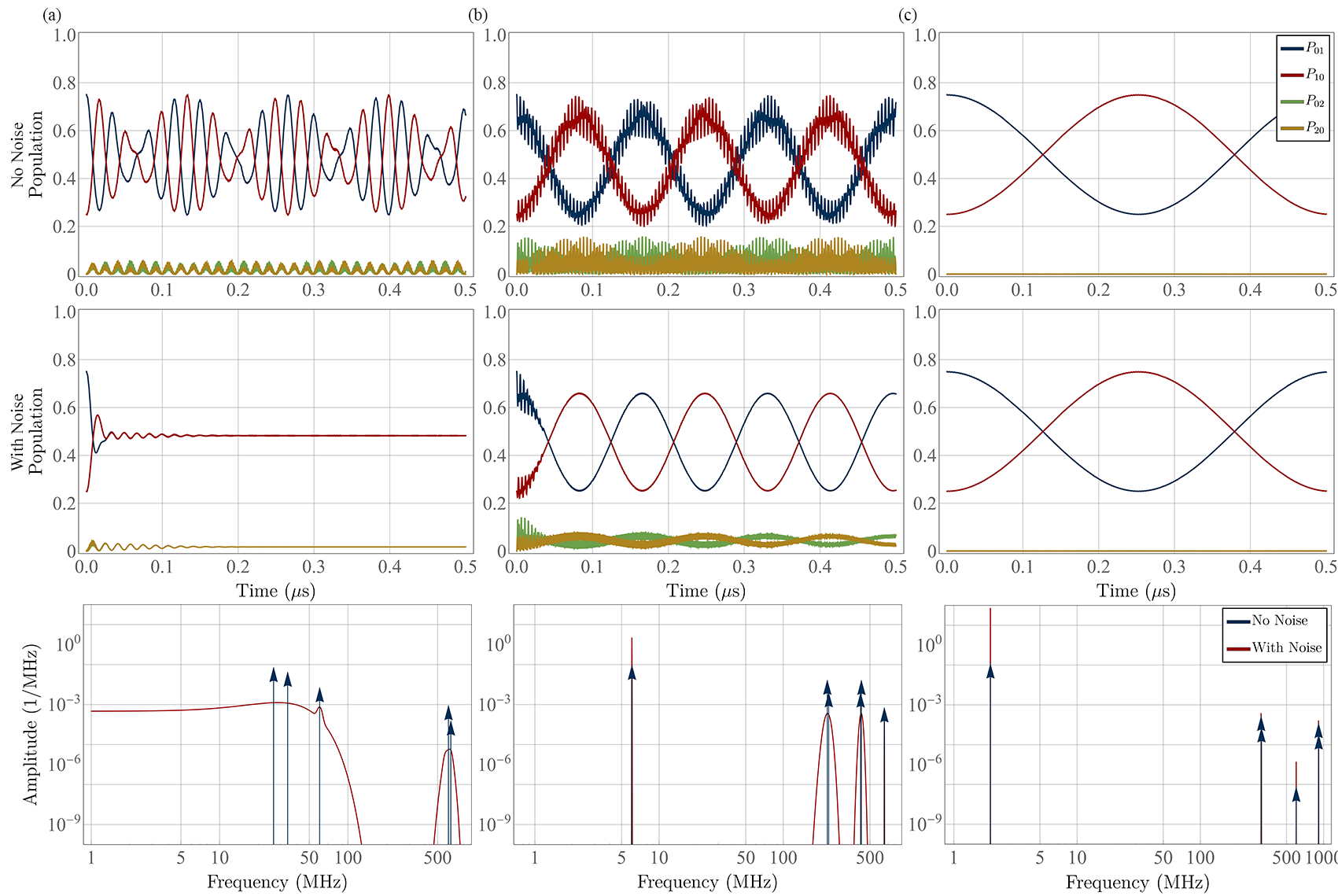}
    \caption{Time evolution for the two qubit system with parameters marked in Figure~\ref{RelaxTime}. The system is initialized to state $\ket{\Psi(0)}=\frac{\sqrt{3}}{2}\ket{01}+\frac{1}{2}\ket{10}$. The top most plots show the time evolution of the system absent any noise while the middle plots include charge noise. The bottom row are the time evolution of $P_{01}$ in the frequency domain. Upward arrows indicate Dirac delta functions. (a) $B=\SI{0.796}{T}$, $E_z-E_c=\SI{3.13}{V/cm}$. This is the location of the single-qubit sweet spot. Without noise, we have gate times of approximately $\SI{150}{ns}$. With noise, we can see that biasing the system to the single-qubit sweet spot is nonideal and quickly decays to its equilibrium state. (b) $B=\SI{0.806}{T}$, $E_z-E_c=\SI{0.95}{V/cm}$.  Moving to the blue band in Figure~\ref{RelaxTime} while trying to maintain fast gate times yields better coherence times but still low fidelity due to the charge leakage. (c) $B=\SI{0.771}{T}$, $E_z-E_c=\SI{0}{}$.  Moving the flip-flop and charge qubits further off-resonance minimizes charge leakage and thus we get excellent coherence times with gate times of approximately $\SI{250}{ns}$.}
    \label{2Q_Evolution}
\end{figure*}

    The parameter space for this system can be roughly divided into three regions based on how the charge coupling strength compares with the qubit's separation from charge excitation (and leakage): (a) $g_c \approx \delta$, (b) weakly detuned $g_c > \delta$, and (c) highly detuned $g_c < \delta$.  Notably, the single-qubit sweet spots are located approximately within the first region, casting doubts about their viability when two qubits are coupled. Below we explore each of these parameter regimes in more detail.

    The noiseless evolution of the population transfer from $\ket{01}$ to $\ket{01}$ state is a combination of primarily two swapping processes described in section~\ref{subsection-dipole}. As shown in Fig.~\ref{2Q_Evolution}, when both qubits are tuned to the individual flip-flop sweet spots, a beating pattern in the two-qubit states develops, indicative of the two processes operating at frequencies $(2g_l \pm 3g_f)/2$ with similar amplitudes. Since $g_l \approx 10g_f$, the slower beat frequency is then equal to $3g_f/2$ while the faster oscillation frequency is given by $g_l$.  The latter is what allows the qubit to leak into charge-excited states, as hinted at by the oscillating population $P_{20}$ and $P_{02}$ in the leakage states with the same frequency.

    When noise is introduced into the system, the relaxation rate is approximately $\gamma_1/2$, which turns out to be the dominant factor for two-qubit decoherence. In other words, the dominant mechanism for decoherence here is dephasing during the $\ket{02}\leftrightarrow \ket{20}$ portion of the overall indirect process. This decoherence channel is absent in the direct process described earlier. At the single-qubit sweet spots the quality factor is given by,
    \begin{equation}
        Q_a = \omega_a/\Gamma_a = 3g_f/\gamma_1 \,.
    \end{equation}
    Using the parameters given in Figs.~\ref{RelaxTime} and~\ref{2Q_Evolution}, we obtain $Q_a = 0.2$ when both qubits are at their respective single-qubit sweet spots. This abysmal quality factor shows that the single-flip-flop-qubit sweet spot is not useful during a two-qubit gate based on electric dipole coupling.  In essence, the dipolar coupling does not commute with single-qubit Hamiltonians, so that the two-qubit system experiences the electrical noise differently from that of the individual single qubits, rendering the single-qubit sweet spots irrelevant for a coupled two-qubit system.
    %
    %

    It is in the region of $g_c \sim \delta$ that the amplitude of the leakage mediated process is the greatest. By moving away from this region, we reduce the amplitude of the less coherent indirect process, and can expect better overall coherence.

    Figure \ref{RelaxTime} shows how two-qubit relaxation depends on the dot-donor detuning and the applied magnetic field. The most prominent bluer region (longer relaxation times) on the figure is along $\epsilon = 0$, which is essentially the charge qubit sweet spot.  Notice that the individual qubit sweet spot corresponds to point $a$ on the figure, which is in a region of faster relaxation and is thus not favorable for two-qubit coherence, consistent with our discussion above.

    To take advantage of the regime with longer relaxation times, we tune the system to two spots in the blue areas of Fig.~\ref{RelaxTime}, specifically the spots labeled b and c. At point b, the two-qubit dynamics is again a combination of the direct and indirect processes, with the indirect process contributing a smaller but still nontrivial amount to the overall time evolution. This is shown in Fig.~\ref{2Q_Evolution}(b) where we can see clearly the high and low frequency oscillations making up the $\ket{01}\leftrightarrow\ket{10}$ transitions along with the increase in leakage population. These two frequencies are
    \begin{subequations}
        \begin{align}
            \omega_\textrm{slow} &= 2g_f\delta^2/(g_c^2-\delta^2) \,, \\
            \omega_\textrm{fast} &= (g_c-\delta) \,,
        \end{align}
    \end{subequations}
    with decay rates of
    \begin{subequations}
        \begin{align}
            \Gamma_\textrm{slow} &= 2\frac{\delta g_f}{g_c^2}\left( 2\gamma_1 - \frac{g_l}{g_f}\gamma_2 \right) \,, \\
            \Gamma_\textrm{fast} &= \gamma_1 \,.
        \end{align}
    \end{subequations}
    Here we have used the relationship $g_l^2 = g_cg_f$. The amplitude of the faster process is about $2g_f/g_c$. When both processes are present, the two-qubit dynamics generally experiences a fast dropoff due to the decay of the indirect process followed by a slow decay from the direct iSWAP process, yielding an overall process that has relatively high coherence from a normal experimental perspective, but leading to a low-fidelity quantum gate. The individual quality factors for the two are
    \begin{subequations}
        \begin{align}
            Q_{b,\textrm{slow}} &= g_fg_c^2\delta/(g_c^2-\delta^2)(2g_f\gamma_1 - g_l\gamma_2) \\
            Q_{b,\textrm{fast}} &= (g_c-\delta)/\gamma_1
        \end{align}
    \end{subequations}
    Notice that the quality factor of the slow process diverges when $2g_f\gamma_1 = g_l\gamma_2$. This is the cause of the narrow blue band in figure~\ref{RelaxTime}.  What is important for a quantum gate, however, is an overall quality factor that accounts for the faster decay rate along with the slower gate time, yielding a factor of
    \begin{equation}
        Q_b = \frac{2g_f\delta^2}{\gamma_1(g_c^2-\delta^2)} \,.
    \end{equation}
    When tuned to be at point b in Fig.~\ref{RelaxTime}, the quality factor is only about $1$.  Clearly, what we need for a high-quality quantum gate is a regime where we can turn off the indirect process via the charge excited state and the associated fast decoherence.
    %
    %

    As shown in Fig.~\ref{2Q_Evolution}(c),the leakage into the charge excited states can be strongly suppressed by increasing the detuning $\delta$ between spin and charge excitation, which in turn leads to very long coherence times for the two-qubit system. The gate frequency in this regime is
    \begin{equation}
        \omega_c = 2g_f\left(\frac{\delta^2}{\delta^2-g_c^2}\right) \,,
    \end{equation}
    with a decoherence rate of
    \begin{equation}
        \Gamma_c = \frac{2g_l\gamma_2}{\delta}-\frac{4\gamma_1g_cg_l^2}{\delta^3} \,,
    \end{equation}
    yielding an overall qualify factor of
    \begin{equation}
        Q_c = \frac{\delta^5 g_f}{ g_l(\delta^2-g_c^2)(\delta^2\gamma_2-2g_cg_l\gamma_1)} \,.
    \end{equation}
    The quality factor peaks when $\delta^2\gamma_2 = 2g_cg_l\gamma_1$, corresponding to the narrow blue line in figure~\ref{RelaxTime}. 


    Given an initial condition of $\ket{\Psi(0)}=\ket{01}$, the equilibrium leakage population is given by
    \begin{equation}
        P_\textrm{leak} = \frac{1}{4}(\sin^2\phi_++\sin^2\phi_-)
    \end{equation}
    where $\tan\phi_\pm=(\pm2g_l)/(-\delta\mp(g_c-g_f))$. 

    This is not to say we can achieve perfect coherence with high fidelity. By increasing the detuning between the spin and charge states, we also decrease the energy difference between the $\ket{01}$, $\ket{10}$, $\ket{02}$, $\ket{20}$ energy manifold with the $\ket{00}$ and $\ket{11}$ manifolds. Noise induced transitions between these manifolds, previously neglected, become more important. The relaxation caused by these transitions occurs on the time scale of tens of microseconds (still longer than the gate times by more than an order of magnitude), putting an upper limit to the overall quality factor of the gates. Additional details on this slower decay channel can be found in Appendix F.
\section{Conclusion}
Here we have performed a comprehensive study of decoherence of single and coupled flip-flop qubits under the influence of charge noise.

The single flip-flop qubit is weakly coupled to charge noise via the charge qubit. Under certain experimental parameters, we can completely eliminate dephasing noise on the flip-flop qubit, resulting in incredibly long coherence times at this flip-flop sweet spot.

The sweet spot does not help so much when we try to perform multiqubit gates however. 
With the flip-flop qubit coupling mediated by the dipole coupling of their charge-qubit components, we identify two channels of information transfer: a direct coupling between qubit states $\ket{01}$ and $\ket{10}$, and an indirect channel via a charge-excited state, which loses its coherence quickly in the presence of charge noise.  As such, the excellent coherence property at the sweet spot is lost.  In essence, the dipole coupling does not commute with the single-qubit Hamiltonian, and the single-qubit sweet spot is lost in the modified state spectrum.

To fight against the charge leakage while two qubits are dipole coupled, we need to turn off the charge excitation as much as we can.  This can be accomplished by increasing the detuning between the charge qubits and the spin qubits, so that dipole coupling is less likely to cause a spin-charge exchange between the two flip-flop qubits. By tuning system parameters (charge qubit detuning, tunnel coupling, magnetic field), we find that we can increase the quality factor of our gates up to $10^2$, with reasonably fast gate and vastly improved two-qubit coherence.

In essence, the second order flip-flop sweet spot only protects the flip-flop qubit from charge noise, but does not protect the charge qubit part specifically. When two flip-flop qubits are coupled, any operations that rely heavily on the charge qubit will be limited by the charge qubit coherence.  As such, the presence of dipole coupling makes the charge qubit sweet spots (instead of the flip-flop qubit sweet spots) better places to be in the parameter space.  In short, the key is to keep the charge qubits in their ground states as much as possible, even if virtual excitations are inevitable during the process.

In more general systems where coherent qubits are coupled by less coherent objects, our work has shown that we can indeed take advantage of these noisy channels to couple qubits together as long as care is taken to minimize leakage into states that are more susceptible to the dominiant noise noise.

\section*{Acknowledgements}
This work is partially supported by US Army Research Office (ARO) through Grant No. W911NF1710257.

\bibliography{references}

\begin{thebibliography}{61}%
\makeatletter
\providecommand \@ifxundefined [1]{%
 \@ifx{#1\undefined}
}%
\providecommand \@ifnum [1]{%
 \ifnum #1\expandafter \@firstoftwo
 \else \expandafter \@secondoftwo
 \fi
}%
\providecommand \@ifx [1]{%
 \ifx #1\expandafter \@firstoftwo
 \else \expandafter \@secondoftwo
 \fi
}%
\providecommand \natexlab [1]{#1}%
\providecommand \enquote  [1]{``#1''}%
\providecommand \bibnamefont  [1]{#1}%
\providecommand \bibfnamefont [1]{#1}%
\providecommand \citenamefont [1]{#1}%
\providecommand \href@noop [0]{\@secondoftwo}%
\providecommand \href [0]{\begingroup \@sanitize@url \@href}%
\providecommand \@href[1]{\@@startlink{#1}\@@href}%
\providecommand \@@href[1]{\endgroup#1\@@endlink}%
\providecommand \@sanitize@url [0]{\catcode `\\12\catcode `\$12\catcode
  `\&12\catcode `\#12\catcode `\^12\catcode `\_12\catcode `\%12\relax}%
\providecommand \@@startlink[1]{}%
\providecommand \@@endlink[0]{}%
\providecommand \url  [0]{\begingroup\@sanitize@url \@url }%
\providecommand \@url [1]{\endgroup\@href {#1}{\urlprefix }}%
\providecommand \urlprefix  [0]{URL }%
\providecommand \Eprint [0]{\href }%
\providecommand \doibase [0]{https://doi.org/}%
\providecommand \selectlanguage [0]{\@gobble}%
\providecommand \bibinfo  [0]{\@secondoftwo}%
\providecommand \bibfield  [0]{\@secondoftwo}%
\providecommand \translation [1]{[#1]}%
\providecommand \BibitemOpen [0]{}%
\providecommand \bibitemStop [0]{}%
\providecommand \bibitemNoStop [0]{.\EOS\space}%
\providecommand \EOS [0]{\spacefactor3000\relax}%
\providecommand \BibitemShut  [1]{\csname bibitem#1\endcsname}%
\let\auto@bib@innerbib\@empty
\bibitem [{\citenamefont {Tosi}\ \emph {et~al.}(2017)\citenamefont {Tosi},
  \citenamefont {Mohiyaddin}, \citenamefont {Schmitt}, \citenamefont {Tenberg},
  \citenamefont {Rahman}, \citenamefont {Klimeck},\ and\ \citenamefont
  {Morello}}]{tosi_silicon_2017}%
  \BibitemOpen
  \bibfield  {author} {\bibinfo {author} {\bibfnamefont {G.}~\bibnamefont
  {Tosi}}, \bibinfo {author} {\bibfnamefont {F.~A.}\ \bibnamefont
  {Mohiyaddin}}, \bibinfo {author} {\bibfnamefont {V.}~\bibnamefont {Schmitt}},
  \bibinfo {author} {\bibfnamefont {S.}~\bibnamefont {Tenberg}}, \bibinfo
  {author} {\bibfnamefont {R.}~\bibnamefont {Rahman}}, \bibinfo {author}
  {\bibfnamefont {G.}~\bibnamefont {Klimeck}},\ and\ \bibinfo {author}
  {\bibfnamefont {A.}~\bibnamefont {Morello}},\ }\bibfield  {title} {\bibinfo
  {title} {Silicon quantum processor with robust long-distance qubit
  couplings},\ }\href {https://doi.org/10.1038/s41467-017-00378-x} {\bibfield
  {journal} {\bibinfo  {journal} {Nature Communications}\ }\textbf {\bibinfo
  {volume} {8}},\ \bibinfo {pages} {450} (\bibinfo {year} {2017})}\BibitemShut
  {NoStop}%
\bibitem [{\citenamefont {DiVincenzo}(2000)}]{divincenzo_physical_2000}%
  \BibitemOpen
  \bibfield  {author} {\bibinfo {author} {\bibfnamefont {D.~P.}\ \bibnamefont
  {DiVincenzo}},\ }\bibfield  {title} {\bibinfo {title} {The physical
  implementation of quantum computation},\ }\href
  {https://doi.org/10.1002/1521-3978(200009)48:9/11<771::aid-prop771>3.0.co;2-e}
  {\bibfield  {journal} {\bibinfo  {journal} {Fortschritte der Physik}\
  }\textbf {\bibinfo {volume} {48}},\ \bibinfo {pages} {771–783} (\bibinfo
  {year} {2000})}\BibitemShut {NoStop}%
\bibitem [{\citenamefont {Hollenberg}\ \emph {et~al.}(2004)\citenamefont
  {Hollenberg}, \citenamefont {Dzurak}, \citenamefont {Wellard}, \citenamefont
  {Hamilton}, \citenamefont {Reilly}, \citenamefont {Milburn},\ and\
  \citenamefont {Clark}}]{PhysRevB.69.113301}%
  \BibitemOpen
  \bibfield  {author} {\bibinfo {author} {\bibfnamefont {L.~C.~L.}\
  \bibnamefont {Hollenberg}}, \bibinfo {author} {\bibfnamefont {A.~S.}\
  \bibnamefont {Dzurak}}, \bibinfo {author} {\bibfnamefont {C.}~\bibnamefont
  {Wellard}}, \bibinfo {author} {\bibfnamefont {A.~R.}\ \bibnamefont
  {Hamilton}}, \bibinfo {author} {\bibfnamefont {D.~J.}\ \bibnamefont
  {Reilly}}, \bibinfo {author} {\bibfnamefont {G.~J.}\ \bibnamefont
  {Milburn}},\ and\ \bibinfo {author} {\bibfnamefont {R.~G.}\ \bibnamefont
  {Clark}},\ }\bibfield  {title} {\bibinfo {title} {Charge-based quantum
  computing using single donors in semiconductors},\ }\href
  {https://doi.org/10.1103/PhysRevB.69.113301} {\bibfield  {journal} {\bibinfo
  {journal} {Phys. Rev. B}\ }\textbf {\bibinfo {volume} {69}},\ \bibinfo
  {pages} {113301} (\bibinfo {year} {2004})}\BibitemShut {NoStop}%
\bibitem [{\citenamefont {Stavrou}\ and\ \citenamefont
  {Hu}(2005)}]{stavrou_charge_2005}%
  \BibitemOpen
  \bibfield  {author} {\bibinfo {author} {\bibfnamefont {V.~N.}\ \bibnamefont
  {Stavrou}}\ and\ \bibinfo {author} {\bibfnamefont {X.}~\bibnamefont {Hu}},\
  }\bibfield  {title} {\bibinfo {title} {Charge decoherence in laterally
  coupled quantum dots due to electron-phonon interactions},\ }\href
  {https://doi.org/10.1103/PhysRevB.72.075362} {\bibfield  {journal} {\bibinfo
  {journal} {Physical Review B}\ }\textbf {\bibinfo {volume} {72}},\ \bibinfo
  {pages} {075362} (\bibinfo {year} {2005})}\BibitemShut {NoStop}%
\bibitem [{\citenamefont {Petersson}\ \emph {et~al.}(2010)\citenamefont
  {Petersson}, \citenamefont {Petta}, \citenamefont {Lu},\ and\ \citenamefont
  {Gossard}}]{PhysRevLett.105.246804}%
  \BibitemOpen
  \bibfield  {author} {\bibinfo {author} {\bibfnamefont {K.~D.}\ \bibnamefont
  {Petersson}}, \bibinfo {author} {\bibfnamefont {J.~R.}\ \bibnamefont
  {Petta}}, \bibinfo {author} {\bibfnamefont {H.}~\bibnamefont {Lu}},\ and\
  \bibinfo {author} {\bibfnamefont {A.~C.}\ \bibnamefont {Gossard}},\
  }\bibfield  {title} {\bibinfo {title} {Quantum coherence in a one-electron
  semiconductor charge qubit},\ }\href
  {https://doi.org/10.1103/PhysRevLett.105.246804} {\bibfield  {journal}
  {\bibinfo  {journal} {Phys. Rev. Lett.}\ }\textbf {\bibinfo {volume} {105}},\
  \bibinfo {pages} {246804} (\bibinfo {year} {2010})}\BibitemShut {NoStop}%
\bibitem [{\citenamefont {Hayashi}\ \emph {et~al.}(2003)\citenamefont
  {Hayashi}, \citenamefont {Fujisawa}, \citenamefont {Cheong}, \citenamefont
  {Jeong},\ and\ \citenamefont {Hirayama}}]{PhysRevLett.91.226804}%
  \BibitemOpen
  \bibfield  {author} {\bibinfo {author} {\bibfnamefont {T.}~\bibnamefont
  {Hayashi}}, \bibinfo {author} {\bibfnamefont {T.}~\bibnamefont {Fujisawa}},
  \bibinfo {author} {\bibfnamefont {H.~D.}\ \bibnamefont {Cheong}}, \bibinfo
  {author} {\bibfnamefont {Y.~H.}\ \bibnamefont {Jeong}},\ and\ \bibinfo
  {author} {\bibfnamefont {Y.}~\bibnamefont {Hirayama}},\ }\bibfield  {title}
  {\bibinfo {title} {Coherent manipulation of electronic states in a double
  quantum dot},\ }\href {https://doi.org/10.1103/PhysRevLett.91.226804}
  {\bibfield  {journal} {\bibinfo  {journal} {Phys. Rev. Lett.}\ }\textbf
  {\bibinfo {volume} {91}},\ \bibinfo {pages} {226804} (\bibinfo {year}
  {2003})}\BibitemShut {NoStop}%
\bibitem [{\citenamefont {Tyryshkin}\ \emph {et~al.}(2012)\citenamefont
  {Tyryshkin}, \citenamefont {Tojo}, \citenamefont {Morton}, \citenamefont
  {Riemann}, \citenamefont {Abrosimov}, \citenamefont {Becker}, \citenamefont
  {Pohl}, \citenamefont {Schenkel}, \citenamefont {Thewalt}, \citenamefont
  {Itoh},\ and\ \citenamefont {Lyon}}]{tyryshkin_electron_2012}%
  \BibitemOpen
  \bibfield  {author} {\bibinfo {author} {\bibfnamefont {A.~M.}\ \bibnamefont
  {Tyryshkin}}, \bibinfo {author} {\bibfnamefont {S.}~\bibnamefont {Tojo}},
  \bibinfo {author} {\bibfnamefont {J.~J.~L.}\ \bibnamefont {Morton}}, \bibinfo
  {author} {\bibfnamefont {H.}~\bibnamefont {Riemann}}, \bibinfo {author}
  {\bibfnamefont {N.~V.}\ \bibnamefont {Abrosimov}}, \bibinfo {author}
  {\bibfnamefont {P.}~\bibnamefont {Becker}}, \bibinfo {author} {\bibfnamefont
  {H.-J.}\ \bibnamefont {Pohl}}, \bibinfo {author} {\bibfnamefont
  {T.}~\bibnamefont {Schenkel}}, \bibinfo {author} {\bibfnamefont {M.~L.~W.}\
  \bibnamefont {Thewalt}}, \bibinfo {author} {\bibfnamefont {K.~M.}\
  \bibnamefont {Itoh}},\ and\ \bibinfo {author} {\bibfnamefont {S.~A.}\
  \bibnamefont {Lyon}},\ }\bibfield  {title} {\bibinfo {title} {Electron spin
  coherence exceeding seconds in high-purity silicon},\ }\href
  {https://doi.org/10.1038/nmat3182} {\bibfield  {journal} {\bibinfo  {journal}
  {Nature Materials}\ }\textbf {\bibinfo {volume} {11}},\ \bibinfo {pages}
  {143} (\bibinfo {year} {2012})},\ \bibinfo {note} {number: 2 Publisher:
  Nature Publishing Group}\BibitemShut {NoStop}%
\bibitem [{\citenamefont {Zwanenburg}\ \emph {et~al.}(2013)\citenamefont
  {Zwanenburg}, \citenamefont {Dzurak}, \citenamefont {Morello}, \citenamefont
  {Simmons}, \citenamefont {Hollenberg}, \citenamefont {Klimeck}, \citenamefont
  {Rogge}, \citenamefont {Coppersmith},\ and\ \citenamefont
  {Eriksson}}]{zwanenburg_silicon_2013}%
  \BibitemOpen
  \bibfield  {author} {\bibinfo {author} {\bibfnamefont {F.~A.}\ \bibnamefont
  {Zwanenburg}}, \bibinfo {author} {\bibfnamefont {A.~S.}\ \bibnamefont
  {Dzurak}}, \bibinfo {author} {\bibfnamefont {A.}~\bibnamefont {Morello}},
  \bibinfo {author} {\bibfnamefont {M.~Y.}\ \bibnamefont {Simmons}}, \bibinfo
  {author} {\bibfnamefont {L.~C.~L.}\ \bibnamefont {Hollenberg}}, \bibinfo
  {author} {\bibfnamefont {G.}~\bibnamefont {Klimeck}}, \bibinfo {author}
  {\bibfnamefont {S.}~\bibnamefont {Rogge}}, \bibinfo {author} {\bibfnamefont
  {S.~N.}\ \bibnamefont {Coppersmith}},\ and\ \bibinfo {author} {\bibfnamefont
  {M.~A.}\ \bibnamefont {Eriksson}},\ }\bibfield  {title} {\bibinfo {title}
  {Silicon quantum electronics},\ }\href
  {https://doi.org/10.1103/RevModPhys.85.961} {\bibfield  {journal} {\bibinfo
  {journal} {Reviews of Modern Physics}\ }\textbf {\bibinfo {volume} {85}},\
  \bibinfo {pages} {961} (\bibinfo {year} {2013})}\BibitemShut {NoStop}%
\bibitem [{\citenamefont {Pla}\ \emph {et~al.}(2013)\citenamefont {Pla},
  \citenamefont {Tan}, \citenamefont {Dehollain}, \citenamefont {Lim},
  \citenamefont {Morton}, \citenamefont {Zwanenburg}, \citenamefont {Jamieson},
  \citenamefont {Dzurak},\ and\ \citenamefont
  {Morello}}]{pla_high-fidelity_2013}%
  \BibitemOpen
  \bibfield  {author} {\bibinfo {author} {\bibfnamefont {J.~J.}\ \bibnamefont
  {Pla}}, \bibinfo {author} {\bibfnamefont {K.~Y.}\ \bibnamefont {Tan}},
  \bibinfo {author} {\bibfnamefont {J.~P.}\ \bibnamefont {Dehollain}}, \bibinfo
  {author} {\bibfnamefont {W.~H.}\ \bibnamefont {Lim}}, \bibinfo {author}
  {\bibfnamefont {J.~J.~L.}\ \bibnamefont {Morton}}, \bibinfo {author}
  {\bibfnamefont {F.~A.}\ \bibnamefont {Zwanenburg}}, \bibinfo {author}
  {\bibfnamefont {D.~N.}\ \bibnamefont {Jamieson}}, \bibinfo {author}
  {\bibfnamefont {A.~S.}\ \bibnamefont {Dzurak}},\ and\ \bibinfo {author}
  {\bibfnamefont {A.}~\bibnamefont {Morello}},\ }\bibfield  {title} {\bibinfo
  {title} {High-fidelity readout and control of a nuclear spin qubit in
  silicon},\ }\href {https://doi.org/10.1038/nature12011} {\bibfield  {journal}
  {\bibinfo  {journal} {Nature}\ }\textbf {\bibinfo {volume} {496}},\ \bibinfo
  {pages} {334} (\bibinfo {year} {2013})},\ \bibinfo {note} {number: 7445
  Publisher: Nature Publishing Group}\BibitemShut {NoStop}%
\bibitem [{\citenamefont {Veldhorst}\ \emph {et~al.}(2014)\citenamefont
  {Veldhorst}, \citenamefont {Hwang}, \citenamefont {Yang}, \citenamefont
  {Leenstra}, \citenamefont {de~Ronde}, \citenamefont {Dehollain},
  \citenamefont {Muhonen}, \citenamefont {Hudson}, \citenamefont {Itoh},
  \citenamefont {Morello},\ and\ \citenamefont
  {Dzurak}}]{veldhorst_addressable_2014}%
  \BibitemOpen
  \bibfield  {author} {\bibinfo {author} {\bibfnamefont {M.}~\bibnamefont
  {Veldhorst}}, \bibinfo {author} {\bibfnamefont {J.~C.~C.}\ \bibnamefont
  {Hwang}}, \bibinfo {author} {\bibfnamefont {C.~H.}\ \bibnamefont {Yang}},
  \bibinfo {author} {\bibfnamefont {A.~W.}\ \bibnamefont {Leenstra}}, \bibinfo
  {author} {\bibfnamefont {B.}~\bibnamefont {de~Ronde}}, \bibinfo {author}
  {\bibfnamefont {J.~P.}\ \bibnamefont {Dehollain}}, \bibinfo {author}
  {\bibfnamefont {J.~T.}\ \bibnamefont {Muhonen}}, \bibinfo {author}
  {\bibfnamefont {F.~E.}\ \bibnamefont {Hudson}}, \bibinfo {author}
  {\bibfnamefont {K.~M.}\ \bibnamefont {Itoh}}, \bibinfo {author}
  {\bibfnamefont {A.}~\bibnamefont {Morello}},\ and\ \bibinfo {author}
  {\bibfnamefont {A.~S.}\ \bibnamefont {Dzurak}},\ }\bibfield  {title}
  {\bibinfo {title} {An addressable quantum dot qubit with fault-tolerant
  control-fidelity},\ }\href {https://doi.org/10.1038/nnano.2014.216}
  {\bibfield  {journal} {\bibinfo  {journal} {Nature Nanotechnology}\ }\textbf
  {\bibinfo {volume} {9}},\ \bibinfo {pages} {981} (\bibinfo {year} {2014})},\
  \bibinfo {note} {number: 12 Publisher: Nature Publishing Group}\BibitemShut
  {NoStop}%
\bibitem [{\citenamefont {Pla}\ \emph {et~al.}(2012)\citenamefont {Pla},
  \citenamefont {Tan}, \citenamefont {Dehollain}, \citenamefont {Lim},
  \citenamefont {Morton}, \citenamefont {Jamieson}, \citenamefont {Dzurak},\
  and\ \citenamefont {Morello}}]{pla_single-atom_2012}%
  \BibitemOpen
  \bibfield  {author} {\bibinfo {author} {\bibfnamefont {J.~J.}\ \bibnamefont
  {Pla}}, \bibinfo {author} {\bibfnamefont {K.~Y.}\ \bibnamefont {Tan}},
  \bibinfo {author} {\bibfnamefont {J.~P.}\ \bibnamefont {Dehollain}}, \bibinfo
  {author} {\bibfnamefont {W.~H.}\ \bibnamefont {Lim}}, \bibinfo {author}
  {\bibfnamefont {J.~J.~L.}\ \bibnamefont {Morton}}, \bibinfo {author}
  {\bibfnamefont {D.~N.}\ \bibnamefont {Jamieson}}, \bibinfo {author}
  {\bibfnamefont {A.~S.}\ \bibnamefont {Dzurak}},\ and\ \bibinfo {author}
  {\bibfnamefont {A.}~\bibnamefont {Morello}},\ }\bibfield  {title} {\bibinfo
  {title} {A single-atom electron spin qubit in silicon},\ }\href
  {https://doi.org/10.1038/nature11449} {\bibfield  {journal} {\bibinfo
  {journal} {Nature}\ }\textbf {\bibinfo {volume} {489}},\ \bibinfo {pages}
  {541} (\bibinfo {year} {2012})},\ \bibinfo {note} {number: 7417 Publisher:
  Nature Publishing Group}\BibitemShut {NoStop}%
\bibitem [{\citenamefont {Xiao}\ \emph {et~al.}(2010)\citenamefont {Xiao},
  \citenamefont {House},\ and\ \citenamefont {Jiang}}]{xiao_measurement_2010}%
  \BibitemOpen
  \bibfield  {author} {\bibinfo {author} {\bibfnamefont {M.}~\bibnamefont
  {Xiao}}, \bibinfo {author} {\bibfnamefont {M.~G.}\ \bibnamefont {House}},\
  and\ \bibinfo {author} {\bibfnamefont {H.~W.}\ \bibnamefont {Jiang}},\
  }\bibfield  {title} {\bibinfo {title} {Measurement of the {Spin} {Relaxation}
  {Time} of {Single} {Electrons} in a {Silicon}
  {Metal}-{Oxide}-{Semiconductor}-{Based} {Quantum} {Dot}},\ }\href
  {https://doi.org/10.1103/PhysRevLett.104.096801} {\bibfield  {journal}
  {\bibinfo  {journal} {Physical Review Letters}\ }\textbf {\bibinfo {volume}
  {104}},\ \bibinfo {pages} {096801} (\bibinfo {year} {2010})},\ \bibinfo
  {note} {publisher: American Physical Society}\BibitemShut {NoStop}%
\bibitem [{\citenamefont {Loss}\ and\ \citenamefont
  {DiVincenzo}(1998)}]{loss_quantum_1998}%
  \BibitemOpen
  \bibfield  {author} {\bibinfo {author} {\bibfnamefont {D.}~\bibnamefont
  {Loss}}\ and\ \bibinfo {author} {\bibfnamefont {D.~P.}\ \bibnamefont
  {DiVincenzo}},\ }\bibfield  {title} {\bibinfo {title} {Quantum computation
  with quantum dots},\ }\href {https://doi.org/10.1103/PhysRevA.57.120}
  {\bibfield  {journal} {\bibinfo  {journal} {Physical Review A}\ }\textbf
  {\bibinfo {volume} {57}},\ \bibinfo {pages} {120} (\bibinfo {year} {1998})},\
  \bibinfo {note} {publisher: American Physical Society}\BibitemShut {NoStop}%
\bibitem [{\citenamefont {Kane}(1998)}]{kane_silicon-based_1998}%
  \BibitemOpen
  \bibfield  {author} {\bibinfo {author} {\bibfnamefont {B.~E.}\ \bibnamefont
  {Kane}},\ }\bibfield  {title} {\bibinfo {title} {A silicon-based nuclear spin
  quantum computer},\ }\href {https://doi.org/10.1038/30156} {\bibfield
  {journal} {\bibinfo  {journal} {Nature}\ }\textbf {\bibinfo {volume} {393}},\
  \bibinfo {pages} {133} (\bibinfo {year} {1998})}\BibitemShut {NoStop}%
\bibitem [{\citenamefont {Petta}\ \emph {et~al.}(2005)\citenamefont {Petta},
  \citenamefont {Johnson}, \citenamefont {Taylor}, \citenamefont {Laird},
  \citenamefont {Yacoby}, \citenamefont {Lukin}, \citenamefont {Marcus},
  \citenamefont {Hanson},\ and\ \citenamefont {Gossard}}]{petta_coherent_2005}%
  \BibitemOpen
  \bibfield  {author} {\bibinfo {author} {\bibfnamefont {J.~R.}\ \bibnamefont
  {Petta}}, \bibinfo {author} {\bibfnamefont {A.~C.}\ \bibnamefont {Johnson}},
  \bibinfo {author} {\bibfnamefont {J.~M.}\ \bibnamefont {Taylor}}, \bibinfo
  {author} {\bibfnamefont {E.~A.}\ \bibnamefont {Laird}}, \bibinfo {author}
  {\bibfnamefont {A.}~\bibnamefont {Yacoby}}, \bibinfo {author} {\bibfnamefont
  {M.~D.}\ \bibnamefont {Lukin}}, \bibinfo {author} {\bibfnamefont {C.~M.}\
  \bibnamefont {Marcus}}, \bibinfo {author} {\bibfnamefont {M.~P.}\
  \bibnamefont {Hanson}},\ and\ \bibinfo {author} {\bibfnamefont {A.~C.}\
  \bibnamefont {Gossard}},\ }\bibfield  {title} {\bibinfo {title} {Coherent
  {Manipulation} of {Coupled} {Electron} {Spins} in {Semiconductor} {Quantum}
  {Dots}},\ }\href {https://doi.org/10.1126/science.1116955} {\bibfield
  {journal} {\bibinfo  {journal} {Science}\ }\textbf {\bibinfo {volume}
  {309}},\ \bibinfo {pages} {2180} (\bibinfo {year} {2005})},\ \bibinfo {note}
  {publisher: American Association for the Advancement of Science Section:
  Research Article}\BibitemShut {NoStop}%
\bibitem [{\citenamefont {Takeda}\ \emph {et~al.}(2020)\citenamefont {Takeda},
  \citenamefont {Noiri}, \citenamefont {Yoneda}, \citenamefont {Nakajima},\
  and\ \citenamefont {Tarucha}}]{takeda_resonantly_2020}%
  \BibitemOpen
  \bibfield  {author} {\bibinfo {author} {\bibfnamefont {K.}~\bibnamefont
  {Takeda}}, \bibinfo {author} {\bibfnamefont {A.}~\bibnamefont {Noiri}},
  \bibinfo {author} {\bibfnamefont {J.}~\bibnamefont {Yoneda}}, \bibinfo
  {author} {\bibfnamefont {T.}~\bibnamefont {Nakajima}},\ and\ \bibinfo
  {author} {\bibfnamefont {S.}~\bibnamefont {Tarucha}},\ }\bibfield  {title}
  {\bibinfo {title} {Resonantly {Driven} {Singlet}-{Triplet} {Spin} {Qubit} in
  {Silicon}},\ }\href {https://doi.org/10.1103/PhysRevLett.124.117701}
  {\bibfield  {journal} {\bibinfo  {journal} {Physical Review Letters}\
  }\textbf {\bibinfo {volume} {124}},\ \bibinfo {pages} {117701} (\bibinfo
  {year} {2020})},\ \bibinfo {note} {publisher: American Physical
  Society}\BibitemShut {NoStop}%
\bibitem [{\citenamefont {Yoneda}\ \emph {et~al.}(2018)\citenamefont {Yoneda},
  \citenamefont {Takeda}, \citenamefont {Otsuka}, \citenamefont {Nakajima},
  \citenamefont {Delbecq}, \citenamefont {Allison}, \citenamefont {Honda},
  \citenamefont {Kodera}, \citenamefont {Oda}, \citenamefont {Hoshi},
  \citenamefont {Usami}, \citenamefont {Itoh},\ and\ \citenamefont
  {Tarucha}}]{yoneda_quantum-dot_2018}%
  \BibitemOpen
  \bibfield  {author} {\bibinfo {author} {\bibfnamefont {J.}~\bibnamefont
  {Yoneda}}, \bibinfo {author} {\bibfnamefont {K.}~\bibnamefont {Takeda}},
  \bibinfo {author} {\bibfnamefont {T.}~\bibnamefont {Otsuka}}, \bibinfo
  {author} {\bibfnamefont {T.}~\bibnamefont {Nakajima}}, \bibinfo {author}
  {\bibfnamefont {M.~R.}\ \bibnamefont {Delbecq}}, \bibinfo {author}
  {\bibfnamefont {G.}~\bibnamefont {Allison}}, \bibinfo {author} {\bibfnamefont
  {T.}~\bibnamefont {Honda}}, \bibinfo {author} {\bibfnamefont
  {T.}~\bibnamefont {Kodera}}, \bibinfo {author} {\bibfnamefont
  {S.}~\bibnamefont {Oda}}, \bibinfo {author} {\bibfnamefont {Y.}~\bibnamefont
  {Hoshi}}, \bibinfo {author} {\bibfnamefont {N.}~\bibnamefont {Usami}},
  \bibinfo {author} {\bibfnamefont {K.~M.}\ \bibnamefont {Itoh}},\ and\
  \bibinfo {author} {\bibfnamefont {S.}~\bibnamefont {Tarucha}},\ }\bibfield
  {title} {\bibinfo {title} {A quantum-dot spin qubit with coherence limited by
  charge noise and fidelity higher than 99.9\%},\ }\href
  {https://doi.org/10.1038/s41565-017-0014-x} {\bibfield  {journal} {\bibinfo
  {journal} {Nature Nanotechnology}\ }\textbf {\bibinfo {volume} {13}},\
  \bibinfo {pages} {102} (\bibinfo {year} {2018})}\BibitemShut {NoStop}%
\bibitem [{\citenamefont {Asaad}\ \emph {et~al.}(2020)\citenamefont {Asaad},
  \citenamefont {Mourik}, \citenamefont {Joecker}, \citenamefont {Johnson},
  \citenamefont {Baczewski}, \citenamefont {Firgau}, \citenamefont {Mądzik},
  \citenamefont {Schmitt}, \citenamefont {Pla}, \citenamefont {Hudson},
  \citenamefont {Itoh}, \citenamefont {McCallum}, \citenamefont {Dzurak},
  \citenamefont {Laucht},\ and\ \citenamefont {Morello}}]{asaad_coherent_2020}%
  \BibitemOpen
  \bibfield  {author} {\bibinfo {author} {\bibfnamefont {S.}~\bibnamefont
  {Asaad}}, \bibinfo {author} {\bibfnamefont {V.}~\bibnamefont {Mourik}},
  \bibinfo {author} {\bibfnamefont {B.}~\bibnamefont {Joecker}}, \bibinfo
  {author} {\bibfnamefont {M.~A.~I.}\ \bibnamefont {Johnson}}, \bibinfo
  {author} {\bibfnamefont {A.~D.}\ \bibnamefont {Baczewski}}, \bibinfo {author}
  {\bibfnamefont {H.~R.}\ \bibnamefont {Firgau}}, \bibinfo {author}
  {\bibfnamefont {M.~T.}\ \bibnamefont {Mądzik}}, \bibinfo {author}
  {\bibfnamefont {V.}~\bibnamefont {Schmitt}}, \bibinfo {author} {\bibfnamefont
  {J.~J.}\ \bibnamefont {Pla}}, \bibinfo {author} {\bibfnamefont {F.~E.}\
  \bibnamefont {Hudson}}, \bibinfo {author} {\bibfnamefont {K.~M.}\
  \bibnamefont {Itoh}}, \bibinfo {author} {\bibfnamefont {J.~C.}\ \bibnamefont
  {McCallum}}, \bibinfo {author} {\bibfnamefont {A.~S.}\ \bibnamefont
  {Dzurak}}, \bibinfo {author} {\bibfnamefont {A.}~\bibnamefont {Laucht}},\
  and\ \bibinfo {author} {\bibfnamefont {A.}~\bibnamefont {Morello}},\
  }\bibfield  {title} {\bibinfo {title} {Coherent electrical control of a
  single high-spin nucleus in silicon},\ }\href
  {https://doi.org/10.1038/s41586-020-2057-7} {\bibfield  {journal} {\bibinfo
  {journal} {Nature}\ }\textbf {\bibinfo {volume} {579}},\ \bibinfo {pages}
  {205} (\bibinfo {year} {2020})},\ \bibinfo {note} {number: 7798 Publisher:
  Nature Publishing Group}\BibitemShut {NoStop}%
\bibitem [{\citenamefont {Sigillito}\ \emph {et~al.}(2019)\citenamefont
  {Sigillito}, \citenamefont {Loy}, \citenamefont {Zajac}, \citenamefont
  {Gullans}, \citenamefont {Edge},\ and\ \citenamefont
  {Petta}}]{sigillito_site-selective_2019}%
  \BibitemOpen
  \bibfield  {author} {\bibinfo {author} {\bibfnamefont {A.}~\bibnamefont
  {Sigillito}}, \bibinfo {author} {\bibfnamefont {J.}~\bibnamefont {Loy}},
  \bibinfo {author} {\bibfnamefont {D.}~\bibnamefont {Zajac}}, \bibinfo
  {author} {\bibfnamefont {M.}~\bibnamefont {Gullans}}, \bibinfo {author}
  {\bibfnamefont {L.}~\bibnamefont {Edge}},\ and\ \bibinfo {author}
  {\bibfnamefont {J.}~\bibnamefont {Petta}},\ }\bibfield  {title} {\bibinfo
  {title} {Site-{Selective} {Quantum} {Control} in an {Isotopically} {Enriched}
  {$^{28}\mathrm{Si}/{\mathrm{Si}}_{0.7}{\mathrm{Ge}}_{0.3}$} {Quadruple}
  {Quantum} {Dot}},\ }\href {https://doi.org/10.1103/PhysRevApplied.11.061006}
  {\bibfield  {journal} {\bibinfo  {journal} {Physical Review Applied}\
  }\textbf {\bibinfo {volume} {11}},\ \bibinfo {pages} {061006} (\bibinfo
  {year} {2019})},\ \bibinfo {note} {publisher: American Physical
  Society}\BibitemShut {NoStop}%
\bibitem [{\citenamefont {Simmons}\ \emph {et~al.}(2011)\citenamefont
  {Simmons}, \citenamefont {Prance}, \citenamefont {Van~Bael}, \citenamefont
  {Koh}, \citenamefont {Shi}, \citenamefont {Savage}, \citenamefont {Lagally},
  \citenamefont {Joynt}, \citenamefont {Friesen}, \citenamefont {Coppersmith},\
  and\ \citenamefont {Eriksson}}]{simmons_tunable_2011}%
  \BibitemOpen
  \bibfield  {author} {\bibinfo {author} {\bibfnamefont {C.~B.}\ \bibnamefont
  {Simmons}}, \bibinfo {author} {\bibfnamefont {J.~R.}\ \bibnamefont {Prance}},
  \bibinfo {author} {\bibfnamefont {B.~J.}\ \bibnamefont {Van~Bael}}, \bibinfo
  {author} {\bibfnamefont {T.~S.}\ \bibnamefont {Koh}}, \bibinfo {author}
  {\bibfnamefont {Z.}~\bibnamefont {Shi}}, \bibinfo {author} {\bibfnamefont
  {D.~E.}\ \bibnamefont {Savage}}, \bibinfo {author} {\bibfnamefont {M.~G.}\
  \bibnamefont {Lagally}}, \bibinfo {author} {\bibfnamefont {R.}~\bibnamefont
  {Joynt}}, \bibinfo {author} {\bibfnamefont {M.}~\bibnamefont {Friesen}},
  \bibinfo {author} {\bibfnamefont {S.~N.}\ \bibnamefont {Coppersmith}},\ and\
  \bibinfo {author} {\bibfnamefont {M.~A.}\ \bibnamefont {Eriksson}},\
  }\bibfield  {title} {\bibinfo {title} {Tunable {Spin} {Loading} and
  \$\{{T}\}\_\{1\}\$ of a {Silicon} {Spin} {Qubit} {Measured} by
  {Single}-{Shot} {Readout}},\ }\href
  {https://doi.org/10.1103/PhysRevLett.106.156804} {\bibfield  {journal}
  {\bibinfo  {journal} {Physical Review Letters}\ }\textbf {\bibinfo {volume}
  {106}},\ \bibinfo {pages} {156804} (\bibinfo {year} {2011})},\ \bibinfo
  {note} {publisher: American Physical Society}\BibitemShut {NoStop}%
\bibitem [{\citenamefont {Zajac}\ \emph {et~al.}(2016)\citenamefont {Zajac},
  \citenamefont {Hazard}, \citenamefont {Mi}, \citenamefont {Nielsen},\ and\
  \citenamefont {Petta}}]{zajac_scalable_2016}%
  \BibitemOpen
  \bibfield  {author} {\bibinfo {author} {\bibfnamefont {D.}~\bibnamefont
  {Zajac}}, \bibinfo {author} {\bibfnamefont {T.}~\bibnamefont {Hazard}},
  \bibinfo {author} {\bibfnamefont {X.}~\bibnamefont {Mi}}, \bibinfo {author}
  {\bibfnamefont {E.}~\bibnamefont {Nielsen}},\ and\ \bibinfo {author}
  {\bibfnamefont {J.}~\bibnamefont {Petta}},\ }\bibfield  {title} {\bibinfo
  {title} {Scalable {Gate} {Architecture} for a {One}-{Dimensional} {Array} of
  {Semiconductor} {Spin} {Qubits}},\ }\href
  {https://doi.org/10.1103/PhysRevApplied.6.054013} {\bibfield  {journal}
  {\bibinfo  {journal} {Physical Review Applied}\ }\textbf {\bibinfo {volume}
  {6}},\ \bibinfo {pages} {054013} (\bibinfo {year} {2016})},\ \bibinfo {note}
  {publisher: American Physical Society}\BibitemShut {NoStop}%
\bibitem [{\citenamefont {Watson}\ \emph {et~al.}(2018)\citenamefont {Watson},
  \citenamefont {Philips}, \citenamefont {Kawakami}, \citenamefont {Ward},
  \citenamefont {Scarlino}, \citenamefont {Veldhorst}, \citenamefont {Savage},
  \citenamefont {Lagally}, \citenamefont {Friesen}, \citenamefont
  {Coppersmith}, \citenamefont {Eriksson},\ and\ \citenamefont
  {Vandersypen}}]{watson_programmable_2018}%
  \BibitemOpen
  \bibfield  {author} {\bibinfo {author} {\bibfnamefont {T.~F.}\ \bibnamefont
  {Watson}}, \bibinfo {author} {\bibfnamefont {S.~G.~J.}\ \bibnamefont
  {Philips}}, \bibinfo {author} {\bibfnamefont {E.}~\bibnamefont {Kawakami}},
  \bibinfo {author} {\bibfnamefont {D.~R.}\ \bibnamefont {Ward}}, \bibinfo
  {author} {\bibfnamefont {P.}~\bibnamefont {Scarlino}}, \bibinfo {author}
  {\bibfnamefont {M.}~\bibnamefont {Veldhorst}}, \bibinfo {author}
  {\bibfnamefont {D.~E.}\ \bibnamefont {Savage}}, \bibinfo {author}
  {\bibfnamefont {M.~G.}\ \bibnamefont {Lagally}}, \bibinfo {author}
  {\bibfnamefont {M.}~\bibnamefont {Friesen}}, \bibinfo {author} {\bibfnamefont
  {S.~N.}\ \bibnamefont {Coppersmith}}, \bibinfo {author} {\bibfnamefont
  {M.~A.}\ \bibnamefont {Eriksson}},\ and\ \bibinfo {author} {\bibfnamefont
  {L.~M.~K.}\ \bibnamefont {Vandersypen}},\ }\bibfield  {title} {\bibinfo
  {title} {A programmable two-qubit quantum processor in silicon},\ }\href
  {https://doi.org/10.1038/nature25766} {\bibfield  {journal} {\bibinfo
  {journal} {Nature}\ }\textbf {\bibinfo {volume} {555}},\ \bibinfo {pages}
  {633} (\bibinfo {year} {2018})}\BibitemShut {NoStop}%
\bibitem [{\citenamefont {Veldhorst}\ \emph {et~al.}(2015)\citenamefont
  {Veldhorst}, \citenamefont {Yang}, \citenamefont {Hwang}, \citenamefont
  {Huang}, \citenamefont {Dehollain}, \citenamefont {Muhonen}, \citenamefont
  {Simmons}, \citenamefont {Laucht}, \citenamefont {Hudson}, \citenamefont
  {Itoh}, \citenamefont {Morello},\ and\ \citenamefont
  {Dzurak}}]{veldhorst_two-qubit_2015}%
  \BibitemOpen
  \bibfield  {author} {\bibinfo {author} {\bibfnamefont {M.}~\bibnamefont
  {Veldhorst}}, \bibinfo {author} {\bibfnamefont {C.~H.}\ \bibnamefont {Yang}},
  \bibinfo {author} {\bibfnamefont {J.~C.~C.}\ \bibnamefont {Hwang}}, \bibinfo
  {author} {\bibfnamefont {W.}~\bibnamefont {Huang}}, \bibinfo {author}
  {\bibfnamefont {J.~P.}\ \bibnamefont {Dehollain}}, \bibinfo {author}
  {\bibfnamefont {J.~T.}\ \bibnamefont {Muhonen}}, \bibinfo {author}
  {\bibfnamefont {S.}~\bibnamefont {Simmons}}, \bibinfo {author} {\bibfnamefont
  {A.}~\bibnamefont {Laucht}}, \bibinfo {author} {\bibfnamefont {F.~E.}\
  \bibnamefont {Hudson}}, \bibinfo {author} {\bibfnamefont {K.~M.}\
  \bibnamefont {Itoh}}, \bibinfo {author} {\bibfnamefont {A.}~\bibnamefont
  {Morello}},\ and\ \bibinfo {author} {\bibfnamefont {A.~S.}\ \bibnamefont
  {Dzurak}},\ }\bibfield  {title} {\bibinfo {title} {A two-qubit logic gate in
  silicon},\ }\href {https://doi.org/10.1038/nature15263} {\bibfield  {journal}
  {\bibinfo  {journal} {Nature}\ }\textbf {\bibinfo {volume} {526}},\ \bibinfo
  {pages} {410} (\bibinfo {year} {2015})},\ \bibinfo {note} {number: 7573
  Publisher: Nature Publishing Group}\BibitemShut {NoStop}%
\bibitem [{\citenamefont {Morello}\ \emph {et~al.}(2010)\citenamefont
  {Morello}, \citenamefont {Pla}, \citenamefont {Zwanenburg}, \citenamefont
  {Chan}, \citenamefont {Tan}, \citenamefont {Huebl}, \citenamefont
  {Möttönen}, \citenamefont {Nugroho}, \citenamefont {Yang}, \citenamefont
  {van Donkelaar}, \citenamefont {Alves}, \citenamefont {Jamieson},
  \citenamefont {Escott}, \citenamefont {Hollenberg}, \citenamefont {Clark},\
  and\ \citenamefont {Dzurak}}]{morello_single-shot_2010}%
  \BibitemOpen
  \bibfield  {author} {\bibinfo {author} {\bibfnamefont {A.}~\bibnamefont
  {Morello}}, \bibinfo {author} {\bibfnamefont {J.~J.}\ \bibnamefont {Pla}},
  \bibinfo {author} {\bibfnamefont {F.~A.}\ \bibnamefont {Zwanenburg}},
  \bibinfo {author} {\bibfnamefont {K.~W.}\ \bibnamefont {Chan}}, \bibinfo
  {author} {\bibfnamefont {K.~Y.}\ \bibnamefont {Tan}}, \bibinfo {author}
  {\bibfnamefont {H.}~\bibnamefont {Huebl}}, \bibinfo {author} {\bibfnamefont
  {M.}~\bibnamefont {Möttönen}}, \bibinfo {author} {\bibfnamefont {C.~D.}\
  \bibnamefont {Nugroho}}, \bibinfo {author} {\bibfnamefont {C.}~\bibnamefont
  {Yang}}, \bibinfo {author} {\bibfnamefont {J.~A.}\ \bibnamefont {van
  Donkelaar}}, \bibinfo {author} {\bibfnamefont {A.~D.~C.}\ \bibnamefont
  {Alves}}, \bibinfo {author} {\bibfnamefont {D.~N.}\ \bibnamefont {Jamieson}},
  \bibinfo {author} {\bibfnamefont {C.~C.}\ \bibnamefont {Escott}}, \bibinfo
  {author} {\bibfnamefont {L.~C.~L.}\ \bibnamefont {Hollenberg}}, \bibinfo
  {author} {\bibfnamefont {R.~G.}\ \bibnamefont {Clark}},\ and\ \bibinfo
  {author} {\bibfnamefont {A.~S.}\ \bibnamefont {Dzurak}},\ }\bibfield  {title}
  {\bibinfo {title} {Single-shot readout of an electron spin in silicon},\
  }\href {https://doi.org/10.1038/nature09392} {\bibfield  {journal} {\bibinfo
  {journal} {Nature}\ }\textbf {\bibinfo {volume} {467}},\ \bibinfo {pages}
  {687} (\bibinfo {year} {2010})}\BibitemShut {NoStop}%
\bibitem [{\citenamefont {Zheng}\ \emph {et~al.}(2019)\citenamefont {Zheng},
  \citenamefont {Samkharadze}, \citenamefont {Noordam}, \citenamefont {Kalhor},
  \citenamefont {Brousse}, \citenamefont {Sammak}, \citenamefont {Scappucci},\
  and\ \citenamefont {Vandersypen}}]{zheng_rapid_2019}%
  \BibitemOpen
  \bibfield  {author} {\bibinfo {author} {\bibfnamefont {G.}~\bibnamefont
  {Zheng}}, \bibinfo {author} {\bibfnamefont {N.}~\bibnamefont {Samkharadze}},
  \bibinfo {author} {\bibfnamefont {M.~L.}\ \bibnamefont {Noordam}}, \bibinfo
  {author} {\bibfnamefont {N.}~\bibnamefont {Kalhor}}, \bibinfo {author}
  {\bibfnamefont {D.}~\bibnamefont {Brousse}}, \bibinfo {author} {\bibfnamefont
  {A.}~\bibnamefont {Sammak}}, \bibinfo {author} {\bibfnamefont
  {G.}~\bibnamefont {Scappucci}},\ and\ \bibinfo {author} {\bibfnamefont
  {L.~M.~K.}\ \bibnamefont {Vandersypen}},\ }\bibfield  {title} {\bibinfo
  {title} {Rapid gate-based spin read-out in silicon using an on-chip
  resonator},\ }\href {https://doi.org/10.1038/s41565-019-0488-9} {\bibfield
  {journal} {\bibinfo  {journal} {Nature Nanotechnology}\ }\textbf {\bibinfo
  {volume} {14}},\ \bibinfo {pages} {742} (\bibinfo {year} {2019})}\BibitemShut
  {NoStop}%
\bibitem [{\citenamefont {Keith}\ \emph {et~al.}(2019)\citenamefont {Keith},
  \citenamefont {House}, \citenamefont {Donnelly}, \citenamefont {Watson},
  \citenamefont {Weber},\ and\ \citenamefont
  {Simmons}}]{keith_single-shot_2019}%
  \BibitemOpen
  \bibfield  {author} {\bibinfo {author} {\bibfnamefont {D.}~\bibnamefont
  {Keith}}, \bibinfo {author} {\bibfnamefont {M.}~\bibnamefont {House}},
  \bibinfo {author} {\bibfnamefont {M.}~\bibnamefont {Donnelly}}, \bibinfo
  {author} {\bibfnamefont {T.}~\bibnamefont {Watson}}, \bibinfo {author}
  {\bibfnamefont {B.}~\bibnamefont {Weber}},\ and\ \bibinfo {author}
  {\bibfnamefont {M.}~\bibnamefont {Simmons}},\ }\bibfield  {title} {\bibinfo
  {title} {Single-{Shot} {Spin} {Readout} in {Semiconductors} {Near} the
  {Shot}-{Noise} {Sensitivity} {Limit}},\ }\href
  {https://doi.org/10.1103/PhysRevX.9.041003} {\bibfield  {journal} {\bibinfo
  {journal} {Physical Review X}\ }\textbf {\bibinfo {volume} {9}},\ \bibinfo
  {pages} {041003} (\bibinfo {year} {2019})},\ \bibinfo {note} {publisher:
  American Physical Society}\BibitemShut {NoStop}%
\bibitem [{\citenamefont {Zhao}\ \emph {et~al.}(2019)\citenamefont {Zhao},
  \citenamefont {Tanttu}, \citenamefont {Tan}, \citenamefont {Hensen},
  \citenamefont {Chan}, \citenamefont {Hwang}, \citenamefont {Leon},
  \citenamefont {Yang}, \citenamefont {Gilbert}, \citenamefont {Hudson},
  \citenamefont {Itoh}, \citenamefont {Kiselev}, \citenamefont {Ladd},
  \citenamefont {Morello}, \citenamefont {Laucht},\ and\ \citenamefont
  {Dzurak}}]{zhao_single-spin_2019}%
  \BibitemOpen
  \bibfield  {author} {\bibinfo {author} {\bibfnamefont {R.}~\bibnamefont
  {Zhao}}, \bibinfo {author} {\bibfnamefont {T.}~\bibnamefont {Tanttu}},
  \bibinfo {author} {\bibfnamefont {K.~Y.}\ \bibnamefont {Tan}}, \bibinfo
  {author} {\bibfnamefont {B.}~\bibnamefont {Hensen}}, \bibinfo {author}
  {\bibfnamefont {K.~W.}\ \bibnamefont {Chan}}, \bibinfo {author}
  {\bibfnamefont {J.~C.~C.}\ \bibnamefont {Hwang}}, \bibinfo {author}
  {\bibfnamefont {R.~C.~C.}\ \bibnamefont {Leon}}, \bibinfo {author}
  {\bibfnamefont {C.~H.}\ \bibnamefont {Yang}}, \bibinfo {author}
  {\bibfnamefont {W.}~\bibnamefont {Gilbert}}, \bibinfo {author} {\bibfnamefont
  {F.~E.}\ \bibnamefont {Hudson}}, \bibinfo {author} {\bibfnamefont {K.~M.}\
  \bibnamefont {Itoh}}, \bibinfo {author} {\bibfnamefont {A.~A.}\ \bibnamefont
  {Kiselev}}, \bibinfo {author} {\bibfnamefont {T.~D.}\ \bibnamefont {Ladd}},
  \bibinfo {author} {\bibfnamefont {A.}~\bibnamefont {Morello}}, \bibinfo
  {author} {\bibfnamefont {A.}~\bibnamefont {Laucht}},\ and\ \bibinfo {author}
  {\bibfnamefont {A.~S.}\ \bibnamefont {Dzurak}},\ }\bibfield  {title}
  {\bibinfo {title} {Single-spin qubits in isotopically enriched silicon at low
  magnetic field},\ }\href {https://doi.org/10.1038/s41467-019-13416-7}
  {\bibfield  {journal} {\bibinfo  {journal} {Nature Communications}\ }\textbf
  {\bibinfo {volume} {10}},\ \bibinfo {pages} {5500} (\bibinfo {year}
  {2019})}\BibitemShut {NoStop}%
\bibitem [{\citenamefont {West}\ \emph {et~al.}(2019)\citenamefont {West},
  \citenamefont {Hensen}, \citenamefont {Jouan}, \citenamefont {Tanttu},
  \citenamefont {Yang}, \citenamefont {Rossi}, \citenamefont {Gonzalez-Zalba},
  \citenamefont {Hudson}, \citenamefont {Morello}, \citenamefont {Reilly},\
  and\ \citenamefont {Dzurak}}]{west_gate-based_2019}%
  \BibitemOpen
  \bibfield  {author} {\bibinfo {author} {\bibfnamefont {A.}~\bibnamefont
  {West}}, \bibinfo {author} {\bibfnamefont {B.}~\bibnamefont {Hensen}},
  \bibinfo {author} {\bibfnamefont {A.}~\bibnamefont {Jouan}}, \bibinfo
  {author} {\bibfnamefont {T.}~\bibnamefont {Tanttu}}, \bibinfo {author}
  {\bibfnamefont {C.-H.}\ \bibnamefont {Yang}}, \bibinfo {author}
  {\bibfnamefont {A.}~\bibnamefont {Rossi}}, \bibinfo {author} {\bibfnamefont
  {M.~F.}\ \bibnamefont {Gonzalez-Zalba}}, \bibinfo {author} {\bibfnamefont
  {F.}~\bibnamefont {Hudson}}, \bibinfo {author} {\bibfnamefont
  {A.}~\bibnamefont {Morello}}, \bibinfo {author} {\bibfnamefont {D.~J.}\
  \bibnamefont {Reilly}},\ and\ \bibinfo {author} {\bibfnamefont {A.~S.}\
  \bibnamefont {Dzurak}},\ }\bibfield  {title} {\bibinfo {title} {Gate-based
  single-shot readout of spins in silicon},\ }\href
  {https://doi.org/10.1038/s41565-019-0400-7} {\bibfield  {journal} {\bibinfo
  {journal} {Nature Nanotechnology}\ }\textbf {\bibinfo {volume} {14}},\
  \bibinfo {pages} {437} (\bibinfo {year} {2019})},\ \bibinfo {note} {number: 5
  Publisher: Nature Publishing Group}\BibitemShut {NoStop}%
\bibitem [{\citenamefont {Crippa}\ \emph {et~al.}(2019)\citenamefont {Crippa},
  \citenamefont {Ezzouch}, \citenamefont {Aprá}, \citenamefont {Amisse},
  \citenamefont {Laviéville}, \citenamefont {Hutin}, \citenamefont {Bertrand},
  \citenamefont {Vinet}, \citenamefont {Urdampilleta}, \citenamefont {Meunier},
  \citenamefont {Sanquer}, \citenamefont {Jehl}, \citenamefont {Maurand},\ and\
  \citenamefont {De~Franceschi}}]{crippa_gate-reflectometry_2019}%
  \BibitemOpen
  \bibfield  {author} {\bibinfo {author} {\bibfnamefont {A.}~\bibnamefont
  {Crippa}}, \bibinfo {author} {\bibfnamefont {R.}~\bibnamefont {Ezzouch}},
  \bibinfo {author} {\bibfnamefont {A.}~\bibnamefont {Aprá}}, \bibinfo
  {author} {\bibfnamefont {A.}~\bibnamefont {Amisse}}, \bibinfo {author}
  {\bibfnamefont {R.}~\bibnamefont {Laviéville}}, \bibinfo {author}
  {\bibfnamefont {L.}~\bibnamefont {Hutin}}, \bibinfo {author} {\bibfnamefont
  {B.}~\bibnamefont {Bertrand}}, \bibinfo {author} {\bibfnamefont
  {M.}~\bibnamefont {Vinet}}, \bibinfo {author} {\bibfnamefont
  {M.}~\bibnamefont {Urdampilleta}}, \bibinfo {author} {\bibfnamefont
  {T.}~\bibnamefont {Meunier}}, \bibinfo {author} {\bibfnamefont
  {M.}~\bibnamefont {Sanquer}}, \bibinfo {author} {\bibfnamefont
  {X.}~\bibnamefont {Jehl}}, \bibinfo {author} {\bibfnamefont {R.}~\bibnamefont
  {Maurand}},\ and\ \bibinfo {author} {\bibfnamefont {S.}~\bibnamefont
  {De~Franceschi}},\ }\bibfield  {title} {\bibinfo {title} {Gate-reflectometry
  dispersive readout and coherent control of a spin qubit in silicon},\ }\href
  {https://doi.org/10.1038/s41467-019-10848-z} {\bibfield  {journal} {\bibinfo
  {journal} {Nature Communications}\ }\textbf {\bibinfo {volume} {10}},\
  \bibinfo {pages} {2776} (\bibinfo {year} {2019})},\ \bibinfo {note} {number:
  1 Publisher: Nature Publishing Group}\BibitemShut {NoStop}%
\bibitem [{\citenamefont {Gonzalez-Zalba}\ \emph {et~al.}(2015)\citenamefont
  {Gonzalez-Zalba}, \citenamefont {Barraud}, \citenamefont {Ferguson},\ and\
  \citenamefont {Betz}}]{gonzalez-zalba_probing_2015}%
  \BibitemOpen
  \bibfield  {author} {\bibinfo {author} {\bibfnamefont {M.~F.}\ \bibnamefont
  {Gonzalez-Zalba}}, \bibinfo {author} {\bibfnamefont {S.}~\bibnamefont
  {Barraud}}, \bibinfo {author} {\bibfnamefont {A.~J.}\ \bibnamefont
  {Ferguson}},\ and\ \bibinfo {author} {\bibfnamefont {A.~C.}\ \bibnamefont
  {Betz}},\ }\bibfield  {title} {\bibinfo {title} {Probing the limits of
  gate-based charge sensing},\ }\href {https://doi.org/10.1038/ncomms7084}
  {\bibfield  {journal} {\bibinfo  {journal} {Nature Communications}\ }\textbf
  {\bibinfo {volume} {6}},\ \bibinfo {pages} {6084} (\bibinfo {year} {2015})},\
  \bibinfo {note} {number: 1 Publisher: Nature Publishing Group}\BibitemShut
  {NoStop}%
\bibitem [{\citenamefont {Hu}(2019)}]{hu_fast_2019}%
  \BibitemOpen
  \bibfield  {author} {\bibinfo {author} {\bibfnamefont {X.}~\bibnamefont
  {Hu}},\ }\bibfield  {title} {\bibinfo {title} {Fast and space-efficient spin
  sensing},\ }\href {https://doi.org/10.1038/s41565-019-0516-9} {\bibfield
  {journal} {\bibinfo  {journal} {Nature Nanotechnology}\ }\textbf {\bibinfo
  {volume} {14}},\ \bibinfo {pages} {735} (\bibinfo {year} {2019})},\ \bibinfo
  {note} {number: 8 Publisher: Nature Publishing Group}\BibitemShut {NoStop}%
\bibitem [{\citenamefont {Hu}\ and\ \citenamefont
  {Das~Sarma}(2006)}]{hu_charge-fluctuation-induced_2006}%
  \BibitemOpen
  \bibfield  {author} {\bibinfo {author} {\bibfnamefont {X.}~\bibnamefont
  {Hu}}\ and\ \bibinfo {author} {\bibfnamefont {S.}~\bibnamefont {Das~Sarma}},\
  }\bibfield  {title} {\bibinfo {title} {Charge-{Fluctuation}-{Induced}
  {Dephasing} of {Exchange}-{Coupled} {Spin} {Qubits}},\ }\href
  {https://doi.org/10.1103/PhysRevLett.96.100501} {\bibfield  {journal}
  {\bibinfo  {journal} {Physical Review Letters}\ }\textbf {\bibinfo {volume}
  {96}},\ \bibinfo {pages} {100501} (\bibinfo {year} {2006})}\BibitemShut
  {NoStop}%
\bibitem [{\citenamefont {Struck}\ \emph {et~al.}(2019)\citenamefont {Struck},
  \citenamefont {Hollmann}, \citenamefont {Schauer}, \citenamefont {Fedorets},
  \citenamefont {Schmidbauer}, \citenamefont {Sawano}, \citenamefont {Riemann},
  \citenamefont {Abrosimov}, \citenamefont {Cywinski}, \citenamefont
  {Bougeard},\ and\ \citenamefont {Schreiber}}]{struck_low-frequency_2019}%
  \BibitemOpen
  \bibfield  {author} {\bibinfo {author} {\bibfnamefont {T.}~\bibnamefont
  {Struck}}, \bibinfo {author} {\bibfnamefont {A.}~\bibnamefont {Hollmann}},
  \bibinfo {author} {\bibfnamefont {F.}~\bibnamefont {Schauer}}, \bibinfo
  {author} {\bibfnamefont {O.}~\bibnamefont {Fedorets}}, \bibinfo {author}
  {\bibfnamefont {A.}~\bibnamefont {Schmidbauer}}, \bibinfo {author}
  {\bibfnamefont {K.}~\bibnamefont {Sawano}}, \bibinfo {author} {\bibfnamefont
  {H.}~\bibnamefont {Riemann}}, \bibinfo {author} {\bibfnamefont {N.~V.}\
  \bibnamefont {Abrosimov}}, \bibinfo {author} {\bibfnamefont {L.}~\bibnamefont
  {Cywinski}}, \bibinfo {author} {\bibfnamefont {D.}~\bibnamefont {Bougeard}},\
  and\ \bibinfo {author} {\bibfnamefont {L.~R.}\ \bibnamefont {Schreiber}},\
  }\bibfield  {title} {\bibinfo {title} {Low-frequency spin qubit detuning
  noise in highly purified {$^{28}\mathrm{Si}/{\mathrm{Si}}{\mathrm{Ge}}$}},\
  }\bibfield  {journal} {\bibinfo  {journal} {arXiv:1909.11397 [cond-mat,
  physics:quant-ph]}\ }\href {https://doi.org/10.1038/s41534-020-0276-2}
  {10.1038/s41534-020-0276-2} (\bibinfo {year} {2019}),\ \bibinfo {note}
  {arXiv: 1909.11397}\BibitemShut {NoStop}%
\bibitem [{\citenamefont {Huang}\ and\ \citenamefont
  {Hu}(2020)}]{huang_impact_2020}%
  \BibitemOpen
  \bibfield  {author} {\bibinfo {author} {\bibfnamefont {P.}~\bibnamefont
  {Huang}}\ and\ \bibinfo {author} {\bibfnamefont {X.}~\bibnamefont {Hu}},\
  }\bibfield  {title} {\bibinfo {title} {Impact of
  \${\textbackslash}mathcal\{{T}\}\$-symmetry on decoherence and control for an
  electron spin in a synthetic spin-orbit field},\ }\href
  {http://arxiv.org/abs/2008.04671} {\bibfield  {journal} {\bibinfo  {journal}
  {arXiv:2008.04671 [cond-mat, physics:quant-ph]}\ } (\bibinfo {year}
  {2020})},\ \bibinfo {note} {arXiv: 2008.04671}\BibitemShut {NoStop}%
\bibitem [{\citenamefont {Shulman}\ \emph {et~al.}(2012)\citenamefont
  {Shulman}, \citenamefont {Dial}, \citenamefont {Harvey}, \citenamefont
  {Bluhm}, \citenamefont {Umansky},\ and\ \citenamefont
  {Yacoby}}]{shulman_demonstration_2012}%
  \BibitemOpen
  \bibfield  {author} {\bibinfo {author} {\bibfnamefont {M.~D.}\ \bibnamefont
  {Shulman}}, \bibinfo {author} {\bibfnamefont {O.~E.}\ \bibnamefont {Dial}},
  \bibinfo {author} {\bibfnamefont {S.~P.}\ \bibnamefont {Harvey}}, \bibinfo
  {author} {\bibfnamefont {H.}~\bibnamefont {Bluhm}}, \bibinfo {author}
  {\bibfnamefont {V.}~\bibnamefont {Umansky}},\ and\ \bibinfo {author}
  {\bibfnamefont {A.}~\bibnamefont {Yacoby}},\ }\bibfield  {title} {\bibinfo
  {title} {Demonstration of {Entanglement} of {Electrostatically} {Coupled}
  {Singlet}-{Triplet} {Qubits}},\ }\href
  {https://doi.org/10.1126/science.1217692} {\bibfield  {journal} {\bibinfo
  {journal} {Science}\ }\textbf {\bibinfo {volume} {336}},\ \bibinfo {pages}
  {202} (\bibinfo {year} {2012})},\ \bibinfo {note} {arXiv:
  1202.1828}\BibitemShut {NoStop}%
\bibitem [{\citenamefont {Mi}\ \emph {et~al.}(2017{\natexlab{a}})\citenamefont
  {Mi}, \citenamefont {Cady}, \citenamefont {Zajac}, \citenamefont {Stehlik},
  \citenamefont {Edge},\ and\ \citenamefont {Petta}}]{mi_circuit_2017}%
  \BibitemOpen
  \bibfield  {author} {\bibinfo {author} {\bibfnamefont {X.}~\bibnamefont
  {Mi}}, \bibinfo {author} {\bibfnamefont {J.~V.}\ \bibnamefont {Cady}},
  \bibinfo {author} {\bibfnamefont {D.~M.}\ \bibnamefont {Zajac}}, \bibinfo
  {author} {\bibfnamefont {J.}~\bibnamefont {Stehlik}}, \bibinfo {author}
  {\bibfnamefont {L.~F.}\ \bibnamefont {Edge}},\ and\ \bibinfo {author}
  {\bibfnamefont {J.~R.}\ \bibnamefont {Petta}},\ }\bibfield  {title} {\bibinfo
  {title} {Circuit quantum electrodynamics architecture for gate-defined
  quantum dots in silicon},\ }\href {https://doi.org/10.1063/1.4974536}
  {\bibfield  {journal} {\bibinfo  {journal} {Applied Physics Letters}\
  }\textbf {\bibinfo {volume} {110}},\ \bibinfo {pages} {043502} (\bibinfo
  {year} {2017}{\natexlab{a}})},\ \bibinfo {note} {publisher: American
  Institute of Physics}\BibitemShut {NoStop}%
\bibitem [{\citenamefont {Mi}\ \emph {et~al.}(2018)\citenamefont {Mi},
  \citenamefont {Benito}, \citenamefont {Putz}, \citenamefont {Zajac},
  \citenamefont {Taylor}, \citenamefont {Burkard},\ and\ \citenamefont
  {Petta}}]{mi_coherent_2018}%
  \BibitemOpen
  \bibfield  {author} {\bibinfo {author} {\bibfnamefont {X.}~\bibnamefont
  {Mi}}, \bibinfo {author} {\bibfnamefont {M.}~\bibnamefont {Benito}}, \bibinfo
  {author} {\bibfnamefont {S.}~\bibnamefont {Putz}}, \bibinfo {author}
  {\bibfnamefont {D.~M.}\ \bibnamefont {Zajac}}, \bibinfo {author}
  {\bibfnamefont {J.~M.}\ \bibnamefont {Taylor}}, \bibinfo {author}
  {\bibfnamefont {G.}~\bibnamefont {Burkard}},\ and\ \bibinfo {author}
  {\bibfnamefont {J.~R.}\ \bibnamefont {Petta}},\ }\bibfield  {title} {\bibinfo
  {title} {A coherent spin–photon interface in silicon},\ }\href
  {https://doi.org/10.1038/nature25769} {\bibfield  {journal} {\bibinfo
  {journal} {Nature}\ }\textbf {\bibinfo {volume} {555}},\ \bibinfo {pages}
  {599} (\bibinfo {year} {2018})},\ \bibinfo {note} {number: 7698 Publisher:
  Nature Publishing Group}\BibitemShut {NoStop}%
\bibitem [{\citenamefont {Samkharadze}\ \emph {et~al.}(2018)\citenamefont
  {Samkharadze}, \citenamefont {Zheng}, \citenamefont {Kalhor}, \citenamefont
  {Brousse}, \citenamefont {Sammak}, \citenamefont {Mendes}, \citenamefont
  {Blais}, \citenamefont {Scappucci},\ and\ \citenamefont
  {Vandersypen}}]{samkharadze_strong_2018}%
  \BibitemOpen
  \bibfield  {author} {\bibinfo {author} {\bibfnamefont {N.}~\bibnamefont
  {Samkharadze}}, \bibinfo {author} {\bibfnamefont {G.}~\bibnamefont {Zheng}},
  \bibinfo {author} {\bibfnamefont {N.}~\bibnamefont {Kalhor}}, \bibinfo
  {author} {\bibfnamefont {D.}~\bibnamefont {Brousse}}, \bibinfo {author}
  {\bibfnamefont {A.}~\bibnamefont {Sammak}}, \bibinfo {author} {\bibfnamefont
  {U.~C.}\ \bibnamefont {Mendes}}, \bibinfo {author} {\bibfnamefont
  {A.}~\bibnamefont {Blais}}, \bibinfo {author} {\bibfnamefont
  {G.}~\bibnamefont {Scappucci}},\ and\ \bibinfo {author} {\bibfnamefont
  {L.~M.~K.}\ \bibnamefont {Vandersypen}},\ }\bibfield  {title} {\bibinfo
  {title} {Strong spin-photon coupling in silicon},\ }\href
  {https://doi.org/10.1126/science.aar4054} {\bibfield  {journal} {\bibinfo
  {journal} {Science}\ }\textbf {\bibinfo {volume} {359}},\ \bibinfo {pages}
  {1123} (\bibinfo {year} {2018})}\BibitemShut {NoStop}%
\bibitem [{\citenamefont {Borjans}\ \emph
  {et~al.}(2020{\natexlab{a}})\citenamefont {Borjans}, \citenamefont {Croot},
  \citenamefont {Putz}, \citenamefont {Mi}, \citenamefont {Quinn},
  \citenamefont {Pan}, \citenamefont {Kerckhoff}, \citenamefont {Pritchett},
  \citenamefont {Jackson}, \citenamefont {Edge}, \citenamefont {Ross},
  \citenamefont {Ladd}, \citenamefont {Borselli}, \citenamefont {Gyure},\ and\
  \citenamefont {Petta}}]{borjans_split-gate_2020}%
  \BibitemOpen
  \bibfield  {author} {\bibinfo {author} {\bibfnamefont {F.}~\bibnamefont
  {Borjans}}, \bibinfo {author} {\bibfnamefont {X.}~\bibnamefont {Croot}},
  \bibinfo {author} {\bibfnamefont {S.}~\bibnamefont {Putz}}, \bibinfo {author}
  {\bibfnamefont {X.}~\bibnamefont {Mi}}, \bibinfo {author} {\bibfnamefont
  {S.~M.}\ \bibnamefont {Quinn}}, \bibinfo {author} {\bibfnamefont
  {A.}~\bibnamefont {Pan}}, \bibinfo {author} {\bibfnamefont {J.}~\bibnamefont
  {Kerckhoff}}, \bibinfo {author} {\bibfnamefont {E.~J.}\ \bibnamefont
  {Pritchett}}, \bibinfo {author} {\bibfnamefont {C.~A.}\ \bibnamefont
  {Jackson}}, \bibinfo {author} {\bibfnamefont {L.~F.}\ \bibnamefont {Edge}},
  \bibinfo {author} {\bibfnamefont {R.~S.}\ \bibnamefont {Ross}}, \bibinfo
  {author} {\bibfnamefont {T.~D.}\ \bibnamefont {Ladd}}, \bibinfo {author}
  {\bibfnamefont {M.~G.}\ \bibnamefont {Borselli}}, \bibinfo {author}
  {\bibfnamefont {M.~F.}\ \bibnamefont {Gyure}},\ and\ \bibinfo {author}
  {\bibfnamefont {J.~R.}\ \bibnamefont {Petta}},\ }\bibfield  {title} {\bibinfo
  {title} {Split-{Gate} {Cavity} {Coupler} for {Silicon} {Circuit} {Quantum}
  {Electrodynamics}},\ }\href {http://arxiv.org/abs/2003.01088} {\bibfield
  {journal} {\bibinfo  {journal} {arXiv:2003.01088 [cond-mat,
  physics:quant-ph]}\ } (\bibinfo {year} {2020}{\natexlab{a}})},\ \bibinfo
  {note} {arXiv: 2003.01088}\BibitemShut {NoStop}%
\bibitem [{\citenamefont {Borjans}\ \emph
  {et~al.}(2020{\natexlab{b}})\citenamefont {Borjans}, \citenamefont {Croot},
  \citenamefont {Mi}, \citenamefont {Gullans},\ and\ \citenamefont
  {Petta}}]{borjans_long-range_2020}%
  \BibitemOpen
  \bibfield  {author} {\bibinfo {author} {\bibfnamefont {F.}~\bibnamefont
  {Borjans}}, \bibinfo {author} {\bibfnamefont {X.~G.}\ \bibnamefont {Croot}},
  \bibinfo {author} {\bibfnamefont {X.}~\bibnamefont {Mi}}, \bibinfo {author}
  {\bibfnamefont {M.~J.}\ \bibnamefont {Gullans}},\ and\ \bibinfo {author}
  {\bibfnamefont {J.~R.}\ \bibnamefont {Petta}},\ }\bibfield  {title} {\bibinfo
  {title} {Long-{Range} {Microwave} {Mediated} {Interactions} {Between}
  {Electron} {Spins}},\ }\href {https://doi.org/10.1038/s41586-019-1867-y}
  {\bibfield  {journal} {\bibinfo  {journal} {Nature}\ }\textbf {\bibinfo
  {volume} {577}},\ \bibinfo {pages} {195} (\bibinfo {year}
  {2020}{\natexlab{b}})},\ \bibinfo {note} {arXiv: 1905.00776}\BibitemShut
  {NoStop}%
\bibitem [{\citenamefont {Harvey-Collard}\ \emph {et~al.}(2017)\citenamefont
  {Harvey-Collard}, \citenamefont {Jacobson}, \citenamefont {Rudolph},
  \citenamefont {Dominguez}, \citenamefont {Ten~Eyck}, \citenamefont {Wendt},
  \citenamefont {Pluym}, \citenamefont {Gamble}, \citenamefont {Lilly},
  \citenamefont {Pioro-Ladrière},\ and\ \citenamefont
  {Carroll}}]{harvey-collard_coherent_2017}%
  \BibitemOpen
  \bibfield  {author} {\bibinfo {author} {\bibfnamefont {P.}~\bibnamefont
  {Harvey-Collard}}, \bibinfo {author} {\bibfnamefont {N.~T.}\ \bibnamefont
  {Jacobson}}, \bibinfo {author} {\bibfnamefont {M.}~\bibnamefont {Rudolph}},
  \bibinfo {author} {\bibfnamefont {J.}~\bibnamefont {Dominguez}}, \bibinfo
  {author} {\bibfnamefont {G.~A.}\ \bibnamefont {Ten~Eyck}}, \bibinfo {author}
  {\bibfnamefont {J.~R.}\ \bibnamefont {Wendt}}, \bibinfo {author}
  {\bibfnamefont {T.}~\bibnamefont {Pluym}}, \bibinfo {author} {\bibfnamefont
  {J.~K.}\ \bibnamefont {Gamble}}, \bibinfo {author} {\bibfnamefont {M.~P.}\
  \bibnamefont {Lilly}}, \bibinfo {author} {\bibfnamefont {M.}~\bibnamefont
  {Pioro-Ladrière}},\ and\ \bibinfo {author} {\bibfnamefont {M.~S.}\
  \bibnamefont {Carroll}},\ }\bibfield  {title} {\bibinfo {title} {Coherent
  coupling between a quantum dot and a donor in silicon},\ }\href
  {https://doi.org/10.1038/s41467-017-01113-2} {\bibfield  {journal} {\bibinfo
  {journal} {Nature Communications}\ }\textbf {\bibinfo {volume} {8}},\
  \bibinfo {pages} {1} (\bibinfo {year} {2017})}\BibitemShut {NoStop}%
\bibitem [{\citenamefont {Calderón}\ \emph {et~al.}(2006)\citenamefont
  {Calderón}, \citenamefont {Koiller}, \citenamefont {Hu},\ and\ \citenamefont
  {Das~Sarma}}]{calderon_quantum_2006}%
  \BibitemOpen
  \bibfield  {author} {\bibinfo {author} {\bibfnamefont {M.~J.}\ \bibnamefont
  {Calderón}}, \bibinfo {author} {\bibfnamefont {B.}~\bibnamefont {Koiller}},
  \bibinfo {author} {\bibfnamefont {X.}~\bibnamefont {Hu}},\ and\ \bibinfo
  {author} {\bibfnamefont {S.}~\bibnamefont {Das~Sarma}},\ }\bibfield  {title}
  {\bibinfo {title} {Quantum {Control} of {Donor} {Electrons} at the {SiO2}
  {Interface}},\ }\href {https://doi.org/10.1103/PhysRevLett.96.096802}
  {\bibfield  {journal} {\bibinfo  {journal} {Physical Review Letters}\
  }\textbf {\bibinfo {volume} {96}},\ \bibinfo {pages} {096802} (\bibinfo
  {year} {2006})}\BibitemShut {NoStop}%
\bibitem [{\citenamefont {Schoenfield}\ \emph {et~al.}(2017)\citenamefont
  {Schoenfield}, \citenamefont {Freeman},\ and\ \citenamefont
  {Jiang}}]{Schoenfield2017}%
  \BibitemOpen
  \bibfield  {author} {\bibinfo {author} {\bibfnamefont {J.~S.}\ \bibnamefont
  {Schoenfield}}, \bibinfo {author} {\bibfnamefont {B.~M.}\ \bibnamefont
  {Freeman}},\ and\ \bibinfo {author} {\bibfnamefont {H.}~\bibnamefont
  {Jiang}},\ }\bibfield  {title} {\bibinfo {title} {Coherent manipulation of
  valley states at multiple charge configurations of a silicon quantum dot
  device},\ }\href {https://doi.org/10.1038/s41467-017-00073-x} {\bibfield
  {journal} {\bibinfo  {journal} {Nature Communications}\ }\textbf {\bibinfo
  {volume} {8}},\ \bibinfo {pages} {64} (\bibinfo {year} {2017})}\BibitemShut
  {NoStop}%
\bibitem [{\citenamefont {Li}\ \emph {et~al.}(2015)\citenamefont {Li},
  \citenamefont {Cao}, \citenamefont {Yu}, \citenamefont {Xiao}, \citenamefont
  {Guo}, \citenamefont {Jiang},\ and\ \citenamefont
  {Guo}}]{li_conditional_2015}%
  \BibitemOpen
  \bibfield  {author} {\bibinfo {author} {\bibfnamefont {H.-O.}\ \bibnamefont
  {Li}}, \bibinfo {author} {\bibfnamefont {G.}~\bibnamefont {Cao}}, \bibinfo
  {author} {\bibfnamefont {G.-D.}\ \bibnamefont {Yu}}, \bibinfo {author}
  {\bibfnamefont {M.}~\bibnamefont {Xiao}}, \bibinfo {author} {\bibfnamefont
  {G.-C.}\ \bibnamefont {Guo}}, \bibinfo {author} {\bibfnamefont {H.-W.}\
  \bibnamefont {Jiang}},\ and\ \bibinfo {author} {\bibfnamefont {G.-P.}\
  \bibnamefont {Guo}},\ }\bibfield  {title} {\bibinfo {title} {Conditional
  rotation of two strongly coupled semiconductor charge qubits},\ }\href
  {https://doi.org/10.1038/ncomms8681} {\bibfield  {journal} {\bibinfo
  {journal} {Nature Communications}\ }\textbf {\bibinfo {volume} {6}},\
  \bibinfo {pages} {7681} (\bibinfo {year} {2015})},\ \bibinfo {note} {number:
  1 Publisher: Nature Publishing Group}\BibitemShut {NoStop}%
\bibitem [{\citenamefont {Mi}\ \emph {et~al.}(2017{\natexlab{b}})\citenamefont
  {Mi}, \citenamefont {Cady}, \citenamefont {Zajac}, \citenamefont {Deelman},\
  and\ \citenamefont {Petta}}]{mi_strong_2017}%
  \BibitemOpen
  \bibfield  {author} {\bibinfo {author} {\bibfnamefont {X.}~\bibnamefont
  {Mi}}, \bibinfo {author} {\bibfnamefont {J.~V.}\ \bibnamefont {Cady}},
  \bibinfo {author} {\bibfnamefont {D.~M.}\ \bibnamefont {Zajac}}, \bibinfo
  {author} {\bibfnamefont {P.~W.}\ \bibnamefont {Deelman}},\ and\ \bibinfo
  {author} {\bibfnamefont {J.~R.}\ \bibnamefont {Petta}},\ }\bibfield  {title}
  {\bibinfo {title} {Strong coupling of a single electron in silicon to a
  microwave photon},\ }\href {https://doi.org/10.1126/science.aal2469}
  {\bibfield  {journal} {\bibinfo  {journal} {Science}\ }\textbf {\bibinfo
  {volume} {355}},\ \bibinfo {pages} {156} (\bibinfo {year}
  {2017}{\natexlab{b}})},\ \bibinfo {note} {publisher: American Association for
  the Advancement of Science Section: Report}\BibitemShut {NoStop}%
\bibitem [{\citenamefont {Thorgrimsson}\ \emph {et~al.}(2017)\citenamefont
  {Thorgrimsson}, \citenamefont {Kim}, \citenamefont {Yang}, \citenamefont
  {Smith}, \citenamefont {Simmons}, \citenamefont {Ward}, \citenamefont
  {Foote}, \citenamefont {Corrigan}, \citenamefont {Savage}, \citenamefont
  {Lagally}, \citenamefont {Friesen}, \citenamefont {Coppersmith},\ and\
  \citenamefont {Eriksson}}]{thorgrimsson_extending_2017}%
  \BibitemOpen
  \bibfield  {author} {\bibinfo {author} {\bibfnamefont {B.}~\bibnamefont
  {Thorgrimsson}}, \bibinfo {author} {\bibfnamefont {D.}~\bibnamefont {Kim}},
  \bibinfo {author} {\bibfnamefont {Y.-C.}\ \bibnamefont {Yang}}, \bibinfo
  {author} {\bibfnamefont {L.~W.}\ \bibnamefont {Smith}}, \bibinfo {author}
  {\bibfnamefont {C.~B.}\ \bibnamefont {Simmons}}, \bibinfo {author}
  {\bibfnamefont {D.~R.}\ \bibnamefont {Ward}}, \bibinfo {author}
  {\bibfnamefont {R.~H.}\ \bibnamefont {Foote}}, \bibinfo {author}
  {\bibfnamefont {J.}~\bibnamefont {Corrigan}}, \bibinfo {author}
  {\bibfnamefont {D.~E.}\ \bibnamefont {Savage}}, \bibinfo {author}
  {\bibfnamefont {M.~G.}\ \bibnamefont {Lagally}}, \bibinfo {author}
  {\bibfnamefont {M.}~\bibnamefont {Friesen}}, \bibinfo {author} {\bibfnamefont
  {S.~N.}\ \bibnamefont {Coppersmith}},\ and\ \bibinfo {author} {\bibfnamefont
  {M.~A.}\ \bibnamefont {Eriksson}},\ }\bibfield  {title} {\bibinfo {title}
  {Extending the coherence of a quantum dot hybrid qubit},\ }\href
  {https://doi.org/10.1038/s41534-017-0034-2} {\bibfield  {journal} {\bibinfo
  {journal} {npj Quantum Information}\ }\textbf {\bibinfo {volume} {3}},\
  \bibinfo {pages} {1} (\bibinfo {year} {2017})},\ \bibinfo {note} {number: 1
  Publisher: Nature Publishing Group}\BibitemShut {NoStop}%
\bibitem [{\citenamefont {Dehollain}\ \emph {et~al.}(2014)\citenamefont
  {Dehollain}, \citenamefont {Muhonen}, \citenamefont {Tan}, \citenamefont
  {Saraiva}, \citenamefont {Jamieson}, \citenamefont {Dzurak},\ and\
  \citenamefont {Morello}}]{dehollain_single-shot_2014}%
  \BibitemOpen
  \bibfield  {author} {\bibinfo {author} {\bibfnamefont {J.~P.}\ \bibnamefont
  {Dehollain}}, \bibinfo {author} {\bibfnamefont {J.~T.}\ \bibnamefont
  {Muhonen}}, \bibinfo {author} {\bibfnamefont {K.~Y.}\ \bibnamefont {Tan}},
  \bibinfo {author} {\bibfnamefont {A.}~\bibnamefont {Saraiva}}, \bibinfo
  {author} {\bibfnamefont {D.~N.}\ \bibnamefont {Jamieson}}, \bibinfo {author}
  {\bibfnamefont {A.~S.}\ \bibnamefont {Dzurak}},\ and\ \bibinfo {author}
  {\bibfnamefont {A.}~\bibnamefont {Morello}},\ }\bibfield  {title} {\bibinfo
  {title} {Single-{Shot} {Readout} and {Relaxation} of {Singlet} and {Triplet}
  {States} in {Exchange}-{Coupled} {$^{31}\mathrm{P}$} {Electron} {Spins} in
  {Silicon}},\ }\href {https://doi.org/10.1103/PhysRevLett.112.236801}
  {\bibfield  {journal} {\bibinfo  {journal} {Physical Review Letters}\
  }\textbf {\bibinfo {volume} {112}},\ \bibinfo {pages} {236801} (\bibinfo
  {year} {2014})},\ \bibinfo {note} {publisher: American Physical
  Society}\BibitemShut {NoStop}%
\bibitem [{\citenamefont {Boross}\ \emph {et~al.}(2016)\citenamefont {Boross},
  \citenamefont {Széchenyi},\ and\ \citenamefont
  {Pályi}}]{boross_valley-enhanced_2016}%
  \BibitemOpen
  \bibfield  {author} {\bibinfo {author} {\bibfnamefont {P.}~\bibnamefont
  {Boross}}, \bibinfo {author} {\bibfnamefont {G.}~\bibnamefont {Széchenyi}},\
  and\ \bibinfo {author} {\bibfnamefont {A.}~\bibnamefont {Pályi}},\
  }\bibfield  {title} {\bibinfo {title} {Valley-enhanced fast relaxation of
  gate-controlled donor qubits in silicon},\ }\href
  {https://doi.org/10.1088/0957-4484/27/31/314002} {\bibfield  {journal}
  {\bibinfo  {journal} {Nanotechnology}\ }\textbf {\bibinfo {volume} {27}},\
  \bibinfo {pages} {314002} (\bibinfo {year} {2016})}\BibitemShut {NoStop}%
\bibitem [{\citenamefont {Hetényi}\ \emph {et~al.}(2019)\citenamefont
  {Hetényi}, \citenamefont {Boross},\ and\ \citenamefont
  {Pályi}}]{hetenyi_hyperfine-assisted_2019}%
  \BibitemOpen
  \bibfield  {author} {\bibinfo {author} {\bibfnamefont {B.}~\bibnamefont
  {Hetényi}}, \bibinfo {author} {\bibfnamefont {P.}~\bibnamefont {Boross}},\
  and\ \bibinfo {author} {\bibfnamefont {A.}~\bibnamefont {Pályi}},\
  }\bibfield  {title} {\bibinfo {title} {Hyperfine-assisted decoherence of a
  phosphorus nuclear-spin qubit in silicon},\ }\href
  {https://doi.org/10.1103/PhysRevB.100.115435} {\bibfield  {journal} {\bibinfo
   {journal} {Physical Review B}\ }\textbf {\bibinfo {volume} {100}},\ \bibinfo
  {pages} {115435} (\bibinfo {year} {2019})}\BibitemShut {NoStop}%
\bibitem [{\citenamefont {Huang}\ and\ \citenamefont
  {Bryant}(2018)}]{huang_spin_2018}%
  \BibitemOpen
  \bibfield  {author} {\bibinfo {author} {\bibfnamefont {P.}~\bibnamefont
  {Huang}}\ and\ \bibinfo {author} {\bibfnamefont {G.~W.}\ \bibnamefont
  {Bryant}},\ }\bibfield  {title} {\bibinfo {title} {Spin relaxation of a donor
  electron coupled to interface states},\ }\href
  {https://doi.org/10.1103/PhysRevB.98.195307} {\bibfield  {journal} {\bibinfo
  {journal} {Physical Review B}\ }\textbf {\bibinfo {volume} {98}},\ \bibinfo
  {pages} {195307} (\bibinfo {year} {2018})},\ \bibinfo {note} {publisher:
  American Physical Society}\BibitemShut {NoStop}%
\bibitem [{\citenamefont {Hung}\ \emph {et~al.}(2013)\citenamefont {Hung},
  \citenamefont {Cywinski}, \citenamefont {Hu},\ and\ \citenamefont
  {Das~Sarma}}]{hung_hyperfine_2013}%
  \BibitemOpen
  \bibfield  {author} {\bibinfo {author} {\bibfnamefont {J.-T.}\ \bibnamefont
  {Hung}}, \bibinfo {author} {\bibfnamefont {L.}~\bibnamefont {Cywinski}},
  \bibinfo {author} {\bibfnamefont {X.}~\bibnamefont {Hu}},\ and\ \bibinfo
  {author} {\bibfnamefont {S.}~\bibnamefont {Das~Sarma}},\ }\bibfield  {title}
  {\bibinfo {title} {Hyperfine interaction induced dephasing of coupled spin
  qubits in semiconductor double quantum dots},\ }\href
  {https://doi.org/10.1103/PhysRevB.88.085314} {\bibfield  {journal} {\bibinfo
  {journal} {Physical Review B}\ }\textbf {\bibinfo {volume} {88}},\ \bibinfo
  {pages} {085314} (\bibinfo {year} {2013})},\ \bibinfo {note} {publisher:
  American Physical Society}\BibitemShut {NoStop}%
\bibitem [{\citenamefont {Dutta}\ and\ \citenamefont
  {Horn}(1981)}]{RevModPhys.53.497}%
  \BibitemOpen
  \bibfield  {author} {\bibinfo {author} {\bibfnamefont {P.}~\bibnamefont
  {Dutta}}\ and\ \bibinfo {author} {\bibfnamefont {P.~M.}\ \bibnamefont
  {Horn}},\ }\bibfield  {title} {\bibinfo {title} {Low-frequency fluctuations
  in solids: $\frac{1}{f}$ noise},\ }\href
  {https://doi.org/10.1103/RevModPhys.53.497} {\bibfield  {journal} {\bibinfo
  {journal} {Rev. Mod. Phys.}\ }\textbf {\bibinfo {volume} {53}},\ \bibinfo
  {pages} {497} (\bibinfo {year} {1981})}\BibitemShut {NoStop}%
\bibitem [{\citenamefont {Paladino}\ \emph {et~al.}(2014)\citenamefont
  {Paladino}, \citenamefont {Galperin}, \citenamefont {Falci},\ and\
  \citenamefont {Altshuler}}]{paladino_1/f_2014}%
  \BibitemOpen
  \bibfield  {author} {\bibinfo {author} {\bibfnamefont {E.}~\bibnamefont
  {Paladino}}, \bibinfo {author} {\bibfnamefont {Y.}~\bibnamefont {Galperin}},
  \bibinfo {author} {\bibfnamefont {G.}~\bibnamefont {Falci}},\ and\ \bibinfo
  {author} {\bibfnamefont {B.}~\bibnamefont {Altshuler}},\ }\bibfield  {title}
  {\bibinfo {title} {1/f noise: {Implications} for solid-state quantum
  information},\ }\href {https://doi.org/10.1103/RevModPhys.86.361} {\bibfield
  {journal} {\bibinfo  {journal} {Reviews of Modern Physics}\ }\textbf
  {\bibinfo {volume} {86}},\ \bibinfo {pages} {361} (\bibinfo {year}
  {2014})}\BibitemShut {NoStop}%
\bibitem [{\citenamefont {San-Jose}\ \emph {et~al.}(2006)\citenamefont
  {San-Jose}, \citenamefont {Zarand}, \citenamefont {Shnirman},\ and\
  \citenamefont {Sch\"on}}]{PhysRevLett.97.076803}%
  \BibitemOpen
  \bibfield  {author} {\bibinfo {author} {\bibfnamefont {P.}~\bibnamefont
  {San-Jose}}, \bibinfo {author} {\bibfnamefont {G.}~\bibnamefont {Zarand}},
  \bibinfo {author} {\bibfnamefont {A.}~\bibnamefont {Shnirman}},\ and\
  \bibinfo {author} {\bibfnamefont {G.}~\bibnamefont {Sch\"on}},\ }\bibfield
  {title} {\bibinfo {title} {Geometrical spin dephasing in quantum dots},\
  }\href {https://doi.org/10.1103/PhysRevLett.97.076803} {\bibfield  {journal}
  {\bibinfo  {journal} {Phys. Rev. Lett.}\ }\textbf {\bibinfo {volume} {97}},\
  \bibinfo {pages} {076803} (\bibinfo {year} {2006})}\BibitemShut {NoStop}%
\bibitem [{\citenamefont {Connors}\ \emph {et~al.}(2019)\citenamefont
  {Connors}, \citenamefont {Nelson}, \citenamefont {Qiao}, \citenamefont
  {Edge},\ and\ \citenamefont {Nichol}}]{PhysRevB.100.165305}%
  \BibitemOpen
  \bibfield  {author} {\bibinfo {author} {\bibfnamefont {E.~J.}\ \bibnamefont
  {Connors}}, \bibinfo {author} {\bibfnamefont {J.}~\bibnamefont {Nelson}},
  \bibinfo {author} {\bibfnamefont {H.}~\bibnamefont {Qiao}}, \bibinfo {author}
  {\bibfnamefont {L.~F.}\ \bibnamefont {Edge}},\ and\ \bibinfo {author}
  {\bibfnamefont {J.~M.}\ \bibnamefont {Nichol}},\ }\bibfield  {title}
  {\bibinfo {title} {Low-frequency charge noise in si/sige quantum dots},\
  }\href {https://doi.org/10.1103/PhysRevB.100.165305} {\bibfield  {journal}
  {\bibinfo  {journal} {Phys. Rev. B}\ }\textbf {\bibinfo {volume} {100}},\
  \bibinfo {pages} {165305} (\bibinfo {year} {2019})}\BibitemShut {NoStop}%
\bibitem [{\citenamefont {Rahman}\ \emph {et~al.}(2009)\citenamefont {Rahman},
  \citenamefont {Park}, \citenamefont {Boykin}, \citenamefont {Klimeck},
  \citenamefont {Rogge},\ and\ \citenamefont
  {Hollenberg}}]{rahman_gate-induced_2009}%
  \BibitemOpen
  \bibfield  {author} {\bibinfo {author} {\bibfnamefont {R.}~\bibnamefont
  {Rahman}}, \bibinfo {author} {\bibfnamefont {S.~H.}\ \bibnamefont {Park}},
  \bibinfo {author} {\bibfnamefont {T.~B.}\ \bibnamefont {Boykin}}, \bibinfo
  {author} {\bibfnamefont {G.}~\bibnamefont {Klimeck}}, \bibinfo {author}
  {\bibfnamefont {S.}~\bibnamefont {Rogge}},\ and\ \bibinfo {author}
  {\bibfnamefont {L.~C.~L.}\ \bibnamefont {Hollenberg}},\ }\bibfield  {title}
  {\bibinfo {title} {Gate-induced \$g\$-factor control and dimensional
  transition for donors in multivalley semiconductors},\ }\href
  {https://doi.org/10.1103/PhysRevB.80.155301} {\bibfield  {journal} {\bibinfo
  {journal} {Physical Review B}\ }\textbf {\bibinfo {volume} {80}},\ \bibinfo
  {pages} {155301} (\bibinfo {year} {2009})},\ \bibinfo {note} {publisher:
  American Physical Society}\BibitemShut {NoStop}%
\bibitem [{\citenamefont {Lutchyn}\ \emph {et~al.}(2008)\citenamefont
  {Lutchyn}, \citenamefont {Cywinski}, \citenamefont {Nave},\ and\
  \citenamefont {Das~Sarma}}]{lutchyn_quantum_2008}%
  \BibitemOpen
  \bibfield  {author} {\bibinfo {author} {\bibfnamefont {R.~M.}\ \bibnamefont
  {Lutchyn}}, \bibinfo {author} {\bibfnamefont {L.}~\bibnamefont {Cywinski}},
  \bibinfo {author} {\bibfnamefont {C.~P.}\ \bibnamefont {Nave}},\ and\
  \bibinfo {author} {\bibfnamefont {S.}~\bibnamefont {Das~Sarma}},\ }\bibfield
  {title} {\bibinfo {title} {Quantum decoherence of a charge qubit in a
  spin-fermion model},\ }\href {https://doi.org/10.1103/PhysRevB.78.024508}
  {\bibfield  {journal} {\bibinfo  {journal} {Physical Review B}\ }\textbf
  {\bibinfo {volume} {78}},\ \bibinfo {pages} {024508} (\bibinfo {year}
  {2008})},\ \bibinfo {note} {publisher: American Physical Society}\BibitemShut
  {NoStop}%
\bibitem [{\citenamefont {Huang}\ and\ \citenamefont
  {Hu}(2014)}]{huang_electron_2014}%
  \BibitemOpen
  \bibfield  {author} {\bibinfo {author} {\bibfnamefont {P.}~\bibnamefont
  {Huang}}\ and\ \bibinfo {author} {\bibfnamefont {X.}~\bibnamefont {Hu}},\
  }\bibfield  {title} {\bibinfo {title} {Electron spin relaxation due to charge
  noise},\ }\href {https://doi.org/10.1103/PhysRevB.89.195302} {\bibfield
  {journal} {\bibinfo  {journal} {Physical Review B}\ }\textbf {\bibinfo
  {volume} {89}},\ \bibinfo {pages} {195302} (\bibinfo {year}
  {2014})}\BibitemShut {NoStop}%
\bibitem [{\citenamefont {Culcer}\ \emph {et~al.}(2009)\citenamefont {Culcer},
  \citenamefont {Hu},\ and\ \citenamefont {Sarma}}]{culcer_dephasing_2009}%
  \BibitemOpen
  \bibfield  {author} {\bibinfo {author} {\bibfnamefont {D.}~\bibnamefont
  {Culcer}}, \bibinfo {author} {\bibfnamefont {X.}~\bibnamefont {Hu}},\ and\
  \bibinfo {author} {\bibfnamefont {S.~D.}\ \bibnamefont {Sarma}},\ }\bibfield
  {title} {\bibinfo {title} {Dephasing of {Si} spin qubits due to charge
  noise},\ }\href {https://doi.org/10.1063/1.3194778} {\bibfield  {journal}
  {\bibinfo  {journal} {Applied Physics Letters}\ }\textbf {\bibinfo {volume}
  {95}},\ \bibinfo {pages} {073102} (\bibinfo {year} {2009})},\ \bibinfo {note}
  {arXiv: 0906.4555}\BibitemShut {NoStop}%
\bibitem [{\citenamefont {Yang}\ \emph {et~al.}(2019)\citenamefont {Yang},
  \citenamefont {Coppersmith},\ and\ \citenamefont
  {Friesen}}]{yang_high-fidelity_2019}%
  \BibitemOpen
  \bibfield  {author} {\bibinfo {author} {\bibfnamefont {Y.-C.}\ \bibnamefont
  {Yang}}, \bibinfo {author} {\bibfnamefont {S.~N.}\ \bibnamefont
  {Coppersmith}},\ and\ \bibinfo {author} {\bibfnamefont {M.}~\bibnamefont
  {Friesen}},\ }\bibfield  {title} {\bibinfo {title} {High-fidelity
  single-qubit gates in a strongly driven quantum-dot hybrid qubit with 1/f
  charge noise},\ }\href {https://doi.org/10.1103/PhysRevA.100.022337}
  {\bibfield  {journal} {\bibinfo  {journal} {Physical Review A}\ }\textbf
  {\bibinfo {volume} {100}},\ \bibinfo {pages} {022337} (\bibinfo {year}
  {2019})}\BibitemShut {NoStop}%
\bibitem [{\citenamefont {Kubo}(1962)}]{kubo_generalized_1962}%
  \BibitemOpen
  \bibfield  {author} {\bibinfo {author} {\bibfnamefont {R.}~\bibnamefont
  {Kubo}},\ }\bibfield  {title} {\bibinfo {title} {Generalized {Cumulant}
  {Expansion} {Method}},\ }\href {https://doi.org/10.1143/JPSJ.17.1100}
  {\bibfield  {journal} {\bibinfo  {journal} {Journal of the Physical Society
  of Japan}\ }\textbf {\bibinfo {volume} {17}},\ \bibinfo {pages} {1100}
  (\bibinfo {year} {1962})}\BibitemShut {NoStop}%
\end{thebibliography}%

\newpage 
\appendix
\onecolumngrid
\section{One Qubit Eigenbasis}
    The dressed flip-flop qubit energies, up to second order, are:
    \begin{subequations}
    \begin{align}
        E_0 &= \frac{1}{2}(-\omega_0-\omega_B)-\frac{A}{8} \left(1-\cos \eta\right)-\frac{\Delta\omega_B}{4}\left(1+\cos \eta\right)  \nonumber\\
        &\qquad-\frac{A^2}{16\omega_B}\left((1-\cos\eta)^2+\sin^2\eta(\frac{\omega_B}{4\omega_0}+\frac{\omega_B}{\omega_0+\omega_B}-\frac{\omega_B\Delta\omega_B}{A\omega_0}+\frac{\omega_B\Delta\omega_B^2}{A^2\omega_0})\right)\\
        E_1 &= \frac{1}{2}(-\omega_0+\omega_B)-\frac{A}{8} \left(1-\cos \eta\right)+\frac{\Delta\omega_B}{4}\left(1+\cos \eta\right)\nonumber\\
        &\qquad-\frac{A^2}{16\omega_B}\left(-(1-\cos\eta)^2+\sin^2\eta(\frac{\omega_B}{4\omega_0}+\frac{\omega_B}{\omega_0-\omega_B}+\frac{\omega_B\Delta\omega_B}{A\omega_0}+\frac{\omega_B\Delta\omega_B^2}{A^2\omega_0})\right) \\
        E_2 &= \frac{1}{2}( \omega_0-\omega_B)-\frac{A}{8} \left(1+\cos \eta\right)-\frac{\Delta\omega_B}{4}\left(1-\cos \eta\right)\nonumber\\
        &\qquad-\frac{A^2}{16\omega_B}\left((1+\cos\eta)^2+\sin^2\eta(-\frac{\omega_B}{4\omega_0}-\frac{\omega_B}{\omega_0-\omega_B}+\frac{\omega_B\Delta\omega_B}{A\omega_0}-\frac{\omega_B\Delta\omega_B^2}{A^2\omega_0})\right) \\
        E_3 &= \frac{1}{2}( \omega_0+\omega_B)-\frac{A}{8} \left(1+\cos \eta\right)+\frac{\Delta\omega_B}{4}\left(1-\cos \eta\right)\nonumber\\
        &\qquad-\frac{A^2}{16\omega_B}\left(-(1+\cos\eta)^2+\sin^2\eta(-\frac{\omega_B}{4\omega_0}-\frac{\omega_B}{\omega_0+\omega_B}-\frac{\omega_B\Delta\omega_B}{A\omega_0}-\frac{\omega_B\Delta\omega_B^2}{A^2\omega_0})\right)
    \end{align}
\end{subequations}
    and corresponding states, up to first order, are:
    \begin{subequations}
    \begin{align}
        \ket{0} &= \ket{g\downarrow} - \frac{A}{4\omega_B}(1-\cos\eta) \ket{g\uparrow} - \frac{A\sin\eta}{8\omega_0}(1-2\Delta\omega_B/A)\ket{e\downarrow}+ \frac{A\sin\eta}{4(\omega_0+\omega_B)} \ket{e\uparrow} \\
        \ket{1} &= \frac{A}{4\omega_B}(1-\cos\eta)\ket{g\downarrow} + \ket{g\uparrow} + \frac{A\sin\eta}{4(\omega_0-\omega_B)} \ket{e\downarrow} -\frac{A\sin\eta}{8\omega_0}(1+2\Delta\omega_B/A) \ket{e\uparrow} \\
        \ket{2} &= \frac{A\sin\eta}{8\omega_0}(1-2\Delta\omega_B/A)\ket{g\downarrow} -\frac{A\sin\eta}{4(\omega_0-\omega_B)} \ket{g\uparrow} + \ket{e\downarrow} -\frac{A}{4\omega_B}(1+\cos\eta)\ket{e\uparrow} \\
        \ket{3} &= -\frac{A\sin\eta}{4(\omega_0+\omega_B)}\ket{g\downarrow} +\frac{A\sin\eta}{8\omega_0}(1+2\Delta\omega_B/A) \ket{g\uparrow} + \frac{A}{4\omega_B}(1+\cos\eta)\ket{e\downarrow} + \ket{e\uparrow}
    \end{align}
\end{subequations}
\section{$Z$ operator coefficients}
    When expressing the electron position operator in the flip-flop eigenbasis:
    \begin{equation}
        Z = \sum_{ij} z_{ij}\sigma'_i\tau'_j
    \end{equation}
    the relevant coefficients, along with the primary physical mechanism they are responsible for, are, 
    \begin{subequations}
    \begin{align}
        z_{03} &= A^2\omega_0^3\cos\eta\sin^2\eta/4\omega_B(\omega_0^2-\omega_B^2)^2                &\textrm{Spin dephasing}\\
        z_{10} &= \sin\eta + A\cos\eta\sin\eta/4\omega_0                                              &\textrm{Charge flip}\\
        z_{30} &= \cos\eta - A\sin^2\eta/4\omega_0                                                    &\textrm{Charge dephasing}\\
        z_{31} &= A\omega_0\sin^2\eta/2(\omega_0^2-\omega_B^2)                                        &\textrm{Spin flip}\\
        z_{33} &= \Delta\omega_B\sin^2\eta/2\omega_0                                                       &\textrm{Spin and charge dephasing}\\
        z_{11} &= -\frac{A\omega_0\cos\eta\sin\eta}{2(\omega_0^2-\omega_B^2)}                         &\textrm{Charge and spin flip}\\
        z_{22} &= z_{11}\cdot\frac{\omega_0}{\omega_B}                                                     &\textrm{Charge and spin flip} \\
        z_{01} &= -\frac{A\omega_0\Delta\omega_B\cos\eta\sin^2\eta}{4\omega_B(\omega_0^2-\omega_B^2)} &\textrm{Spin flip} \\
        z_{13} &= -\frac{\Delta\omega_B\cos\eta\sin\eta}{2\omega_0}                                        &\textrm{Charge flip}
    \end{align}
\end{subequations}
    Any remaining terms are zero. 
    In our case, since the charge excited states are also our leakage states, all of the charge flip terms are synonymous with leakage.

    $z_{13}$ and $z_{01}$ are generally several orders of magnitude smaller than their counterparts so we will neglect them. Thus, for two-qubit coupling, the dominant terms are $z_{10}$ and $z_{31}$ and to a lesser extent, $z_{11}+z_{22}$. 
    
    The $z_{11}$ and $z_{22}$ terms play an important role when the applied electric field is large. At that point, our charge and spin qubits are very well separated and the flip-flop qubit can be thought as a nearly pure spin qubit. However, since we are more interested in the regions near the flip-flop sweet spot and charge sweet spots, these terms are also relatively small compared to the $z_{10}$ and $z_{31}$ terms and were also neglected in the analytical results in the main work. For a general solution that doesn't omit these terms, see appendix section~\ref{general_model}.

\section{General Time Evolution of the Density Matrix}
    The equation of motion in the interaction picture for a Hamiltonian, $H(t) = H_0 + H_n(t)$, is given by
    \begin{equation}
        i\hbar \frac{d}{dt}\rho^I = \mathcal{L}(t)\rho^I
    \end{equation}
    where $\rho^I=U_0^\dagger\rho U_0$, $H_n^I(t) = U_0^\dagger H_n(t) U_0$, $\mathcal{L}(t)\rho^I \equiv \left[H_n^I(t),\rho^I\right]$. This can be solved by means of cumulant expansion~\cite{yang_high-fidelity_2019,kubo_generalized_1962},
    \begin{equation}
        \langle\rho^I(t)\rangle =  \exp\left(\sum_{n=1}\frac{(-i)^n}{\hbar^n}\int_0^tdt_1 \dots\int_0^{t_n-1}dt_n\langle \mathcal{L}(t_1)\dots\mathcal{L}(t_n)\rangle \right)\rho^I(0)
    \end{equation}
    Further details can be found in reference~\cite{yang_high-fidelity_2019}. For a Hamiltonian written in the form
    \begin{equation*}
        H(t) = H_0 + \sum_i f_i(t)h_i
    \end{equation*}
        the noise-averaged density matrix in Liouville-Fock space can then be expressed up to second order as
    \begin{equation}
        \langle\vec{\rho}(t)\rangle = \sum_{jk} e^{-i\omega_{jk}t}(R\otimes R)(\ket{j}\bra{j}\otimes\ket{k}\bra{k})e^{-\sum_iK_i(t)}(R^{-1}\otimes R^{-1})\vec{\rho}(0)
    \end{equation}
    $K_i(t)$ describes how the noise affects the system and is given by:
    \begin{equation}
        K_i(t) = \frac{1}{\hbar^2}\int_{t_0}^tdt_1\int_{t_0}^{t_1}dt_2 S_i(t_1-t_2) \mathcal{L}_i(t_1) \mathcal{L}_i(t_2)
        \label{Ki}
    \end{equation}
    Substituting the definition for $\mathcal{L}$ into equation~\ref{Ki} and transforming back into the Schroedinger picture yields an expression for $K_i(t)$ in terms of the noise Hamiltonians $h_i' = R^{-1}h_iR$ in the eigenbasis of $H_0$.
    \begin{multline}
        K_i(t) =\frac{1}{\hbar^2}\sum_{abc}(h'_{i,ab}h'_{i,bc}J_i(t,\omega_{ba},\omega_{cb})\ket{a}\bra{c}\otimes 1+h'_{i,ba}h'_{i,cb}J_i(t,\omega_{ab},\omega_{bc}) 1\otimes \ket{a}\bra{c}) \\
    -\frac{1}{\hbar^2}\sum_{abcd}h'_{i,ac}h'_{i,db} (J_i(t,\omega_{ca},\omega_{db})+J_i(t,\omega_{db},\omega_{ca})) \ket{a}\bra{c}\otimes \ket{b}\bra{d}
    \label{eq_Kit}
    \end{multline}
    $J_i(t,\omega_1,\omega_2)$ describes the decay profile of the system and is defined as
    \begin{equation}
        J_i(t,\omega_1,\omega_2) = \int_0^tdt_1\int_0^{t_1}dt_2 S_i(t_1-t_2)e^{i\omega_1t_2}e^{i\omega_2t_2}
    \end{equation}

    \begin{figure}
    \centering
    \includegraphics[width=0.5\columnwidth]{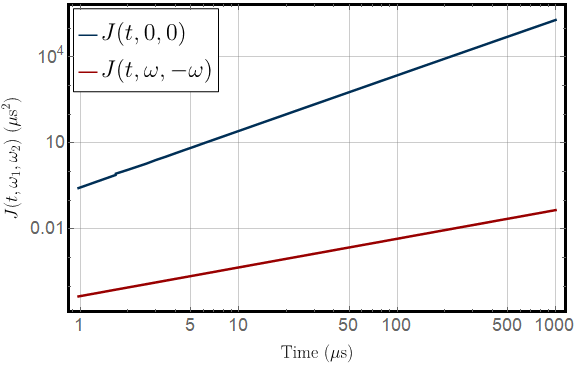}
    \caption{$J(t,\omega_1,\omega_2)$ for $\omega_1=\omega_2=0$ and $\omega_1=-\omega_2=2\pi*\SI{10}{GHz}$ for a $1/f$ noise spectrum. $J(t,0,0)$ is several orders of magnitude larger than $J(t,\omega,-\omega)$ justifying the approximation made in section~\ref{time-evo}.}
    \label{J}
\end{figure} 

    Calculating the time evolution of the system now comes down to simply solving for the eigenenergies and eigenstates of the noiseless Hamiltonian, $H_0$, and using that to obtain the noise Hamiltonians, $h_i'$, in the eigenbasis of the noiseless system.


    \section{Generic Four-Level Model}
    \label{general_model}
    \begin{figure}
    \centering
    \includegraphics[width=0.5\columnwidth]{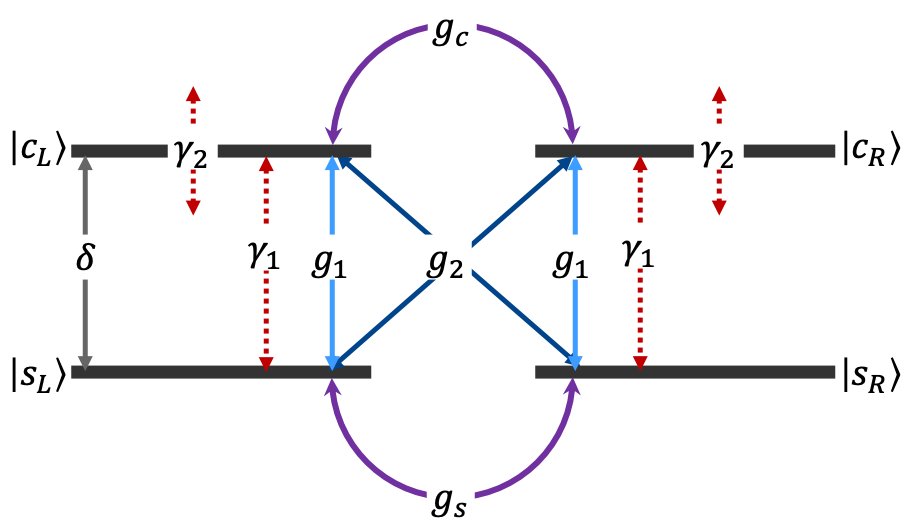}
    \caption{Schematic diagram of a generic four level, doubly degenerate system with energy separation $\delta$. The four states are coupled together by various coupling factors $g_s$, $g_c$, $g_1$, and $g_2$. The system is also subject to classical noise. The noise can induce transitions within each two-level subspace via $\gamma_1$ and are also the source of energy fluctuations via $\gamma_2$.}
    \label{four_level_diagram}
\end{figure}
    In this section, we set out to solve a generic four level system consisting of pairs of degenerate states separated by some energy difference $\delta$ as shown in figure~\ref{four_level_diagram}. We label these states $\ket{s_L}$, $\ket{s_R}$, $\ket{c_L}$, and $\ket{c_R}$. The $\ket{s}$ ($\ket{c}$) states are coupled to each other via $g_{s}$ ($g_{c}$). We also include on-site excitations via $g_1$ and cross-site excitations via $g_2$.
    The Hamiltonian is
    \begin{align}
        H_0&=-\frac{1}{2}\delta \sigma_z + \frac{1}{2}g_s(1+\sigma_z)\tau_x + \frac{1}{2}g_c(1-\sigma_z)\tau_x + g_1\sigma_x + g_2\sigma_x\tau_x =\left( \begin{array}{cccc}
                0 & g_s & g_{1} & g_{2} \\
                g_s & 0 & g_{2} & g_{1} \\
                g_{1} &g_{2} & \delta & g_c \\
                g_{2} & g_{1} & g_c & \delta
         \end{array}\right)\, ,
     \end{align}

    For our work on the two-qubit dipole coupled system, this model corresponds with the spin excited ($\ket{01}$ and $\ket{10}$) and the charge excited ($\ket{02}$ and $\ket{20}$) subspaces. We can also envision this generic model describing other systems. For example, this model can also be used to describe an electron in a symmetric quantum double dot.

    We also define two uncorrelated sources of noise.
    \begin{subequations}
        \begin{align}
            h_L &= \hbar\gamma_1 (\ket{s_L}\bra{c_L} +\ket{c_L}\bra{s_L})-\hbar\gamma_2 \cdot \mathrm{diag} (0,0,2,0)  \\
            h_R &= \hbar\gamma_1 (\ket{s_R}\bra{c_R} +\ket{c_R}\bra{s_R})-\hbar\gamma_2 \cdot \mathrm{diag} (0,0,0,2) \, ,
        \end{align}
    \end{subequations}
    where $\gamma_1$ can cause transitions between the ground and excited states on each side and $\gamma_2$ causes energy fluctuations and ultimately dephasing.

    When matched to our flip-flop system, the two $s$ states would be equivalent to the qubit states $\ket{01}$ and $\ket{10}$ while the two $c$ are the charge leakage states $\ket{02}$ and $\ket{20}$. Assuming symmetric biasing on both donors, $\delta$ is then the difference in energy between the charge and flip-flop qubits, $E_2-E_1$. $g_s$ would then be the dipole induced flip-flop coupling, $V_{dd}z_{31}^2$, $g_{1}$ and $g_{2}$ the leakage coupling strengths, $V_{dd}(z_{11}+z_{22})(z_{03}+z_{30}+z_{33})$ and $V_{dd}z_{31}z_{10}$, respectively, and $g_c$ the charge coupling,$V_{dd}z_{10}^2$. For the flip-flop system, typically $g_c \approx 10g_{2} \approx 100g_s$. At small applied electric field biases, $g_{2}\gg g_{1}$. While large applied fields can cause $g_{1}$ to be larger $g_{2}$, for simplicity we will restrict ourselves to parameter regimes that fall under the former condition and neglect $g_{1}$. At the single qubit sweet spot, $g_c \approx \delta$. The strength of the noise in the flip-flip qubit is $\gamma_2 = \omega_n (z_{30}-z_{03})/2$ and $\gamma_1 = \omega_n (z_{11}+z_{22})/2$.

    Both operational times and decoherence times for this general Hamiltonian can be determined by defining mixing angles $\phi_\pm$.
    \begin{equation}
        \tan\phi_\pm = \frac{-2(g_{1}\pm g_{2})}{-\delta\mp(g_c-g_s)}
    \end{equation}
    \begin{figure}
    \centering
    \includegraphics[width=0.9\columnwidth]{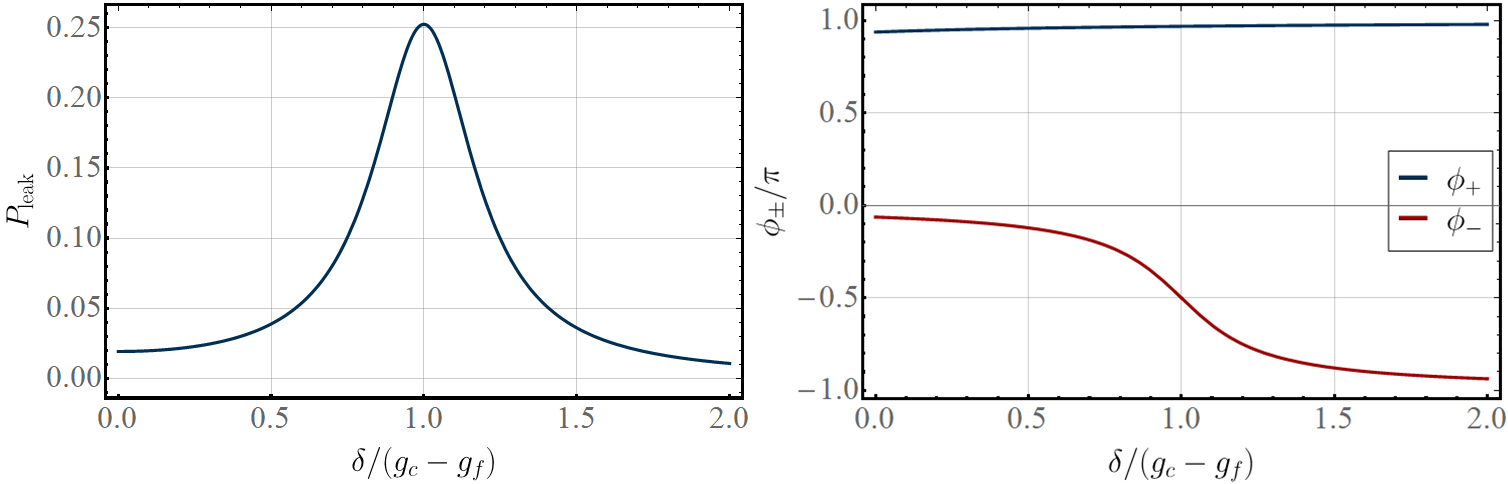}
    \caption{Long term leakage and two-qubit mixing angles $\phi_\pm$ for $g_{2} / (g_c-g_f) = 0.1$ and $g_{1}=0$.}
    \label{phi}
\end{figure}
    Neglecting the effect of the noise, the time evolution of this system is solvable exactly. The rotation matrix to diagonalize $H_0$ is:
    \begin{equation}
        R=\frac{1}{\sqrt{2}}\left( \begin{array}{cccc}
                 \cos\frac{\phi_+}{2} &  \cos\frac{\phi_-}{2} & \sin\frac{\phi_+}{2} &  \sin\frac{\phi_-}{2} \\
                 \cos\frac{\phi_+}{2} & -\cos\frac{\phi_-}{2} & \sin\frac{\phi_+}{2} & -\sin\frac{\phi_-}{2} \\
                -\sin\frac{\phi_+}{2} & -\sin\frac{\phi_-}{2} & \cos\frac{\phi_+}{2} &  \cos\frac{\phi_-}{2} \\
                -\sin\frac{\phi_+}{2} &  \sin\frac{\phi_-}{2} & \cos\frac{\phi_+}{2} & -\cos\frac{\phi_-}{2}
         \end{array}\right)
     \end{equation}
    
    This system evolves with oscillations about the equilibrium state at 6 fundamental frequencies (and their negatives):
    \begin{subequations}
        \begin{align}
            \hbar\omega_a = -\hbar\omega_{-a} &= \frac{1}{2}\delta(\cos\phi_+ - \cos\phi_-) - \frac{1}{2}g_c(2 - \cos\phi_+ - \cos\phi_-) \nonumber \\
            &\qquad + g_{2}(\sin\phi_+ + \sin\phi_-) - \frac{1}{2}g_s(2 + \cos\phi_+ + \cos\phi_-) + g_{1}(\sin\phi_+ - \sin\phi_-) \\
            \hbar\omega_b = -\hbar\omega_{-b} &= \frac{1}{2}\delta(\cos\phi_+ + \cos\phi_-) - \frac{1}{2}g_c(2 - \cos\phi_+ + \cos\phi_-)\nonumber \\
            &\qquad + g_{2}(\sin\phi_+ - \sin\phi_-) - \frac{1}{2}g_s(2 + \cos\phi_+ - \cos\phi_-)+ g_{1}(\sin\phi_+ + \sin\phi_-)\\
            \hbar\omega_c = -\hbar\omega_{-c} &= \frac{1}{2}\delta(\cos\phi_+ + \cos\phi_-) + \frac{1}{2}g_c(2 + \cos\phi_+ - \cos\phi_-) \nonumber \\
            &\qquad+ g_{2}(\sin\phi_+ - \sin\phi_-) + \frac{1}{2}g_s(2 - \cos\phi_+ + \cos\phi_-)+ g_{1}(\sin\phi_+ + \sin\phi_-)\\
            \hbar\omega_d = -\hbar\omega_{-d} &= \frac{1}{2}\delta(\cos\phi_+ - \cos\phi_-) + \frac{1}{2}g_c(2 + \cos\phi_+ + \cos\phi_-) \nonumber \\
            &\qquad+ g_{2}(\sin\phi_+ + \sin\phi_-) + \frac{1}{2}g_s(2 - \cos\phi_+ - \cos\phi_-)+ g_{1}(\sin\phi_+ - \sin\phi_-) \\
            \hbar\omega_e = -\hbar\omega_{-e} &= \delta\cos\phi_+ + g_c\cos\phi_+ + 2g_{2} \sin\phi_+ - g_s\cos\phi_+ + 2g_{1}\sin\phi_+ \\
            \hbar\omega_f = -\hbar\omega_{-f} &= \delta\cos\phi_- - g_c\cos\phi_- - 2g_{2} \sin\phi_- + g_s\cos\phi_- + 2g_{1}\sin\phi_-
        \end{align}
    \end{subequations}
    with their amplitudes depending on which density matrix element of interest. In general, these amplitudes are given by
    \begin{equation}
        C_{ab,jk}=\sum_{mn}R_{aj}R_{jm}^{-1}R_{bk}R_{kn}^{-1}\rho_{mn}(0)
    \end{equation}
    The summation of all terms with $j=k$ yield the long term equilibrium value for matrix element $\rho_{ab}$. Terms where $j\ne k$ can all be associated with one of the above frequencies.

    As an example, if we initialize the system to be in state $\ket{\Psi(0)} = \alpha\ket{f_L} + \beta\ket{f_R}$ with real $\alpha$ and $\beta$, the population of the $\ket{f_L}$ state (i.e. $\rho_{00}(t)$), oscillates at the above frequencies with amplitudes:
    \begin{subequations}
        \begin{align}
            C_{0000,a}=C_{0000,-a} &= \frac{1}{4}(\alpha^2-\beta^2)\cos^2(\phi_+/2)\cos^2(\phi_-/2)\\
            C_{0000,b}=C_{0000,-b} &= \frac{1}{4}(\alpha^2-\beta^2)\cos^2(\phi_+/2)\sin^2(\phi_-/2)\\
            C_{0000,c}=C_{0000,-c} &= \frac{1}{4}(\alpha^2-\beta^2)\sin^2(\phi_+/2)\cos^2(\phi_-/2)\\
            C_{0000,d}=C_{0000,-d} &= \frac{1}{4}(\alpha^2-\beta^2)\sin^2(\phi_+/2)\sin^2(\phi_-/2)\\
            C_{0000,e}=C_{0000,-e} &= \frac{1}{16}(\alpha+\beta)^2\sin^2\phi_+\\
            C_{0000,f}=C_{0000,-f} &= \frac{1}{16}(\alpha-\beta)^2\sin^2\phi_-
        \end{align}
    \end{subequations}

    When noise is included, each of these oscillating terms will decay with rates
    \begin{subequations}
        \begin{align}
            \Gamma_a &= \gamma_2(\cos\phi_+ - \cos\phi_-)/2 + \gamma_1(\sin\phi_+-\sin\phi_-)/2\\
            \Gamma_b &= \gamma_2(\cos\phi_+ + \cos\phi_-)/2 - \gamma_1(\sin\phi_++\sin\phi_-)/2\\
            \Gamma_c &= \gamma_2(\cos\phi_+ + \cos\phi_-)/2 - \gamma_1(\sin\phi_++\sin\phi_-)/2\\
            \Gamma_d &= \gamma_2(\cos\phi_+ - \cos\phi_-)/2 + \gamma_1(\sin\phi_+-\sin\phi_-)/2\\
            \Gamma_e &= \gamma_2\cos\phi_+-\gamma_1\sin\phi_+\\
            \Gamma_f &= \gamma_2\cos\phi_--\gamma_1\sin\phi_-
        \end{align}
    \end{subequations}
    Overall, the time evolution can be written as
    \begin{equation}
        \rho_{ab} = \rho_{ab}(t\rightarrow\infty) + \sum_{j={a..f}}\left(C_{ab,j}\exp(-i\omega_jt)+C_{ab,-j}\exp(i\omega_jt)\right)\exp(-2J(t,0,0)\Gamma_j^2)
    \end{equation}
    Notice that only the amplitudes depend on the particular matrix element while the frequencies and decay rates are common for each.
    
    If the two lower energy states (the $\ket{s}$ states) are part of the qubit logical basis, the amount of leakage (population of the $\ket{c}$ states), we can expect as $t\rightarrow\infty$ is equal to
    \begin{equation}
        P_\textrm{leak}(t\rightarrow\infty) = \frac{1}{4}\left( \alpha\beta(\cos2\phi_--\cos2\phi_+) + \sin^2\phi_++\sin^2\phi_-\right)
        \label{leakage}
    \end{equation}

    \begin{figure}
    \centering
    \includegraphics[width=0.95\columnwidth]{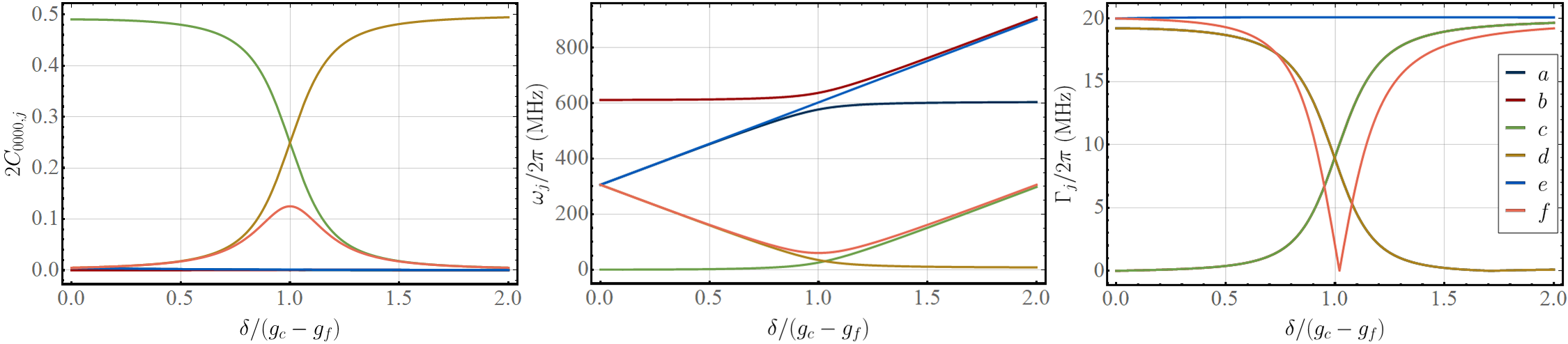}
    \caption{Oscillation amplitude, frequency and decay rate for $g_{2} / (g_c-g_s) = 0.1$, $\gamma_1/2\pi=\SI{20}{MHz}$, $\gamma_2/2\pi=\SI{0}{MHz}$, $(g_c-g_s)/2\pi= \SI{300}{MHz}$, $g_{1}=0$, and initial condition $\ket{\Psi(0)} = \ket{f_L}$. We see that for small $\delta$, the dominant oscillator is $c$ which has small frequency but also a slow decay rate. Similarly, for large $\delta$, the dominant oscillator is $d$, which also has a small frequency and slow decay rate. When $\delta/(g_c-g_s)$ is $1$, we see two main oscillators with the same amplitude. In addition, their frequencies are just slightly off resonance with each other, creating beats in the oscillations. The problem is that at this point, they also have non-zero decay rates. A non-zero $\gamma_2$ breaks the symmetry of the $f$-line and shifts the valley slightly.}
    \label{osc}
\end{figure}

    Now we can look at several different parameter regimes.

    \emph{Case 1: $\delta \gg (g_c-g_s) > g_{2}$, weak coupling}. In the weak coupling regime, the two leakage states are well separated from the qubit states. The two mixing angles both approach $\pi$. In this particular case, there is only one relevant frequency, $\omega_d$, and decay rate $\Gamma_d$, which needs to be expanded in a series to obtain a non-zero value.
    \begin{equation}
        P_{fL}(t) = \frac{1}{2}\left(1 + (\alpha^2-\beta^2)\cos(2g_st/\hbar)\exp\left(-2J(t,0,0)\left(\frac{4(g_c-g_s)g_{2}^2\omega_n}{\delta^3}\right)^2\right)\right)
    \end{equation}
    In this regime, the time is for an $X$ gate is
    \begin{equation}
        T_{g} = \frac{\hbar\pi}{2g_s}
    \end{equation}
    and quality factor
    \begin{equation}
        Q = (\Gamma T_g)^{-1} = \frac{\delta^3g_s}{2\pi\hbar\gamma_2(g_c-g_s)g_{2}^2}
    \end{equation}
    In addition to large quality factor, we also have minimal leakage. The expected long time leakage is
    \begin{equation}
        P_\textrm{leak} = 2g_{2}^2/\delta^2
    \end{equation}

    \emph{Case 2: $\delta = (g_c-g_s) > g_{2}$, resonance}. In this regime, $\phi_+$ still approaches $\pi$, but $\phi_-$ becomes exactly equal to $-\pi/2$. Now the expression for $P_{fL}(t)$ becomes
    \begin{multline}
        P_{fL}(t) = \frac{3}{8} + \frac{1}{8}(\alpha-\beta)^2\cos(2g_{2}t/\hbar)\exp\left( -2J(t,0,0) \gamma_1^2 \right) \\
        + \frac{1}{4}(\alpha^2-\beta^2)\cos((2g_{2}-3g_s)t/2\hbar)\exp\left( -2J(t,0,0) (\gamma_2-\gamma_1)^2/4 \right) \\
        + \frac{1}{4}(\alpha^2-\beta^2)\cos((2g_{2}+3g_s)t/2\hbar)\exp\left( -2J(t,0,0) (\gamma_2+\gamma_1)^2/4 \right)
    \end{multline}
    This expression indicates beats in the evolution with beat frequencies $3g_s$ and $2g_{2}$. The beats will decay away however due to the different decay rates. To estimate a quality factor in this limit, we use the slower beat frequency along with the average for the two decay rates, which is roughly just the larger of $\gamma_2$ and $\omega_2$. The quality factor in this regime is then approximately
    \begin{equation}
        Q = \frac{3g_s}{\pi\hbar\gamma_2}
    \end{equation}
    This does not necessarily mean that in this regime, the gate time is expected to to be the same as in case 1. If $g_{2}$ is sufficiently large compared to $g_s$, we can instead use the faster frequency to determine the gate time depending on the fidelity desired. This regime has a long time leakage of
    \begin{equation}
        P_\textrm{leak} = \frac{1}{4}(1-2\alpha\beta)
    \end{equation}

    \emph{Case 3: $\delta \ll (g_c-g_s) > g_{2}$, strong coupling}. The angle $\phi_+$ moves toward $\tan^{-1}(2g_{2}/(g_c-g_s))$ and $\phi_-$ tends toward $\tan^{-1}(-2g_{2}/(g_c-g_s))$. As a simple approximation for $g_c-g_s>g_{2}$., we use $\phi_+ = \pi$ and $\phi_-=0$ to obtain for the time evolution,
    \begin{equation}
        P_{fL}(t) = \frac{1}{2}\left(1+(\alpha^2-\beta^2)\cos(2g_st/\hbar)\exp\left( -2J(t,0,0)\left( \frac{4g_{2}^2\delta}{(4g_{2}^2 + (g_c-g_s)^2)^{3/2}} \right)^2 \right)\right)
    \end{equation}
    This yields a quality factor of
    \begin{equation}
        Q = \frac{(4g_{2}^2 + (g_c-g_s)^2)^{3/2}g_s}{2\pi\hbar\gamma_2\delta}
    \end{equation}
    While this does look promising, there remains a problem. And that is leakage. Going back to equation~\ref{leakage}, we can see that the leakage is non-zero when we do not use the above simplification. We'll be seeing a leakage of about
    \begin{equation}
        P_\textrm{leak} = \frac{2g_{2}^2}{4g_{2}^2+(g_c-g_s)^2}
    \end{equation}
    This could be much worse as the detuning, $\delta$, is increased and approaches the resonant condition described in case 2.

    \section{Density Matrix Off-Diagonal Elements}
    \begin{figure*}[thb]
    \centering
    \includegraphics[width=\textwidth]{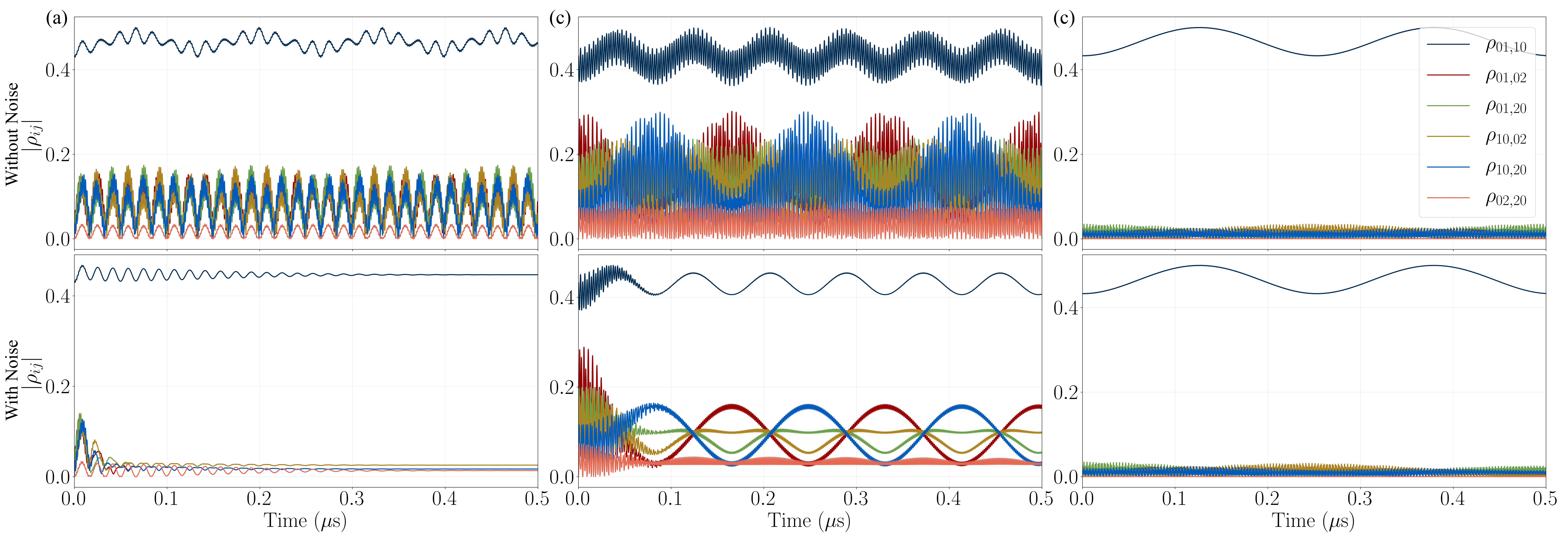}
    \caption{Time evolution for the off-diagonal matrix elements two qubit density matrix with parameters marked in Figure~\ref{RelaxTime}. The system is initialized to state $\ket{\Psi(0)}=\frac{\sqrt{3}}{2}\ket{01}+\frac{1}{2}\ket{10}$. (a) $B=\SI{0.796}{T}$, $E_z-E_c=\SI{3.13}{V/cm}$. (b) $B=\SI{0.806}{T}$, $E_z-E_c=\SI{0.95}{V/cm}$. (c) $B=\SI{0.771}{T}$, $E_z-E_c=\SI{0}{}$.}
    \label{2Q_Evolution_OffDiag}
\end{figure*}

    In this section, we'll look at the off-diagonal element between the states $\ket{01}$ and $\ket{10}$. The amplitudes for this element are:     
    \begin{subequations}
        \begin{align}
            C_{0110,a}=-C_{0110,-a} &= +\frac{1}{4}(\alpha^2-\beta^2)\cos^2(\phi_+/2)\cos^2(\phi_-/2)\\
            C_{0110,b}=-C_{0110,-b} &= +\frac{1}{4}(\alpha^2-\beta^2)\cos^2(\phi_+/2)\sin^2(\phi_-/2)\\
            C_{0110,c}=-C_{0110,-c} &= -\frac{1}{4}(\alpha^2-\beta^2)\sin^2(\phi_+/2)\cos^2(\phi_-/2)\\
            C_{0110,d}=-C_{0110,-d} &= -\frac{1}{4}(\alpha^2-\beta^2)\sin^2(\phi_+/2)\sin^2(\phi_-/2)\\
            C_{0110,e}=+C_{0110,-e} &= +\frac{1}{16}(\alpha+\beta)^2\sin^2\phi_+\\
            C_{0110,f}=+C_{0110,-f} &= -\frac{1}{16}(\alpha-\beta)^2\sin^2\phi_-
        \end{align}
    \end{subequations}
    Due to the opposing signs of the first four terms, these are responsible for the imaginary part of the matrix element. Conversely, the remaining two are responsible for the real part. Since the magnitudes of these terms remain unchanged, the analysis for the relaxation rates in the main work can be similarly applied for the dephasing rates as shown in figure~\ref{2Q_Evolution_OffDiag} which shows similar decay behavior as that shown in figure~\ref{2Q_Evolution}.

\section{Long Time Decay}
    \begin{figure*}[thb]
    \centering
    \includegraphics[width=\textwidth]{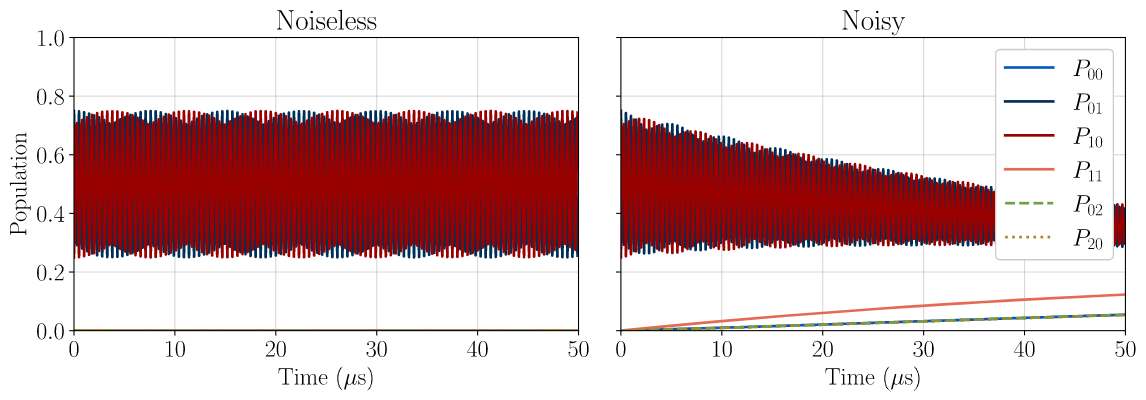}
    \caption{Long time evolution for the two qubit system. The system is initialized to state $\ket{\Psi(0)}=\frac{\sqrt{3}}{2}\ket{01}+\frac{1}{2}\ket{10}$ and parameters set to $B=\SI{0.771}{T}$, $E_z-E_c=\SI{0}{}$, and $V_t = \SI{47.15}{\mu eV}$, the same as point C in Figures~\ref{RelaxTime} and~\ref{2Q_Evolution}. The relaxation at this time scale is due to noise-induced charge and spin flips.}
    \label{2Q_Evolution_LongTime}
\end{figure*}  

In order to study the slow decay due to noise induced transitions, we must relax a couple of previously made approximations. 

First, we include additional terms in the noise Hamiltonians.        
\begin{equation}
    h_{i} = \frac{1}{2}\omega_{ni}\left( z_{30}\sigma_{zi}'+z_{03}\tau_{zi}'+z_{33}\sigma_{zi}'\tau_{zi}'+z_{11}\sigma_{xi}'\tau_{xi}'+z_{22}\sigma_{yi}'\tau_{yi}' +z_{10}\sigma_{xi}' + z_{31}\sigma_{zi}'\tau_{xi}' \right) \,.
\end{equation}
The first five terms are the same as in equation~\ref{eq_noise_hamiltonian} of the main text. The remaining two terms are responsible for the noise induced charge and spin flips, respectively. 

Secondly, in equation~\ref{eq_Kit}, we keep $J(t,\omega_1,\omega_2)$ terms where $\omega_1=-\omega_2\ne 0$ in addition to the ones where $\omega_1=\omega_2=0$. This allows for the noise to directly induce transitions between states rather than just cause dephasing within the system eigenstates.

We calculate an approximate decoherence rate for this decay channel
\begin{equation}
    \frac{1}{T_1} \approx \omega_n^2(z_{10}^2/\omega_0 + z_{31}^2/\omega_B) \,.
\end{equation}

\end{document}